\documentclass[traditabstract]{aa}

\usepackage[pdftex,pdfpagemode={UseOutlines},bookmarks,bookmarksopen,colorlinks,linkcolor={blue},citecolor={blue},urlcolor={red}]{hyperref}
\usepackage{txfonts}
\usepackage[table]{xcolor}
\usepackage{longtable}
\usepackage{pdflscape}
\usepackage[maxfloats=110]{morefloats}
\usepackage{graphicx}
\usepackage{hyperref}
\usepackage{subcaption}
\usepackage{amssymb}
\usepackage{pifont}

\usepackage{natbib}
\bibpunct{(}{)}{;}{a}{}{,} 
\usepackage{epsfig}

\newcommand{\twopartdef}[4]
{
	\left\{
		\begin{array}{ll}
			#1 & \mbox{if } #2 \\ \\
			#3 & \mbox{if } #4
		\end{array}
	\right.
}

\begin{document}
\rowcolors{2}{gray!25}{white}
\title{The EBLM Project\thanks{Based on photometric observations with the SuperWASP and SuperWASP-South instruments and radial velocity measurement from the CORALIE spectrograph, mounted on the Swiss 1.2\,m \textit{Euler} Telescope, located at ESO, La Silla, Chile. The data is publicly available at the \textit{CDS} Strasbourg and on demand to the main author.}
\\ {\Large IV. Spectroscopic orbits of over 100 eclipsing M dwarfs \\masquerading as transiting hot-Jupiters}
}
\author{Amaury H.M.J. Triaud\inst{1}
\and David V. Martin\inst{2}
\and Damien S\'egransan \inst{2}
\and Barry Smalley\inst{3}
\and Pierre F.L. Maxted\inst{3}
\and \\David R. Anderson\inst{3}
\and Fran\c cois Bouchy\inst{2}
\and Andrew Collier Cameron\inst{4}
\and Francesca Faedi\inst{5}
\and Yilen G\'omez Maqueo Chew\inst{6}
\and Leslie Hebb\inst{7}
\and Coel Hellier\inst{3}
\and Maxime Marmier\inst{2}
\and Francesco Pepe\inst{2}
\and Don Pollacco\inst{5}
\and Didier Queloz\inst{8}
\and St\'ephane Udry \inst{2}
\and Richard West\inst{5}
}

\offprints{aht34@cam.ac.uk}

\institute{Institute of Astronomy, Madingley Road, Cambridge CB3 0HA, UK
\and Observatoire Astronomique de l'Universit\'e de Gen\`eve, Chemin des Maillettes 51, CH-1290 Sauverny, Switzerland
\and Astrophysics Group, Keele University, Staffordshire, ST5 5BG, UK
\and SUPA, School of Physics \& Astronomy, University of St Andrews, North Haugh, KY16 9SS, St Andrews, Fife, Scotland, UK
\and Department of Physics, University of Warwick, Coventry CV4 7AL, UK
\and Instituto de Astronom\'ia, Universidad Nacional Aut\'onoma de M\'exico, Ciudad Universitaria, Ciudad de M\'exico, 04510, M\'exico
\and Hobart and William Smith Colleges, Department of Physics, Geneva, NY 14456, USA
\and Cavendish Laboratory, J J Thomson Avenue, Cambridge, CB3 0HE, UK
}

\date{Received date / accepted date}
\authorrunning{Triaud et al.}
\titlerunning{Orbits of EBLM}

\abstract{{ We present 2271 radial velocity measurements taken on 118 single-line binary stars, taken over eight years with the CORALIE spectrograph. The binaries consist of F/G/K primaries and M-dwarf secondaries. They were initially discovered photometrically by the WASP planet survey, as their shallow eclipses mimic a hot-Jupiter transit. The observations we present permit a precise characterisation of the binary orbital elements and mass function. With modelling of the primary star this mass function is converted to a mass of the secondary star. In the future, this spectroscopic work will be combined with precise photometric eclipses to draw an empirical mass/radius relation for the bottom of the mass sequence. This has applications in both stellar astrophysics and the growing number of exoplanet surveys around M-dwarfs. In particular, we have discovered { 34} systems with a secondary mass below $0.2 M_\odot$, and so we will ultimately double the known number of very low-mass stars with well characterised mass and radii.

The quality of our data combined with the amplitude of the Doppler variations mean that we are able to detect eccentricities as small as 0.001 and orbital periods to sub-second precision. Our sample can revisit some earlier work on the tidal evolution of close binaries, extending it to low mass ratios. We find some exceptional binary systems that are eccentric at orbital periods below three days, while our longest circular orbit has a period of 10.4 days. Amongst our systems, we note one remarkable architecture in J1146-42, that boasts three stars within one astronomical unit.

By collating the EBLM binaries with published WASP planets and brown dwarfs, we derive a mass spectrum with twice the resolution of previous work. We compare the WASP/EBLM sample of tightly-bound orbits with work in the literature on more distant companions up to 10 AU. We note that the brown dwarf desert appears wider, as it carves into the planetary domain for our short-period orbits. This would mean that a significantly reduced abundance of planets begins at $\sim 3M_{\rm Jup}$, well before the Deuterium-burning limit. This may shed light on the formation and migration history of massive gas giants.}

 }

\maketitle

\section{Introduction}

Hot-Jupiters are a rare type of exoplanet existing around 0.5 to 1\% of solar-type stars \citep{Mayor:1995uq,Howard:2012qy,Santerne:2016lr}. Their sizes range from $\sim 0.7 R_{\rm Jup}$ to around $ 2 R_{\rm Jup}$ \citep{Sato:2005qo,Anderson:2010fj}, a range in size which also corresponds to late M dwarfs with masses lower than $\sim 0.2 M_\odot$ \citep{Chabrier:1997uq,Baraffe:1998ly,Baraffe:2015lr,Dotter:2008yq,Demory:2009ys,Diaz:2014lr,Hatzes:2015rz,Chen:2016lr}. To further the comparison, gas giants, brown dwarfs and very low-mass stars also have have similar temperatures in addition to similar sizes. Hot-Jupiters' dayside temperatures range { from $\sim 800$ to up to $\sim4,600$ K \citep{Triaud:2015fk,Gaudi:2017bp}}. This means that all those objects share a similar parameter space in colour-magnitude diagrams \citep{Triaud:2014kq}, with several planets re-emitting more flux than some stars. This easily explains how photometric surveys designed to detect transiting gas giants also net within their data a large number of low-mass stellar secondary companions, which we here report upon. All of our systems were identified with the CORALIE spectrograph, while distinguishing which of many candidates provided by the Wide Angle Search for Planets (WASP\footnote{\url{http://wasp-planets.net}.}; \citealt{Pollacco:2006fj}) were indeed planets. Our paper is part of an ongoing investigation on eclipsing binaries with low mass (EBLM) following { three previous instalments, which focused on four specific targets \citep{Triaud:2013lr,Gomez-Maqueo-Chew:2014jk,Boetticher:2017un}.}

{ The} main objective { of the EBLM project} is to empirically measure the mass/radius relationship at the bottom of the main sequence, and compare it to theoretical expectations \citep{Chabrier:1997uq,Baraffe:1998ly,Baraffe:2015lr,Dotter:2008yq}. This can be done by making careful measurements of the ratio of sizes of the two components during eclipse, and of their mass function thanks to radial velocities. Assuming parameters for the primaries, as is often done for exoplanetary studies, we can derive accurate physical parameters for the secondaries. Information about the primaries will soon be refined thanks to {\it GAIA}'s parallaxes \citep{de-Bruijne:2012qy}, as well as possibly thanks to asteroseismologic measurements collected by the forthcoming {\it PLATO} \citep{Rauer:2014qq}.

The photometric identification of an eclipsing low-mass star versus a transiting gas-giant is similar. We will use our sample of low-mass eclipsing secondaries to serve as comparison sample to the gas-giants discovered by WASP. We also aim to revisit the relative frequencies of low-mass stars to planets, to brown dwarfs in order to confirm the presence, extent and dryness of the brown dwarf desert \citep{Marcy:2000yf,Grether:2006kx,Sahlmann:2011qy,Ma:2014sf}. We will compare the stellar and planetary eccentricity/period and eccentricity/mass distributions in order to study tides \citep{Zahn:1989fj,Mathieu:1990fk,Mazeh:1997rw,Terquem:1998jk,Meibom:2005kx,Milliman:2014fj}, investigate whether the statistics on the presence of additional perturbing bodies are similar \citep{Tokovinin:2006la,Knutson:2014fg,Neveu-VanMalle:2016xy}, and find out if the spin--orbit misalignments frequently observed in hot Jupiter systems are also observed in binary stars thanks to the Rossiter--McLaughlin effect \citep{Hale:1994fk,Triaud:2010fr,Albrecht:2014lr,Esposito:2014lr,Lendl:2014yu,Winn:2015lr}.

Another important goal is to warn fellow planet hunters operating within the celestial Southern Hemisphere (HAT-South \citep{Bakos:2013qf}, KELT \citep{Pepper:2012ul}, ASTEP \citep{Crouzet:2010ve}, NGTS \citep{Wheatley:2013zl}, {\it K2} \citep{Howell:2014lq}, {\it TESS} \citep{Ricker:2014qv} where are located many systems most likely to masquerade as hot-Jupiters.  

{ We emphasise that whilst all of these binaries were discovered by WASP photometrically in eclipse, these data are not present in this paper, nor are any secondary radii that may be inferred from them. The low precision and presence of some mis-understood systematics mean that presenting the WASP photometric data on its own would be misleading\footnote{For all of the published hot-Jupiters in WASP and the few EBLM binaries published in \citet{Triaud:2013lr,Gomez-Maqueo-Chew:2014jk,Boetticher:2017un} improved transits/eclipses had been obtained with larger telescopes.}. }However, taking the time to follow-up each of these photometrically is beyond the manpower of our team. While we intend to follow some systems (and have already for a few), we cannot do everything. This release is therefore an opportunity for the community to help characterise the mass/radius relationship of the smallest stars within our Galaxy by collecting high quality photometric data during primary { and secondary eclipses. In particular, the { 34} secondaries with }masses below $0.2 M_{\odot}$ { will ultimately double} the number of known objects in that part of the mass-radius diagram.

Late M dwarfs form the bulk of the stellar population \citep{Chabrier:2003fk,Henry:2006uq}. There exist a number of dedicated surveys to discover planets around small stars, such as MEarth \citep{Nutzman:2008qy} and Apache \citep{Giacobbe:2012qy}, with new surveys such as SPECULOOS coming online shortly \citep{Gillon:2013qy}. As planets are known to orbit these stars, they will likely reveal the most frequent pathway to planet formation. The bottom of the main sequence is also where it is most optimal to discover and to study the atmospheres of Earth-sized worlds \citep[e.g.][]{de-Wit:2016fk,He:2017lq}, with the recently discovered TRAPPIST-1, a particularly suited example \citep{Gillon:2016gh,Gillon:2017yq,Luger:2017fk}.

In the following section, we will describe our instrument, how we selected the targets and how the observations were finally obtained. In section~\ref{sec:data_reduction} we present our data reduction, and the treatment leading to the radial velocities and their uncertainties. Section~\ref{sec:fitting} details how we adjust a Keplerian model to the radial velocities, with section~\ref{sec:model_selection} focusing on our model selection when several appear compatible with the data. Our results appear in section~\ref{sec:results}, and our interpretations in section~\ref{sec:tides}. There are extensive appendices containing tables supporting the main text, as well as a graphical representation of the orbits for each of the 118 systems that we announce here.

\begin{figure}
\captionsetup[subfigure]{labelformat=empty}
\begin{center}
\begin{subfigure}[b]{0.49\textwidth}
\includegraphics[width=\textwidth,trim={0 0 0 0},clip]{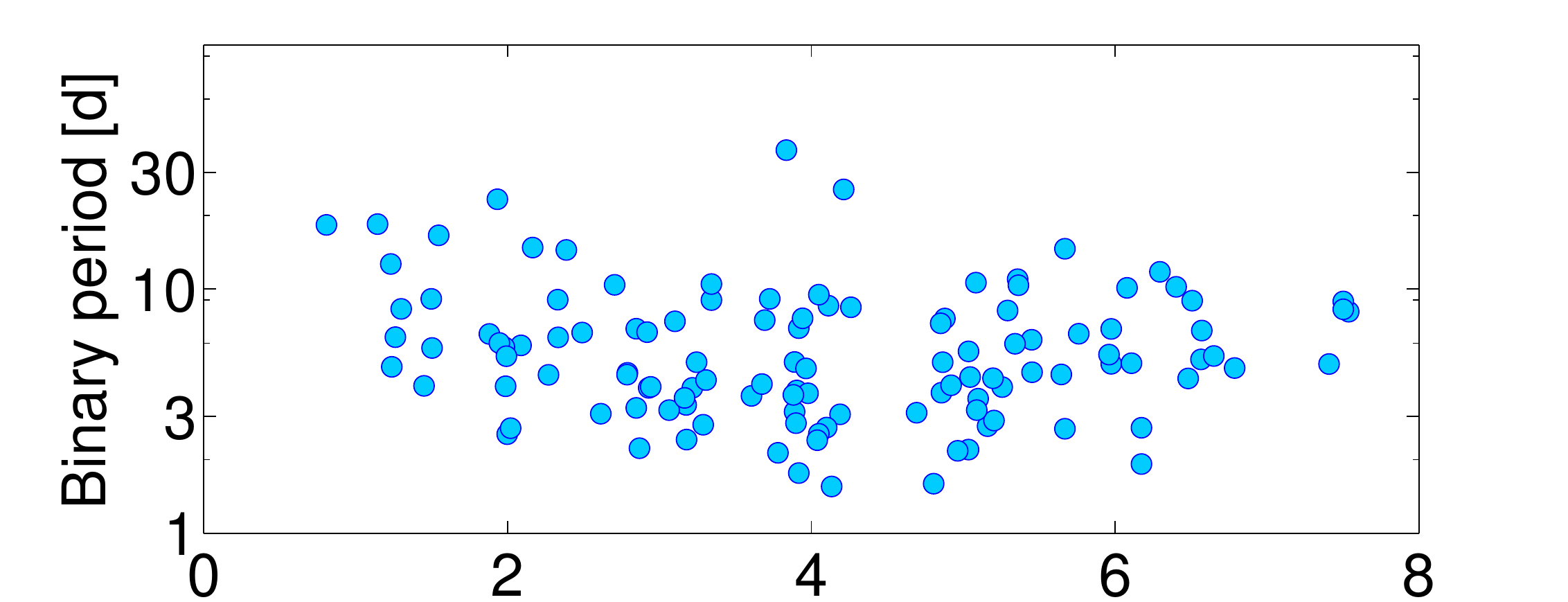}
\end{subfigure}
\begin{subfigure}[b]{0.49\textwidth}
\includegraphics[width=\textwidth,trim={0 0 0 0},clip]{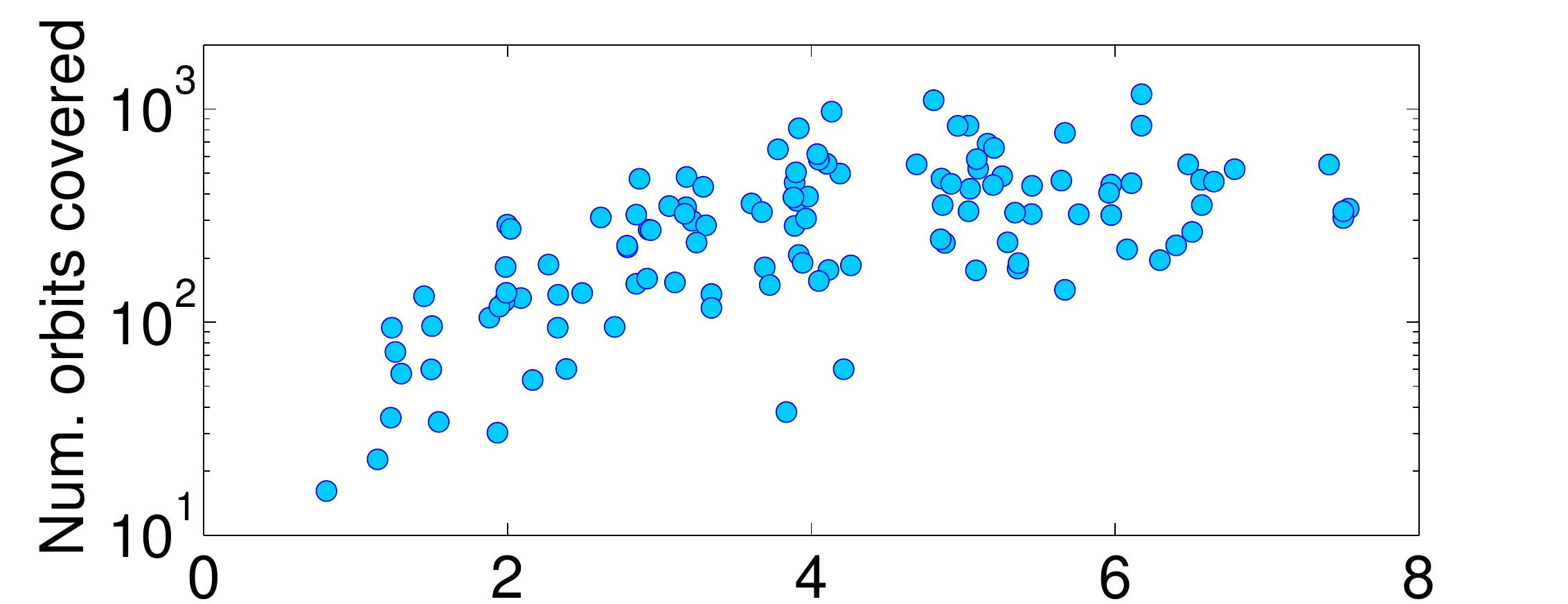}
\end{subfigure}
\begin{subfigure}[b]{0.49\textwidth}
\includegraphics[width=\textwidth,trim={0 0 0 0},clip]{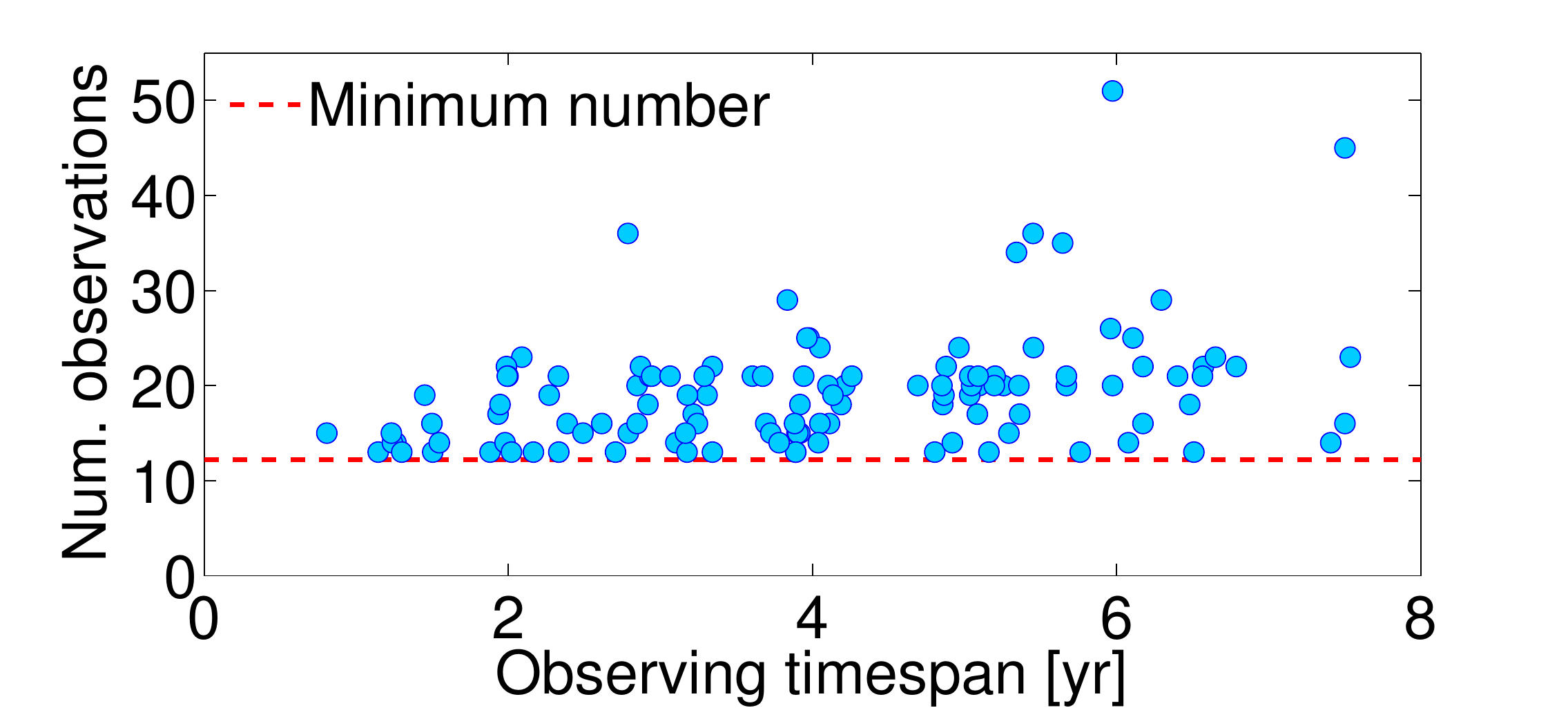}
\end{subfigure}
\caption{Top: timespan of observations over binaries of different periods. The flat distribution means that our sensitivity to tertiary objects is roughly independent of the binary period. Middle: number of orbits covered for each binary. There is no strict minimum although 86\% of binaries have been observed over a timespan covering at least 100 orbits. Bottom: number of observations given to each binary. There is a requirement of at least 13 observations.}
\label{fig:timespan}
\end{center}
\end{figure}

\section{Observational Campaign}\label{sec:obs}

\subsection{The CORALIE instrument}\label{subsec:inst}
CORALIE \citep{Queloz:2001ty}, is a thermally stabilised (but not pressure-stabilised), high-resolution, fibre-fed spectrograph, mounted on the 1.2m {\it Euler} Telescope, a facility belonging to the University of Geneva and installed at ESO's observatory of La Silla, in Chile. The spectrograph was built on ELODIE's design \citep{Baranne:1996qa}, which was installed on the 1.93m at OHP and produced the first radial-velocity detection of an exoplanet \citep{Mayor:1995uq}. The wavelength solution is obtained by simultaneously illuminating a CCD detector with the star, and with a thorium-argon calibration lamp \citep{Lovis:2007ec}. Radial velocities are extracted by cross-correlating the observed spectrum with a numerical mask. The resulting cross-correlation function (CCF) is fitted with a Gaussian profile whose mean corresponds to the radial velocity.
 
In 2007, CORALIE received a major upgrade allowing it to be more efficient and appropriate for the detection of gas-giants orbiting star as faint as V $\sim 13$ \citep{Wilson:2008lr,Segransan:2010vn}. CORALIE has a resolution of order 55\,000. Since 2007, we have announced in excess of a 100 transiting planets in collaboration with WASP \citep[e.g.][]{Turner:2016sf}, with several dozens remaining in preparation. Further improvements to the instrument were conducted in November 2014 (change from circular to octagonal fibres), and in April 2015 (wavelength solution now done using a Fabry-P\'erot). The first of these two operations produced a small offset in the zero-point of the instrument, of order 10 m\,s$^{-1}$, which remains irrelevant for the precision we obtained on the binary star sample but which needs to be accounted for when deriving orbits for stars with planets \citep{Triaud:2016fp}.

\subsection{Target selection and observing campaign}\label{subsec:sample}

\begin{figure}
\captionsetup[subfigure]{labelformat=empty}
\begin{center}
\begin{subfigure}[b]{0.49\textwidth}
\includegraphics[width=\textwidth,trim={0 0 0 0},clip]{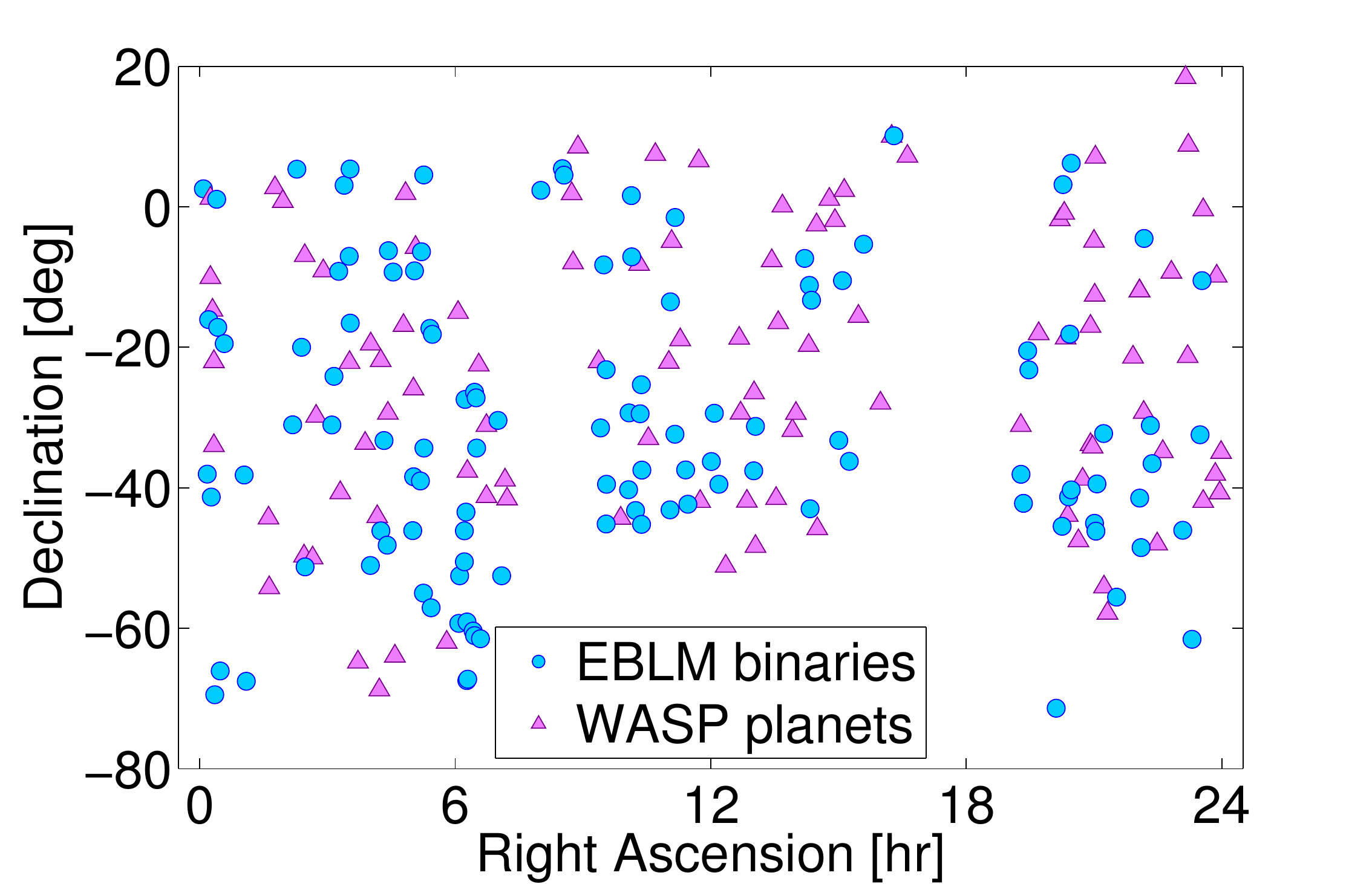}
\end{subfigure}
\begin{subfigure}[b]{0.49\textwidth}
\includegraphics[width=\textwidth,trim={0 0 0 0},clip]{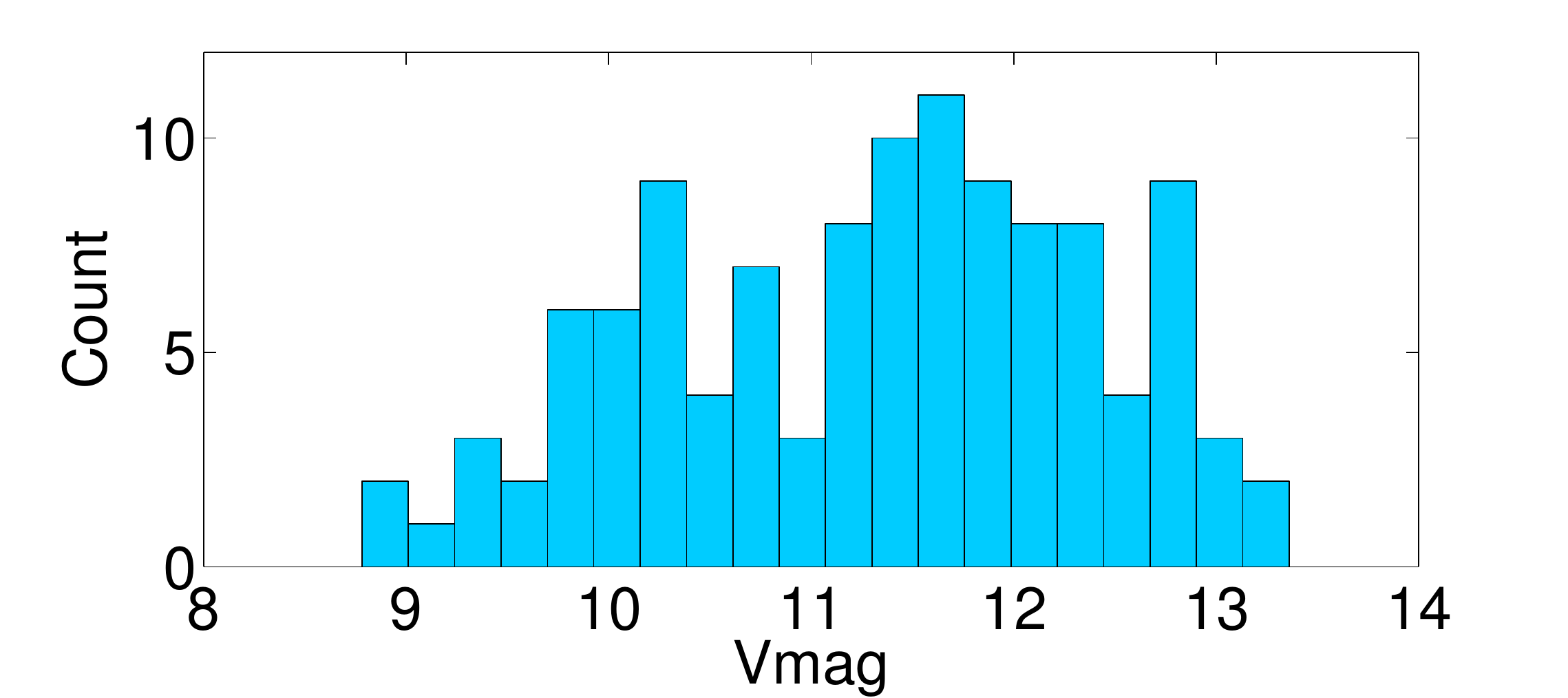}
\end{subfigure}
\caption{Top: coordinates of each of the EBLM binaries (blue circles) and a comparative distribution of the coordinates of the WASP planets (purple triangles). Bottom: histogram of the visual magnitude.}
\label{fig:coordinates}
\end{center}
\end{figure}

Stars showing periodic photometric dimmings consistent with the transit of a hot Jupiter are identified by WASP using an algorithm named {\sc hunter} \citep{Collier-Cameron:2007pb}. An analysis of the stellar colours, of their reduced proper-motion and of the duration of the events permits an exclusion of most giant primaries, as well as a preliminary estimate of the stellar radius ($R_\star$). The depth of the event, ($D = R^2_{\rm p}/R^2_\star$) leads to an estimate of the transiter's size. If its radius is consistent with $R_{\rm p} \leq 2.1 R_{\rm Jup}$ and no ellipsoidal variation is initially detected then the object is kept and becomes a planet candidate.

The spectroscopic validation of candidates with declination $\delta < +10^\circ$ is done using CORALIE \citep{Triaud:2011qy}. We start with two exposures of 1800 seconds, { timed to be near the expected radial velocity maximum and minimum}. Any amount of radial velocity variation is investigated (even if anti-phased) until the nature of the variation is understood (wrong period from WASP, long period binaries, stellar activity, chance alignment with another eclipsing system, EBLM, etc.). These two spectra are taken on every star except if on the first attempt we detect a secondary set of lines, in which case we classify this object as a double-line binary (SB2), which is no longer observed.

If there is a radial-velocity variation of less than 100 km s$^{-1}$ between the two first epochs, but in excess of order 5-6 km s$^{-1}$, we classify the object temporarily as part of the EBLM project. Systems { with} lower variations are followed-up intensively as planetary or brown dwarf candidates, and systems above the criterion are discarded. { An amplitude of 100 km s}$^{-1}$ corresponds approximately to a secondary mass of 0.6 $M_\odot$, for an orbital period of 15 days about a $1 M_\odot$ primary. These requirements therefore contain all the secondaries that we could possibly be interested in.

Figures~\ref{fig:timespan} \& \ref{fig:coordinates} graphically represent several characteristics of this current data release. We decided to include all EBLM candidates for which we had at least 13 radial velocity measurement by 2016-03-14, and where the orbital period derived from radial velocities is consistent with that derived from photometry (i.e. all are confirmed to be eclipsing). 
Thirteen measurements correspond to the bare minimum necessary to adjust up to two Keplerian models through the data, although most of the time this is not needed. Usually, the 11+ measurements that complement the first pair, were obtained at reduced exposures of 600 and 900 seconds (since the semi-amplitudes are much larger than planets), and as high a precision is not required. Many systems received more visits for a variety of reasons including: detection of a tertiary companion, testing whether our uncertainties on periods and eccentricities are robustly determined, and a limited attempt to detect circumbinary objects.
Observations were spread over more than three years for the majority of systems, which are mostly contained between V magnitudes 9 and 13, just like the hot-Jupiters we identified. There is a spread in the timespan between roughly one and eight years, because new targets were provided by WASP for spectroscopic follow-up progressively. The amount of time spent on a given target is roughly independent of the binary period, so as to limit any potential biases. We also present the amount of orbits covered (timespan / $P$), indicating that a large majority of targets have been monitored for over 100 orbits. This is important for identifying stellar trends like activity, as well as radial velocity drifts induced by a tertiary star. The EBLMs are spread almost uniformly across the Southern skies in declination and right ascension, with the exception of the galactic plane ($\alpha \sim 6-7$h and $\alpha \sim 16-17$h), which has not been observed by WASP due to heavy stellar crowding.



\begin{figure}
\captionsetup[subfigure]{labelformat=empty}
\begin{center}
\begin{subfigure}[b]{0.49\textwidth}
\includegraphics[width=\textwidth,trim={0 0 0 0},clip]{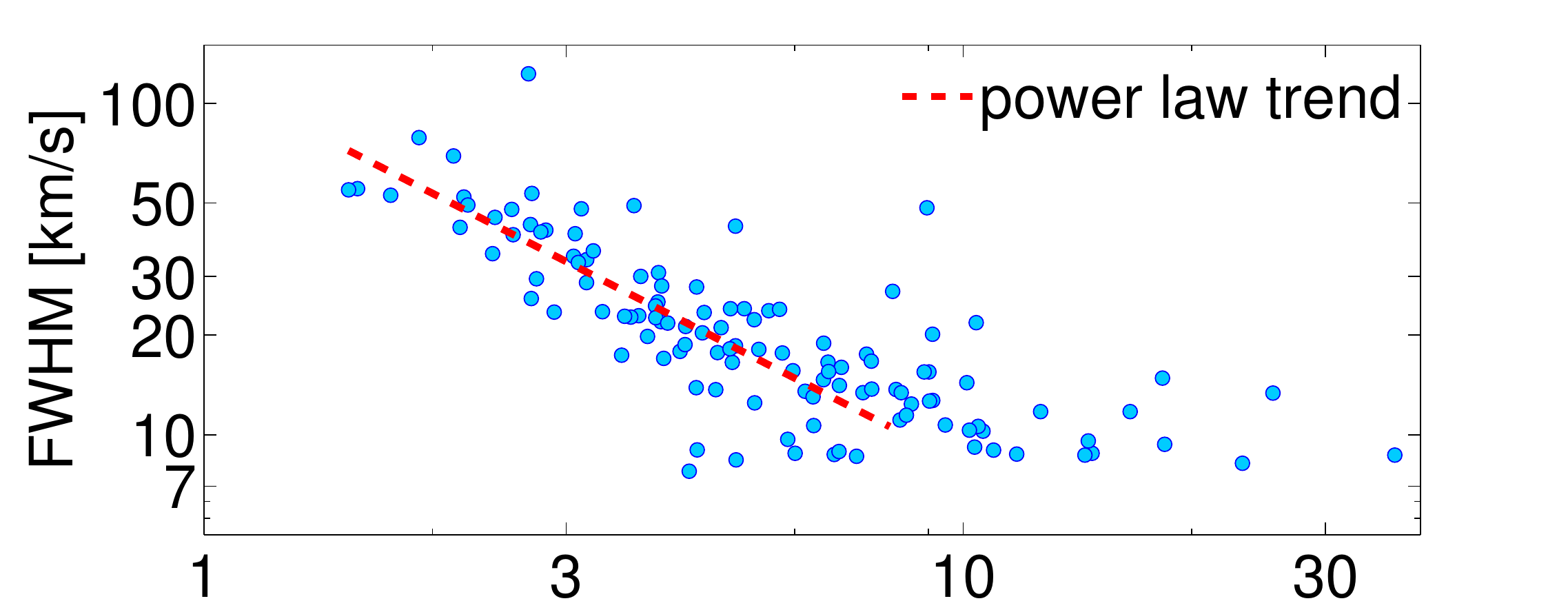}
\end{subfigure}
\begin{subfigure}[b]{0.49\textwidth}
\includegraphics[width=\textwidth,trim={0 0 0 0},clip]{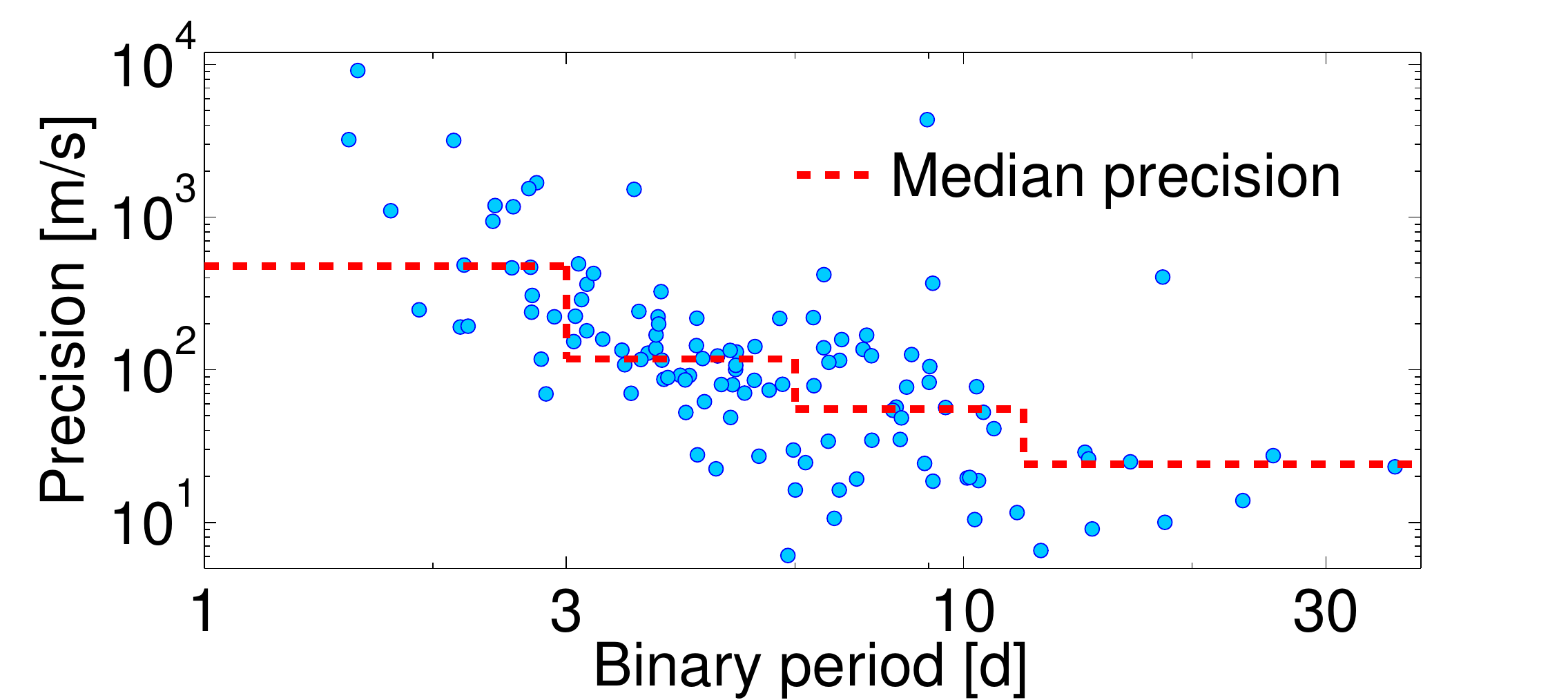}
\end{subfigure}
\caption{Top: FWHM of the CCF as a function of binary period and a red dashed trend fitted to the data for $P<8$ d. Roughly beyond this period the primary stars are rotating slow enough such that the instrumental broadening dominates the rotational broadening, and truncates any potentially continued trend below 7 km s$^{-1}$. Bottom: precision of the radial velocity measurements for each binary as a function of binary period. The red dashed line shows the median precision for all binaries within four coarse period bins, chosen for the later study of triple systems in Sect.~\ref{subsec:triples}.}
\label{fig:precision}
\end{center}
\end{figure}

Across all 118 targets, we get a median precision of 107~m~s$^{-1}$. We calculated these values by taking the median photon noise error on the measurements for each system, and quadratically adding an extra term, $\sigma_{\rm add}$ whose estimation is explained in 
Sect.~\ref{sec:data_reduction}. The precision obtained is not uniform with binary period, as shown in the bottom plot of Fig.~\ref{fig:precision}. The precision of our radial velocity measurements tends to be worse for shorter binary periods, because these stars are forced to rotate synchronously with their orbital periods, owing to tidal forces; this leads to broadened spectral lines. We verify this by plotting the full width at half maximum (FWHM) of the CCF, in the top of Fig.~\ref{fig:precision} where it can be seen to increase with decreasing binary period. The FWHM that we measure has two dominant components, and is defined as:
\begin{equation}
{\rm FWHM} \simeq \sqrt{{\rm FWHM}_{\rm rot}^2 + {\rm FWHM}_{\rm inst}^2},
\end{equation}
where ${\rm FWHM}_{\rm rot}$ is the broadening of the absorption lines (and consequently of the CCF) as caused by the rotation of the star, while ${\rm FWHM}_{\rm inst}$ is the instrumental broadening of CORALIE, which depends on its resolution. For CORALIE, ${\rm FWHM}_{\rm inst}\sim 7$, which sets the minimum observable FWHM in Fig.~\ref{fig:precision}. The FWHM$_{\rm rot}$ increases as the orbital period decreases for objects where the primary star's rotation period is tidally synchronised to the orbital period of its secondary. This effect saturates at FWHM $\sim 7$ below which we cannot reliably measure the { primaries'} rotational broadening. 

\subsection{Determination of the primaries' effective temperatures and masses}\label{subsec:primary_masses}

{ We have used the empirical colour -- effective temperature from
\citet{Boyajian:2013jh}  to estimate the effective temperatures of the
primary stars in these binary systems. We extracted
photometry for each target from the following catalogues -- B$_{\rm T}$ and
V$_{\rm T}$ magnitudes from the Tycho-2 catalogue \citep{Hog:2000uq};
B, V, g$^{\prime}$, r$^{\prime}$ and i$^{\prime}$ magnitudes from data release
9 of the AAVSO Photometric All Sky Survey
\citep[APASS9,][]{Henden:2015ad}; J, H and K$_{\rm s}$ magnitudes from
the Two-micron All Sky Survey (2MASS, \citealt{Skrutskie:2006kx});
i$^{\prime}$, J and K magnitudes from the Deep Near-infrared Southern Sky
Survey \cite[DENIS,][]{DENIS-Consortium:2005fj}. Not all stars have data in all
these catalogues. 

 Our model for the observed photometry has the following parameters --
g$^{\prime}_{0}$, the apparent g$^{\prime}$-band magnitude for the star
corrected for extinction; $T_{\rm eff}$ the effective temperature; E$({\rm
B}-{\rm V})$, the reddening to the system; $\sigma_{\rm ext}$ the additional
systematic error added in quadrature to each measurement to account for
systematic errors. For each trial combination of these parameters we use the
empirical colour -- effective temperature relations by
\citet{Boyajian:2013jh} to predict the apparent magnitudes for the binary
in each of the observed bands. We assume that the contribution from the low
mass companion is negligible at all wavelengths. We used the same transformation between the
Johnson and 2MASS photometric systems as \citet{Boyajian:2013jh}. We used  Cousins
I$_{\rm C}$ as an approximation to the DENIS Gunn i$^{\prime}$ band and the
2MASS K$_{\rm s}$  as an approximation to the DENIS K band \citep[see Fig.
4;][]{Bessell:2005tr}. We used interpolation in Table 3 of
\citet{Bessell:2000zz} to transform the Johnson B, V magnitudes to
Tycho-2 B$_{\rm T}$ and V$_{\rm T}$ magnitudes. We assume that the extinction
in the V band is $3.1\times {\rm E}({\rm B}-{\rm V})$. Extinction in the SDSS
and 2MASS bands is calculated using A$_{\rm r} = 2.770\times {\rm E}({\rm
B}-{\rm V})$ from \cite{Fiorucci:2003nn} and extinction coefficients
  relative to the r$^{\prime}$ band from \citet{Davenport:2014kj}.

We used {\sc emcee} \citep{Foreman-Mackey:2013rm}
to sample the posterior probability distribution for our model parameters. We
used the reddening maps by \citet{Schlafly:2011lk} to estimate the total
line-of-sight extinction to each target, ${\rm E}({\rm B}-{\rm V})_{\rm map}$.
This value is used to impose the following (unnormalized) prior on $\Delta =
{\rm E}({\rm B}-{\rm V}) - {\rm E}({\rm B}-{\rm V})_{\rm map}$: 

\rowcolors{2}{white}{white}

\[ P(\Delta) =
\left\{ \begin{array}{ll} 1 & \Delta \le 0 \\ \exp(-0.5(\Delta/0.034)^2) &
\Delta > 0 \\ \end{array} \right. \]

The constant 0.034 is taken from
\citet{Maxted:2014fa} and is based on a comparison of ${\rm E}({\rm
B}-{\rm V})_{\rm map}$ to ${\rm E}({\rm B}-{\rm V})$ from Str\"{o}mgren
photometry for 150 A-type stars. }

{ Finally, primary stellar masses were then estimated by interpolation with in Table~B.1
of \citet{Gray:2008fj}. The masses we obtain can be found in table~\ref{tab:primary} of this paper, as well as values for $T_{\rm eft}$, and $E(B-V)$. The error for the primary mass is calculated by

\begin{equation}
\sigma_{\rm m_{\rm A}} = \sqrt{\left(\sigma_{T_{\rm eff}}\frac{dm_{\rm A}}{dT_{\rm eff}} \right)^2 + \left(0.06m_{\rm A}\right)^2},
\end{equation}
where $\sigma_{\rm T_{\rm eff}}$ is the error in $T_{\rm eff}$ and $dm_{\rm A}/dT_{\rm eff}$ is calculated from the empirical mass-effective temperature relation \citep{Torres:2010oq} by generating 1000 values of $T_{\rm eff}$ from a normal distribution with a standard deviation of $\sigma T_{\rm eff}$. The factor of $0.06$ accounts for the scatter measured around the mass-effective temperature relation \citep{Torres:2010oq}.}

The determination of our primaries' masses is currently coarse and a finer spectroscopic analysis will be done, to update the values that we provide here.
We provide detailed results on the model parameters so that the masses for the secondaries can be easily updated when this newer, more accurate information on the primaries is finally released. 

{ Finally, the primary stellar radii were also determined based on \citet{Gray:2008fj}. The only purpose of these radii in this paper is to calculate an inclination-based uncertainty in the secondary mass measurement, as will be explained in Sect.~\ref{subsec:mass}.}

\section{Treatment of radial velocity data}\label{sec:data_reduction}

\subsection{Data reduction software}

The spectroscopic data were reduced using the CORALIE Data Reduction Software (DRS). The radial velocity information was obtained by removing the instrumental blaze function and cross-correlating each spectrum with a numerical mask corresponding to the spectral type of the primary. The position of all orders were calibrated at the beginning of the night using a tungsten lamp. 
Masks came in two flavours: G2 and K5. This correlation was compared with the Th-Ar spectrum used as a wavelength-calibration reference (see \citet{Baranne:1996qa} and \citet{Pepe:2002lh} for further information). As the instrument was not pressurised, the wavelength solution changed with variations in atmospheric pressure (approximately equivalent to 100 m s$^{-1}$ mbar$^{-1}$). The simultaneous calibration Th-Ar (now Fabry-P\'erot), on each science frame, accurately corrects instrumental changes. As a precaution, additional calibrations of the wavelength solution were obtained during the night when a drift in excess of 50 m s$^{-1}$ is detected.

The { CORALIE} DRS was built similarly to the DRS for the HARPS, HARPS-North and SOPHIE instruments, and has been shown to achieve remarkable stability, precision and accuracy \citep[e.g.][]{Mayor:2009rw,Molaro:2013lr,Lopez-Morales:2014qv,Motalebi:2015lr} thanks in part to a revision of the reference lines for thorium and argon by \citet{Lovis:2007ec} as well as a better understanding of instrumental systematics \citep[e.g.][]{Dumusque:2015vn}. With a resolving power $R=55\,000$, we obtained a cross-correlation function (CCF) binned in $0.5$\,km\,s$^{-1}$ increments.
The range over which we computed the CCF was adapted to be three times the size of the full width at half maximum (FWHM) of the CCF on each of the spectra. This ensures that wings of the function are well adjusted by the Gaussian model applied to the CCF.

\subsection{Calculating error bars}

Uncertainties on individual data points were estimated by the DRS from photon noise alone. CORALIE was stable to $\sim$6 m/s for many years \citep{Marmier:2014rz}, but a recent change from a circular optical fibre to a octagonal at the end of 2014  improved stability to $\sim$3 m/s \citep{Triaud:2016fp}. Given that the majority of the measurements for the EBLM project were taken before the change of fibre, we systematically added 6 m/s of noise, quadratically, to the $1\sigma$ photon noise uncertainties. In the vast majority of cases the photon noise dominates, so the effect of this correction is minimal.

The majority of spectroscopic observations using CORALIE are for a volume-limited Doppler survey to detect planets around bright, {\it Hipparcos}-selected, low $v\,\sin\,i_\star$ stars \citep{Mayor:2011fj,Marmier:2013lr}, and the confirmation of the WASP transiting planet candidates \citep{Triaud:2011vn,Lendl:2014yu,Neveu-VanMalle:2014rf}. For these two programmes the obtained spectra have high signal-to-noise ratios ($SNR > 15$), whereas for the EBLM project, since we mostly deal with shortened exposure times, we frequently obtained lower signal-to-noise spectra ($SNR \sim 3-7$). 
Furthermore many of our primaries spin rapidly. This decreases the signal-to-noise we obtain on the peak of the CCF, and affects our radial-velocity precision. Our automated error bar estimation was therefore not very well adapted to this new regime of observations for CORALIE, which led to an under-estimation of measurement uncertainties. We corrected the DRS's uncertainties using an indicator called the span of the bisector slope.

The span of the bisector slope (or the bisector thereafter) measures the asymmetry of the CCF, which reflects the asymmetry of all absorption lines \citep{Queloz:2001lr}. It has an uncertainty twice the value of the uncertainty achieved on the radial-velocity \citep{Queloz:2001lr,Figueira:2013lr}. The bisector is traditionally used to test whether any detected low amplitude radial-velocity variation is caused by a translation of the CCF (as expected for a Doppler reflex motion), instead of caused by a change in the shape of the CCF. This can be produced by stellar activity \citep[leading to an anti-correlation;][]{Queloz:2001lr}, or by the Doppler reflex motion of a blended, secondary set of lines \citep[creating a correlation;][]{Santos:2002fk}. Whilst this is important to discover exoplanets whose signal can be similar in amplitude to a line shape variation, in our case, the EBLM project, the orbital motion is large ($>$ than the FWHM of the CCF) in addition to not being subject to any detection problem. In our case, we can use the dispersion of the bisector to calculate the true uncertainty on our radial-velocities and correct any under-estimation produced by the DRS. We therefore computed an additional noise term, $\sigma_\mathrm{add}$, in the following manner:
\begin{equation}
\sigma_\mathrm{add} = \sqrt{  {\delta_\mathrm{bis}^2 \over{4}} - \langle\sigma_\gamma^2\rangle  }
\end{equation}
where $\delta_\mathrm{bis}$ is the rms of the bisector measurements, and $\sigma_\gamma$ is the photon noise error. Once $\sigma_\mathrm{add}$ is estimated, we check the procedure by finding the mean of the bisector measurements and measuring that the dispersion is compatible with a $\chi^2_\mathrm{reduced} = 1$ where the error terms on the bisector have been updated by quadratically adding:

\begin{equation}
\label{eq:bis_add}
\sigma_\mathrm{bis} = 2 \sqrt{\sigma^2_\mathrm{add} + \sigma_\gamma^2}
\end{equation}
similarly, the new errors on the radial velocity measurements become:
\begin{equation}
\sigma_\mathrm{rv} = \sqrt{\sigma^2_\mathrm{add} + \sigma_\gamma^2}
\end{equation}

{ In cases} $\delta_\mathrm{bis}  <  \,\langle\sigma_\gamma^2\rangle$ there is no need for any additional noise term, { and hence} $\sigma_\mathrm{add}$ is set to 0. 

\subsection{Outlier removal}\label{subsec:outliers}

Several steps were taken to remove outliers. First, all observations with a bisector position more than three interquartile ranges below the first quartile or above the third quartile were automatically removed. Observations such as these with significantly different bisector positions are often indicative of the wrong star accidentally being observed or an anomalously low signal-to-noise, generally owing to poor observing conditions. Any bisector variation within the remaining observations was accounted for with the added $\sigma_{\rm bis}$ noise term described in the previous section. After this automated removal procedure a visual inspection was done of all data series. In particular, there was a check for the consistency of the FWHM, as occasionally the wrong star being observed may still result in a coincidentally similar bisector, but different FWHM. Additionally, some targets received Rossiter-McLaughlin observations during eclipses to measure the projected spin-orbit alignment. These results are to be presented in a future paper and are removed from the data analysis in this paper as the Rossiter-McLaughlin anomaly would likely bias the radial-velocity fit if not modelled accurately. For other observations, care was taken to take them out of eclipse.

\section{Orbit fitting}\label{sec:fitting}

{ Orbits are fitted using the {\sc Yorbit} software developed at the University of Geneva. It uses a genetic algorithm to scan a broad parameter space and avoiding falling into local minima. This is coupled with a Markov Chain Monte Carlo to calculate the final orbital solution. Keplerian orbits are fitted independently with no N body interactions between them, although this is something to be developed in the future. This software has been used in particular in many CORALIE and HARPS radial velocity surveys in the past (e.g. \citealt{Mayor:2011fj}) and is discussed in more detail in \citet{Segransan:2010vn} and \citet{Bouchy:2016vn}.}



\begin{table*}
\caption{Models ranked by ascending complexity}
\label{tab:models}
\centering 
\tiny
\begin{tabular}{|ccccccccccccccccc|}
\hline
\rowcolor{gray!50}
name & num. params. & $P_1$ & $K_1$ & $T_{\rm peri,1}$ & $f_1(m)$ & $e_1$ & $\omega_1$  & lin. & quad. & cubic & $P_2$ & $K_2$ & $T_{\rm peri,2}$ & $f_2(m)$ & $e_2$ & $\omega_2$ \\
\hline\hline 
\multicolumn{17}{|c|}{Base models} \\
\hline
k1 (circ) & 4 & \checkmark & \checkmark & \checkmark & \checkmark & $--$ & $--$ & $--$ & $--$ & $--$ & $--$ & $--$ & $--$ & $--$ & $--$ & $--$ \\
k1d1 (circ) & 5 & \checkmark & \checkmark & \checkmark & \checkmark & $--$ & $--$ & \checkmark& $--$ & $--$ & $--$ & $--$ & $--$ & $--$ & $--$ & $--$ \\
k1 (ecc) & 6 & \checkmark & \checkmark & \checkmark & \checkmark & \checkmark & \checkmark & $--$ & $--$ & $--$ & $--$ & $--$ & $--$ & $--$ & $--$ & $--$ \\
k1d2 (circ) & 6 & \checkmark & \checkmark & \checkmark & \checkmark & $--$ & $--$ & \checkmark & \checkmark & $--$ & $--$ & $--$ & $--$ & $--$ & $--$ & $--$ \\
k1d1 (ecc) & 7 & \checkmark & \checkmark & \checkmark & \checkmark & \checkmark & \checkmark & \checkmark & $--$ & $--$ & $--$ & $--$ & $--$ & $--$ & $--$ & $--$ \\
k1d2 (ecc) & 8 & \checkmark & \checkmark & \checkmark & \checkmark & \checkmark & \checkmark & \checkmark & \checkmark & $--$ & $--$ & $--$ & $--$ & $--$ & $--$ & $--$ \\
\hline 
\multicolumn{17}{|c|}{Complex models} \\
\hline
k1d3 (circ) & 7 & \checkmark & \checkmark & \checkmark & \checkmark & $--$ & $--$ & \checkmark & \checkmark & \checkmark & $--$ & $--$ & $--$ & $--$ & $--$ & $--$ \\
k1d3 (ecc) & 9 & \checkmark & \checkmark & \checkmark & \checkmark & \checkmark & \checkmark & \checkmark & \checkmark & \checkmark & $--$ & $--$ & $--$ & $--$ & $--$ & $--$ \\
k2 (circ) & 10 & \checkmark & \checkmark & \checkmark & \checkmark & $--$ & $--$ & $--$ & $--$ & $--$ & \checkmark & \checkmark & \checkmark & \checkmark & \checkmark & \checkmark \\
k2 (ecc) & 12 & \checkmark & \checkmark & \checkmark & \checkmark & \checkmark & \checkmark & $--$ & $--$ & $--$ & \checkmark & \checkmark & \checkmark & \checkmark & \checkmark & \checkmark \\
\hline
\end{tabular}
\end{table*}

\subsection{{ Calculating the secondary mass}}\label{subsec:mass}

A single Keplerian orbit is characterised by six parameters. There is more than one way to paramaterise this problem. The ones provided by {\sc Yorbit} are: period, $P$, semi-amplitude, $K$, eccentricity, $e$, time of periapsis passage, $T_0$, mass function, $f(m)$ and argument of periapsis, $\omega$. For each of these parameters the {\sc Yorbit} calculates $1\sigma$ error bars using 5000 Monte Carlo simulations. These parameters are determined independently of the mass of the primary star, $m_{\rm A}$.

{ Since these are single-lined binaries it is not possible to directly measure the primary and secondary masses. The only mass quantity which we directly measure is the mass function. We must instead use a primary mass inferred from the models described in Sect.~\ref{subsec:primary_masses}. The secondary mass is then calculated from the mass function by solving the following equation numerically:}

\begin{equation}
\label{eq:mB}
f(m_{\rm A},m_{\rm B}) = \frac{\left(m_{\rm B}\sin I\right)^3}{\left(m_{\rm A} + m_{\rm B}\right)^2} = \frac{PK^3}{2\pi G}.
\end{equation}
{ When evaluating Eq.~\ref{eq:mB} we take $I=90^{\circ}$, since our binaries are eclipsing.} To calculate the error of $m_{\rm B}$ we use the fact that our binaries all have small mass ratios, allowing us to simplify Eq.~\ref{eq:mB} to $f(m_{\rm A},m_{\rm B}) \sim m_{\rm B}^3\sin^3I/m_{\rm A}^2$, for which the error calculation becomes:
\begin{equation}
\label{eq:mB_error}
\frac{\delta m_{\rm B}}{m_{\rm B}} = \frac{1}{3}\left(\frac{\delta f(m_{\rm A},m_{\rm B})}{f(m_{\rm A},m_{\rm B})} + 2\frac{\delta m_{\rm A}}{m_{\rm A}} + 3\frac{\delta \sin I}{\sin I} \right).
\end{equation}
{ The error in $\sin I$ stems from us not precisely characterising the eclipse impact parameter and hence inclination using the WASP photometry. Based on possible eclipse geometries, the inclination uncertainty is $\delta \sin I = R_{\rm A}/a$. This is a less than 20\% contribution to the relative uncertainty in $m_{\rm B}$.}

The semi-major axis is calculated using Kepler's { third} law,

\begin{equation}
\label{eq:semi-major_axis}
a = \left( \frac{P^2 G (m_{\rm A} + m_{\rm B})}{4\pi^2}\right)^{1/3},
\end{equation}
and the error is calculated as
\begin{equation}
\label{eq:semi-major_axis_error}
\frac{\delta a}{a} = \frac{1}{3}\left(2\frac{\delta P}{P} + \frac{\delta m_{\rm A} + \delta m_{\rm B}}{m_{\rm A} + m_{\rm B}}\right).
\end{equation}
The precision in the semi-major axis is always significantly worse than that of the period, due to the uncertainty in the stellar masses.

\section{Model selection}\label{sec:model_selection}

We now describe the models which we have fitted to each star and how we choose the most appropriate one.

\subsection{Ten different models applied to the data}\label{subsec:model}

For each system we try to fit various models to the spectroscopic data and calculate the goodness of fit. The models tested are:

\begin{enumerate}
\item k1: a single Keplerian orbit
\item k1d1: a single Keplerian plus a linear drift
\item k1d2: a single Keplerian plus a quadratic drift
\item k1d3: a single Keplerian plus a cubic drift
\item k2: two Keplerians

\end{enumerate}

The drift terms are indicative of an outer third body that is causing the radial velocities to deviate from a single Keplerian over time. Whether we require a linear, quadratic or cubic fit is function of the amplitude of the radial-velocity signal induced by the third body and also the temporal fraction of its orbit covered. When the tertiary orbit is well-covered ($\gtrsim 30\%$), a second Keplerian is generally a better fit. For each of the cases we further tested the goodness of fit with both the binary eccentricity found by {\sc Yorbit}, and with a forced circular orbit.  This means that in total, we adjusted and can test ten different models on our data.

While forcing a circular model, two parameters are dropped: eccentricity, $e$, and argument of periapsis, $\omega$. We denote eccentric and circular models using the parentheses (ecc) and (circ), respectively. The ten models are split into ``base" models - k1, k1d1 and k1d2 - and ``complex" models - k1d3 and k2, where sometimes the number of measurements approaches the number of degrees of freedom. This distinction is used in the model selection procedure. Note that for a two-Keplerian fit we only ever force the inner binary orbit to be circular, not the tertiary body. In Table~\ref{tab:models} we rank all ten models in ascending order of complexity (number of parameters). Ticks are used to indicate the orbital parameters used in each fit, including linear (lin), quadratic (quad) and cubic drift coefficients.

\subsection{Using the BIC to select between models}\label{subsec:bic}

\begin{figure}
\begin{center}
\includegraphics[width=0.49\textwidth,trim={0 0 0 0},clip]{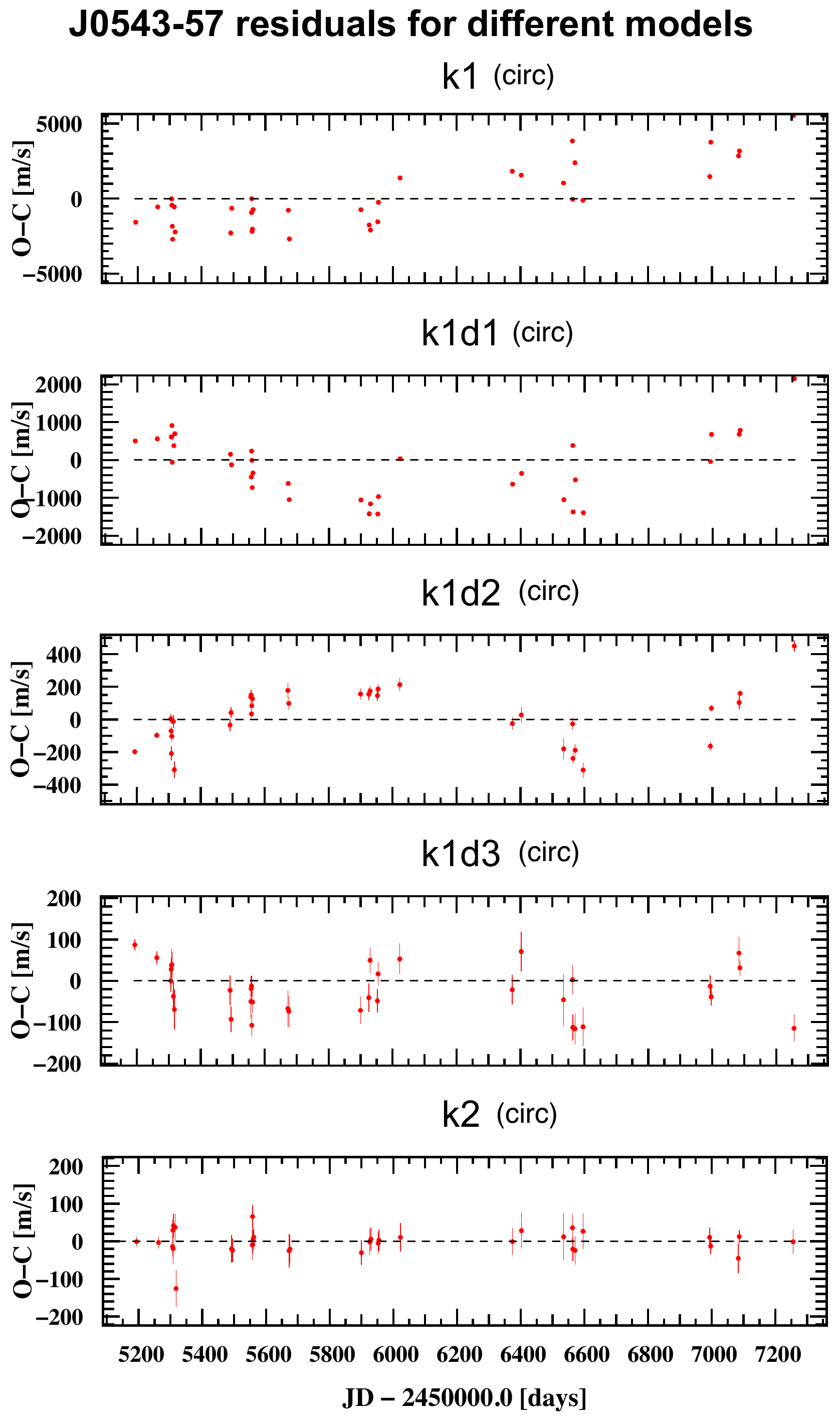}
\caption{The residuals { (O-C)} of the radial velocity fit to J0543-57 of five different models with increasing complexity and improved goodness of fit from top to bottom. A k2 (circ) model was ultimately chosen according to our procedure.}\label{fig:residuals}
\end{center}
\end{figure}

We use the Bayesian Information Criterion \citep[BIC;][]{Schwarz:1978zz} to select the model that provides the optimal balance between goodness of fit and complexity. We assume that the errors in our radial velocity measurements are independent and identically distributed following a normal distribution, so that the BIC can be calculated using

\begin{equation}
\label{eq:BIC}
{\rm BIC} = \chi^2 + k \ln(n_{\rm obs}),
\end{equation}
where $\chi^2$ is the weighted sum of the square of the residuals,  $k$ is the number of model parameters and $n_{\rm obs}$ is the number of observations. The BIC is constructed to naturally penalise models that are unnecessarily complex and not justified by the data (otherwise known as Ockham's razor).  Whenever choosing between one model and the next most complex (in terms of the number of parameters) we demand that the BIC increases by at least 6 in order to justify the added complexity. This is deemed ``strong" evidence in the literature \citep{Kass:1995rt}.

Our model selection procedure follows a forward method, where we start with the simplest model and move up in complexity. The steps are as followed:

\begin{enumerate}

\item Calculate the BIC for the simplest model: a circular single Keplerian
\item Calculate the BIC for subsequent base models with increasing { numbers of parameters}, as denoted by the order in Table.~\ref{tab:models}. 
\item Whenever we want to jump from one model to the { one with the next highest number of parameters}, we demand that the BIC improves (i.e. decreases) by at least 6.
\item { Note that k1 (dcc) and k1d2 (circ) both have six parameters. When choosing between those two models simply the smallest BIC is chosen.}
\item In some situations the next most complex model may only marginally improve the BIC (i.e. not by 6) but the more complex model after that may be a significant improvement. In these exceptional circumstances one may ``jump" to the well-fitting model two ranks of complexity above by improving the BIC by a factor of $2\times6=12$, or in general $n\times 6$ where $n$ is the number of ranks of complexity you want to move up.
\item For the base model chosen according to the BIC we calculate the reduced $\chi^2$ statistic, $\chi_{\rm red}^2 = \chi^2/(n_{\rm obs}-k)$, where $(n_{\rm obs}-k)$ is the number of degrees of freedom. For a good fit to the data we expect $\chi_{\rm red}^2\sim 1$.
\item If this value of $\chi_{\rm red}^2<2$ then we consider the simple model to be a sufficient fit to the data and do not test any others. This conservative approach helps avoid over-fitting.
\item Alternatively, if $\chi_{\rm red}^2>2$ we then test more complex models: k1d3 and k2 (both eccentric and circular). These models are then treated with the same model selection procedure as before. In some cases complex models are tested but a base model is ultimately still chosen.
\item An exception to the above procedure comes in the case of heightened stellar activity. This activity can cause variation in the radial velocity measurements that may be confused for another physical body in the system. This occurs in two cases: J0021-16 and J2025-45. These binaries have $\chi_{\rm red}^2$ for the base models of 3.49 and 7.09, respectively, but we do not test more complex models and instead manually assign an appropriate, simpler model. These individual cases are discussed further in Sect.~\ref{subsec:activity}.

\end{enumerate}

Figure~\ref{fig:residuals} shows our procedure in action on the residuals obtained after removing the most likely parameters for a set model. In this particular case adding a second Keplerian visibly improves the goodness of fit, which also happens in the BIC values.
In Table~\ref{tab:BIC} we show the data pertaining to the model selection. The BIC of the selected model is highlighted in bold font. For most systems the simple models tested yielded a $\chi_{\rm red}^2<2$ and hence no models of further complexity were needed. In Table~\ref{tab:BIC} we count the number of binaries fitted to each of the ten models. In the appendices, Table~\ref{tab:params} contains the orbital parameters for all of the binaries, taken from the chosen model according to the BIC. For the four binaries where a k2 model was selected we provide the orbital parameters of the tertiary body in Table~\ref{tab:triple}.

\subsection{Providing upper limits on undetected nested parameters}\label{subsec:limits}

In reality, no orbit is exactly circular, meaning that the true physical model ought to be eccentric even though statistically that extra degree of complexity in the model is not formally detected. To remedy the issue we provide estimates for upper limits to the eccentricity and the coefficient of linear drift, along with the selected model values, in Table~\ref{tab:params}. The values we provide were estimated using the model of higher complexity on that particular parameter. We provide values at 67\% confidence by taking the fitted value and adding the $1\sigma$ uncertainty. We do the same for the upper limit of the linear drift coefficient for binaries where a single Keplerian fit was chosen.

\section{Results}\label{sec:results}


\subsection{Summary}\label{subsec:summary}

In total we analysed 118 eclipsing binaries. Table~\ref{tab:BIC} shows the number of stars for which each model was selected using the BIC. The results of the model fits to individual stars are given in a series of tables in the appendices to the paper. First, in Table~\ref{tab:BIC} we demonstrate our model selection procedure based on the Bayesian Information Criterion and reduced $\chi^2$ statistic. The chosen model is given in this table and the BIC for that model is highlighted in bold. For most systems the simple models tested yielded a $\chi_{\rm red}^2<2$ and hence no models of further complexity were needed. { The flag column has three different flags: ``drift" indicating that a linear, quadratic or cubic drift was the best fit to the data, ``triple" for the four systems where we fitted two Keplerian orbits to the triple star system and ``active" for the two systems showing signs of stellar activity.}

{ The orbits of the best-fitting models and their residuals are shown in Appendix~\ref{app:orbits_and_residuals}.}

{ Contained in} in Table~\ref{tab:params} are the orbital parameters for all of the binaries. For each parameter we show both the measured value and the uncertainty. The uncertainty is the value inside the brackets and corresponds to the final two digits of the measured value. For example, for J0008+02 $P=4.7222907(63)$ days, which means $P=4.7222907\pm0.0000063$ days. { This table includes the calculated primary and secondary masses. More detailed parameters for the primary stars are shown in Table~\ref{tab:primary}.} The J and V magnitudes come from the NOMAD survey and the R magnitude comes from 2MASS. An exception is that for three targets, J1934-42, J1509-10 and J2353-10, no Vmag was available from NOMAD so it was calculated as a function of the primary mass using models by \citet{Baraffe:2015lr} at an age of 1 Gyr.

For the four targets with characterised tertiary orbits we provide their orbital parameters and plots of the radial velocity fits in Appendix~\ref{app:triple}.

{ Parameters for the secondary stars are shown in Table~\ref{tab:secondary}. The error in the secondary mass is predominantly due to uncertainties in the primary mass and orbital inclination, and not the radial velocity semi-amplitude. Unlike the primary star, which has measured magnitudes, the secondary magnitudes are all calculated using the \citet{Baraffe:2015lr} models. Values for the V, R and J magnitudes are given at ages of 1 Gyr and 5 Gyr. This is because we do not have accurate estimates for the true ages of the systems, although we note that the magnitude difference is small.}

Finally, in Table~\ref{tab:obs} are various observational parameters for the binaries. The { period $P$ and }times { $T_{\rm 0, pri}$ and $T_{\rm 0, sec}$ (for the primary and secondary eclipses)} are taken from the radial velocity fit, not the  WASP photometry. 

\begin{figure}
\captionsetup[subfigure]{labelformat=empty}
\begin{center}
\begin{subfigure}[b]{0.49\textwidth}
\includegraphics[width=\textwidth,trim={0 0 0 0},clip]{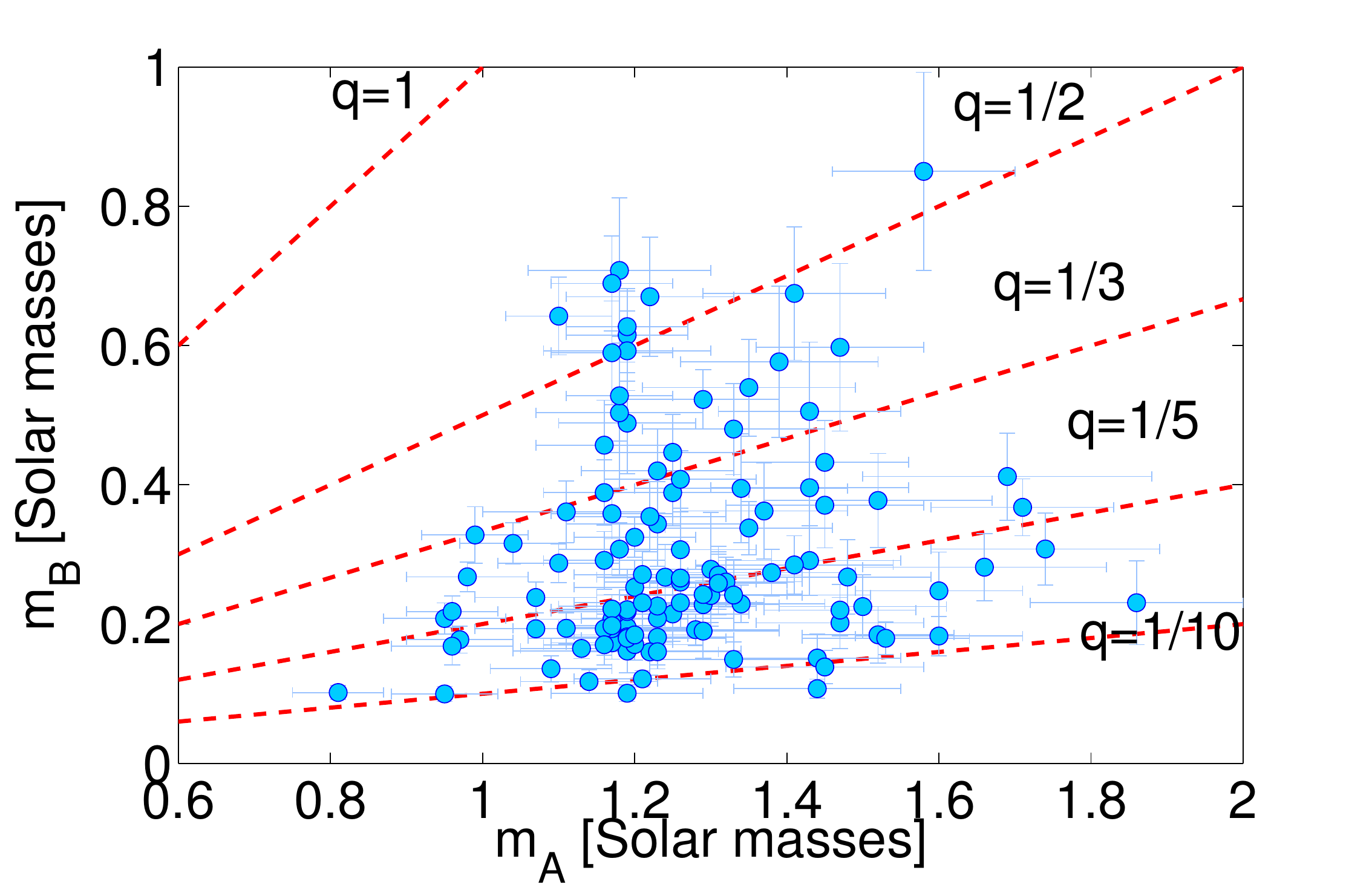}
\end{subfigure}
\begin{subfigure}[b]{0.49\textwidth}
\includegraphics[width=\textwidth,trim={0 0 0 0},clip]{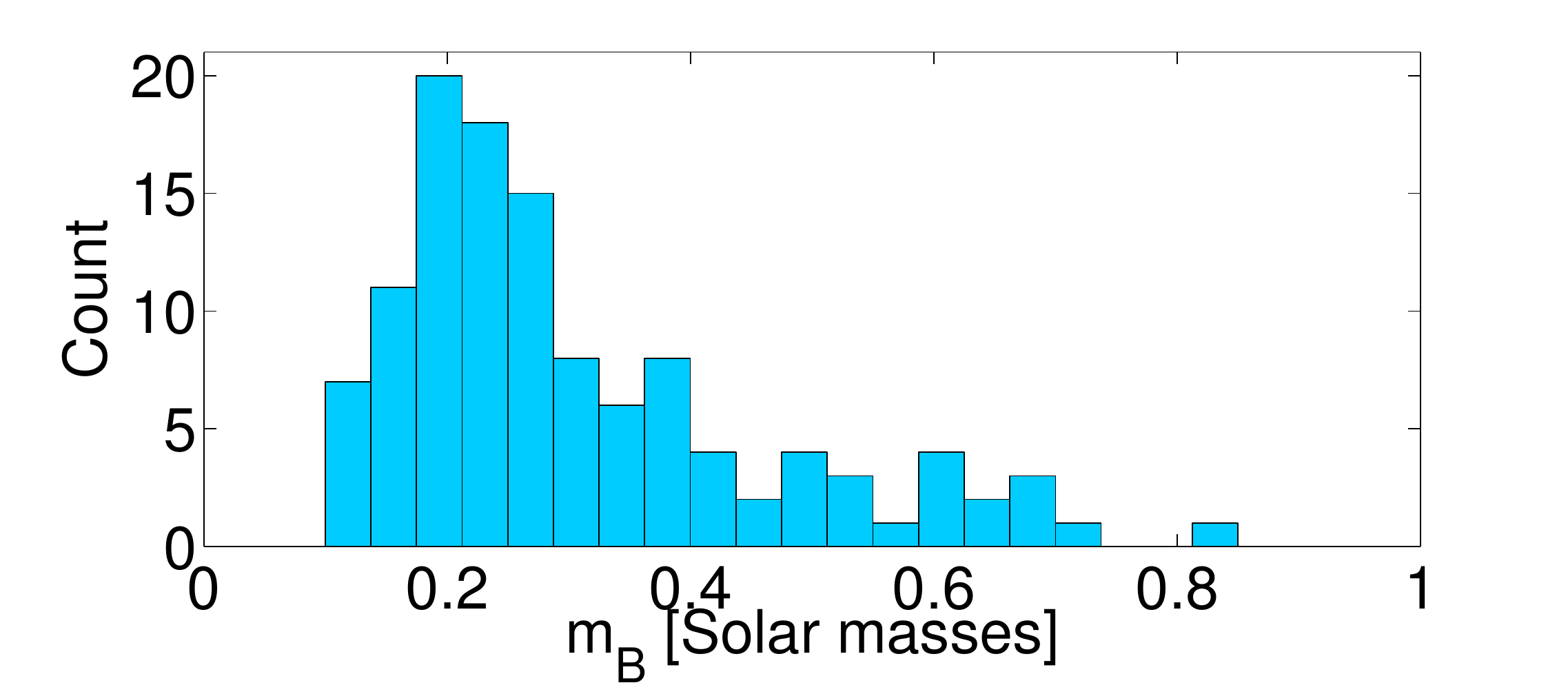}
\end{subfigure}
\begin{subfigure}[b]{0.49\textwidth}
\includegraphics[width=\textwidth,trim={0 0 0 0},clip]{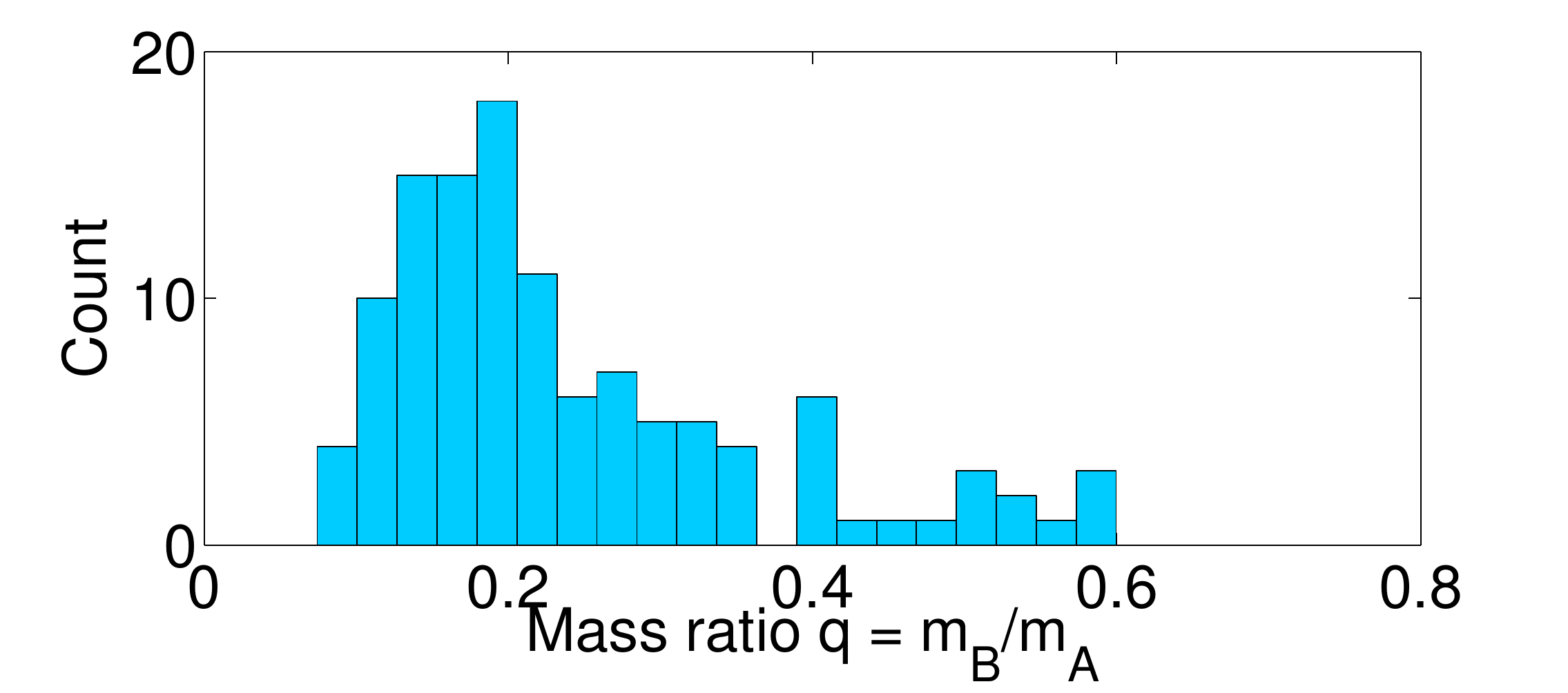}
\end{subfigure}
\caption{Top: mass ratio $q=m_{\rm A}/m_{\rm B}$ for each of the binaries including error bars. The fractional error of the secondary is generally similar to that of the primary, and consequently the absolute error for the secondary is invisibly small on this plot for small $m_{\rm B}$. Red dashed lines correspond to mass ratios of $1$, $1/2$, $1/3$, $1/5$ and $1/10$. Middle: histogram of the secondary mass. Bottom: histogram of the mass ratio.}
\label{fig:mass_ratio}
\end{center}
\end{figure}

\subsection{Primary and secondary masses and magnitudes}\label{subsec:primary_and_secondary}

We now demonstrate visually some of the results in our sample. In Figure~\ref{fig:mass_ratio} we show the primary and secondary masses in our sample. It is seen that { 60} of our binaries (50\% of the sample) have mass ratio $q < 0.2$, and, { 34} (31\%) companions have masses $m_{\rm B} < 0.2 M_\odot$. A consequence of these small mass ratios is that our secondary stars are all between 3.1 and 12.6 magnitudes fainter than the primary stars, and hence we only observe a single-line spectroscopic binary. A histogram of this difference in magnitudes is shown in Fig.~\ref{fig:Vmag_difference_histogram}. 

 Figure~\ref{fig:mass_spec} shows a { combined WASP/EBLM} mass spectrum for objects creating photometric eclipses compatible with sizes $< 2.1 R_{\rm Jup}$. We usually give a ``WASP'' identifier for all sub-stellar objects (planets and brown dwarfs). We collected all objects with WASP identifiers { that are public and were observed} with the CORALIE spectrograph and added all stellar companions in this paper. In overall this means 143 substellar objects and { the 118 stellar companions presented in this paper}. One of our sub-stellar companions, WASP-30 \citep{Triaud:2013lr} falls within the brown dwarf range. 

We compare our preliminary results to the 50pc mass spectrum shown in \citet{Grether:2006kx}. To do this, we normalise their histogram to our number of sub-stellar objects with masses superior to $1 M_{\rm Jup}$. From Fig.~\ref{fig:mass_spec}, we see that our mass spectrum covers a broader range in the planetary masses (although at low masses we are most likely incomplete), and does not cover stars as massive (due to the restriction of our survey). Over the common range between  \citet{Grether:2006kx} and us, we have a resolution that is twice better. The results are broadly consistent, but differ in an interesting way. { The brown dwarf desert derived from the WASP and EBLM results appears to stretch deeper into the planetary domain than the result of \citet{Grether:2006kx}. We find that in our results, massive gas-giants ($\sim 3-13M_{\rm Jup}$) appear less abundant. 

The reason for the discrepancy with the \citet{Grether:2006kx} results is that their work probed planets on wider orbits $a<\lesssim10$ AU, whereas EBLM and WASP are typically sensitive to $a<0.2$AU. 
Whatever process(es) is(are) important in shaping the hot Jupiters population, it(they) favour(s) smaller mass gas-giants. This was also noted in \citet{Udry:2003na}.}


{ The entire EBLM sample contains over 200 binaries (of which only 118 are presented here); the WASP survey is still on-going. Once those results are all published we will revisit this mass spectrum. In particular, in the future we will be able to compare the relative abundance of close hot-Jupiters, brown dwarfs and M-dwarfs. On this occasion, we will also produce a more thorough analysis and debiasing of the spectrum.}

\subsection{Eccentricities}\label{subsec:eccentricities}

\begin{figure*}  
\captionsetup[subfigure]{labelformat=empty}
\begin{center}  
	\begin{subfigure}[b]{0.81\textwidth}
		\includegraphics[width=\textwidth]{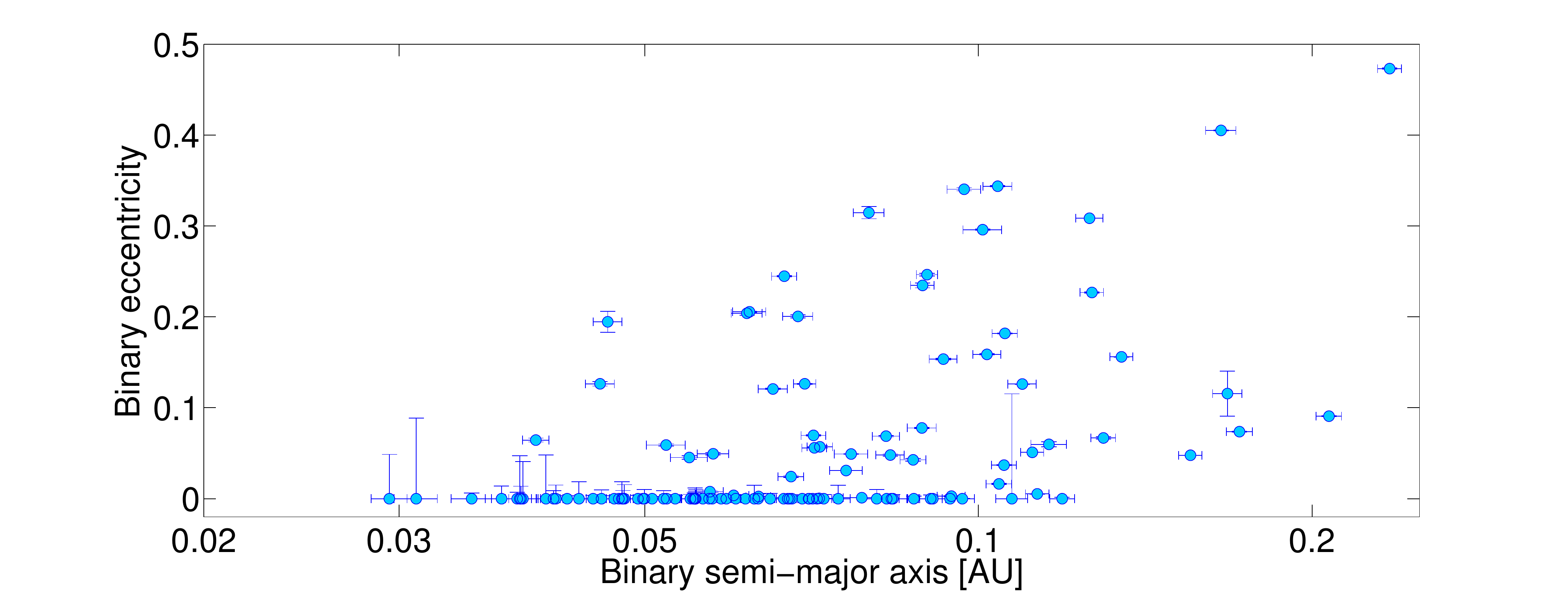}  
	\end{subfigure}
	\begin{subfigure}[b]{0.81\textwidth}
		\includegraphics[width=\textwidth]{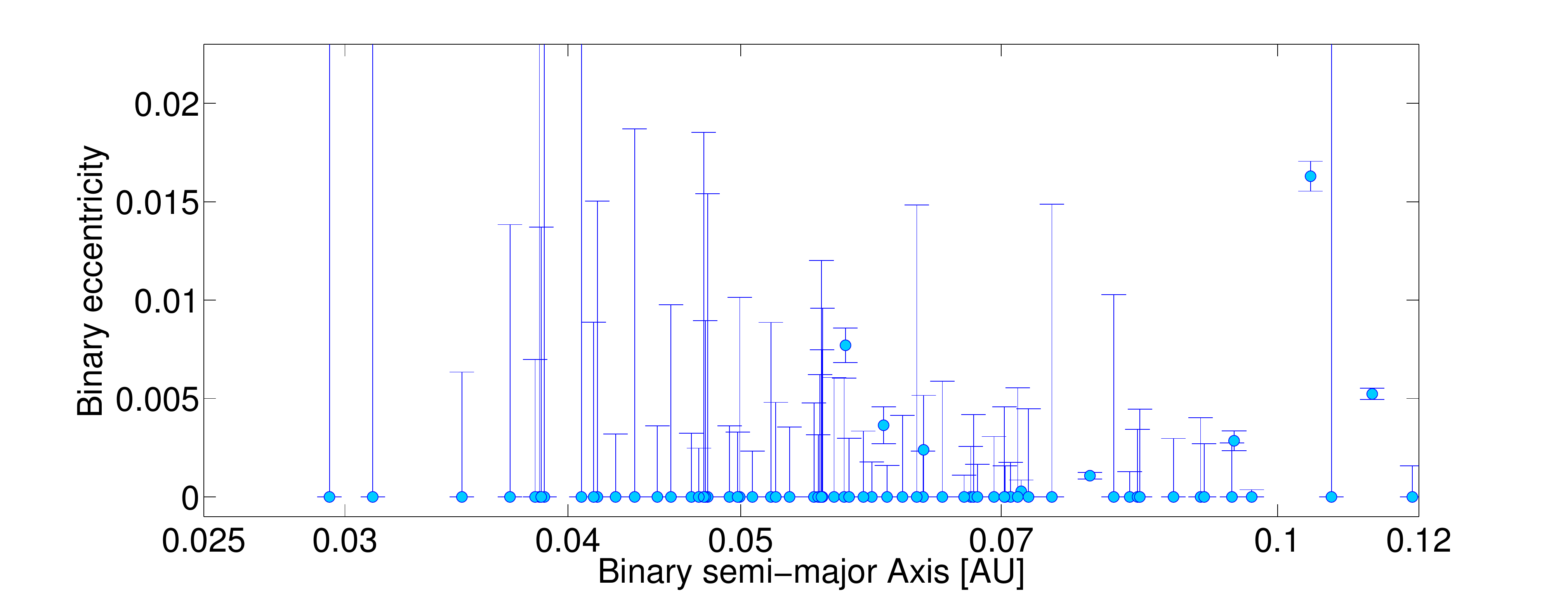}  
	\end{subfigure}	
	\begin{subfigure}[b]{0.81\textwidth}
		\includegraphics[width=\textwidth]{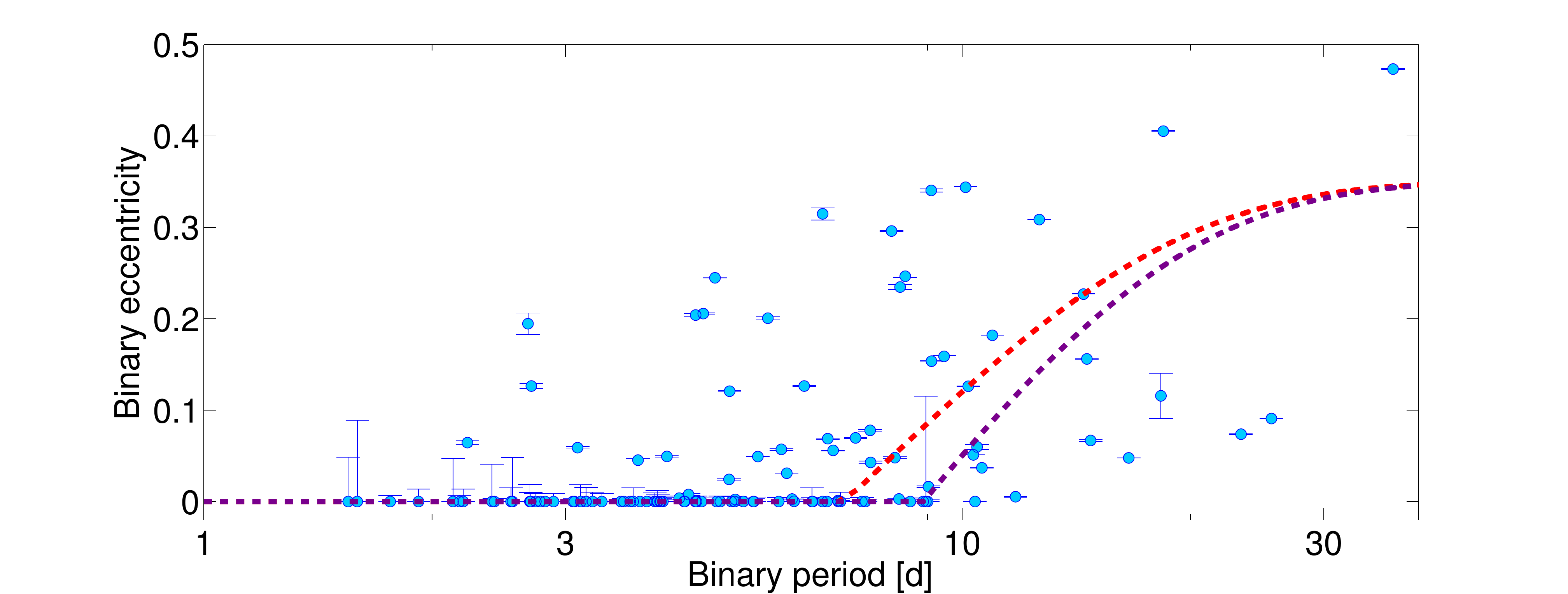}  
	\end{subfigure}
	\begin{subfigure}[b]{0.81\textwidth}
		\includegraphics[width=\textwidth]{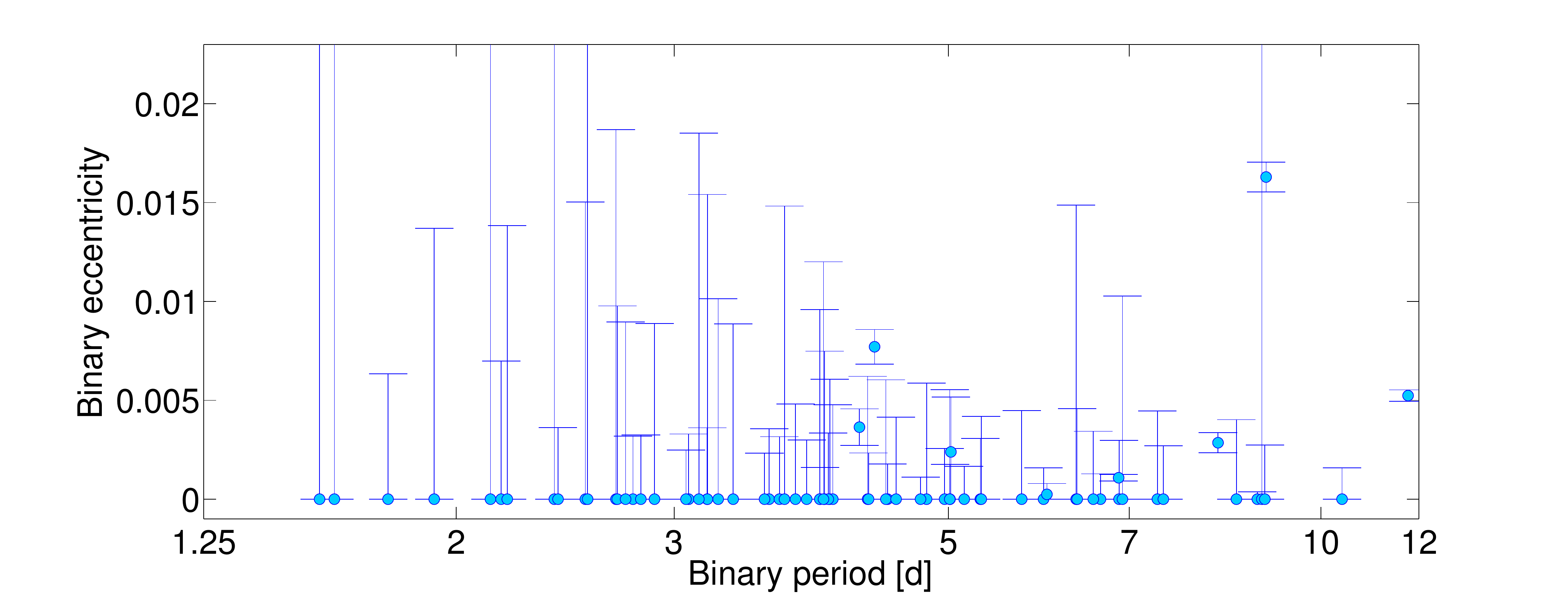}  
	\end{subfigure}
	\caption{The eccentricity of the eclipsing binary as a function of semi-major axis (top two plots) and period (bottom two plots). Plots b and d are zoomed versions of the a and c, respectively, showing the tightest binaries ($P_{\rm bin}<12$ d) with eccentricities compatible with zero. { In the top plot the error in semi-major axis is shown, but this is excluded in the zoomed version for clarity. }In the third figure (period vs eccentricity, not zoomed) we use dashed lines to denote fits using the Meibom \& Mathieu function in Eq.~\ref{eq:mm_fit}. The purple dashed line is a fit to all of the data where $P_{\rm cut}=8.9$ days, whilst the red dashed line is a fit to all binaries with $M_1<1.3 M_{\odot}$ and no sign of a tertiary companion. In this latter case $P_{\rm cut}=7.0$ days. }
\label{fig:eccentricity_vs_a_and_P}  
\end{center}  
\end{figure*}   

\begin{figure}
\begin{center}
\includegraphics[width=0.49\textwidth,trim={0 0 0 0},clip]{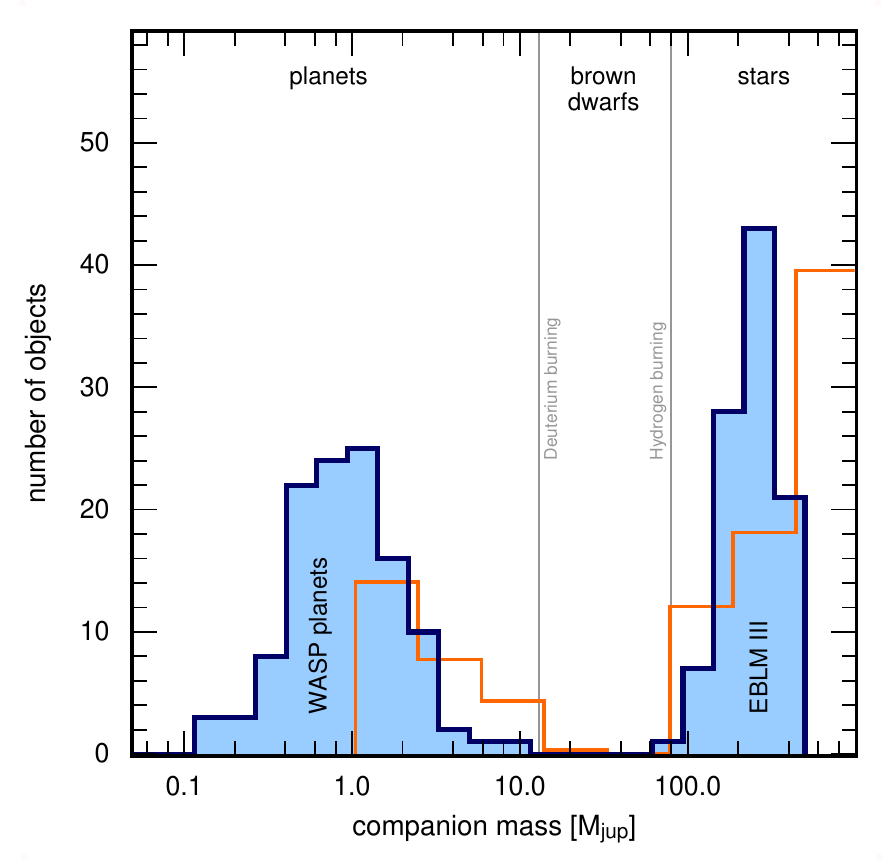}
\caption{{ In dark blue, the observed mass spectrum from the WASP planet survey and the EBLM binaries released in this paper.}. The orange histogram depicts the results from \citet{Grether:2006kx}, normalised { to the number of WASP and EBLM objects heavier than $1M_{\rm Jup}$. The vertical grey lines denote the rough mass limits for Deuterium burning ($13M_{\rm Jup}$) and Hydrogen burning ($80M_{\rm Jup}$). There is an evident deficit of objects between these two limits, which corresponds to the realm of brown dwarfs.}}\label{fig:mass_spec}
\end{center}
\end{figure}

\begin{figure}
\captionsetup[subfigure]{}
\begin{center}
\begin{subfigure}[b]{0.49\textwidth}
\includegraphics[width=\textwidth,trim={0 0 0 0},clip]{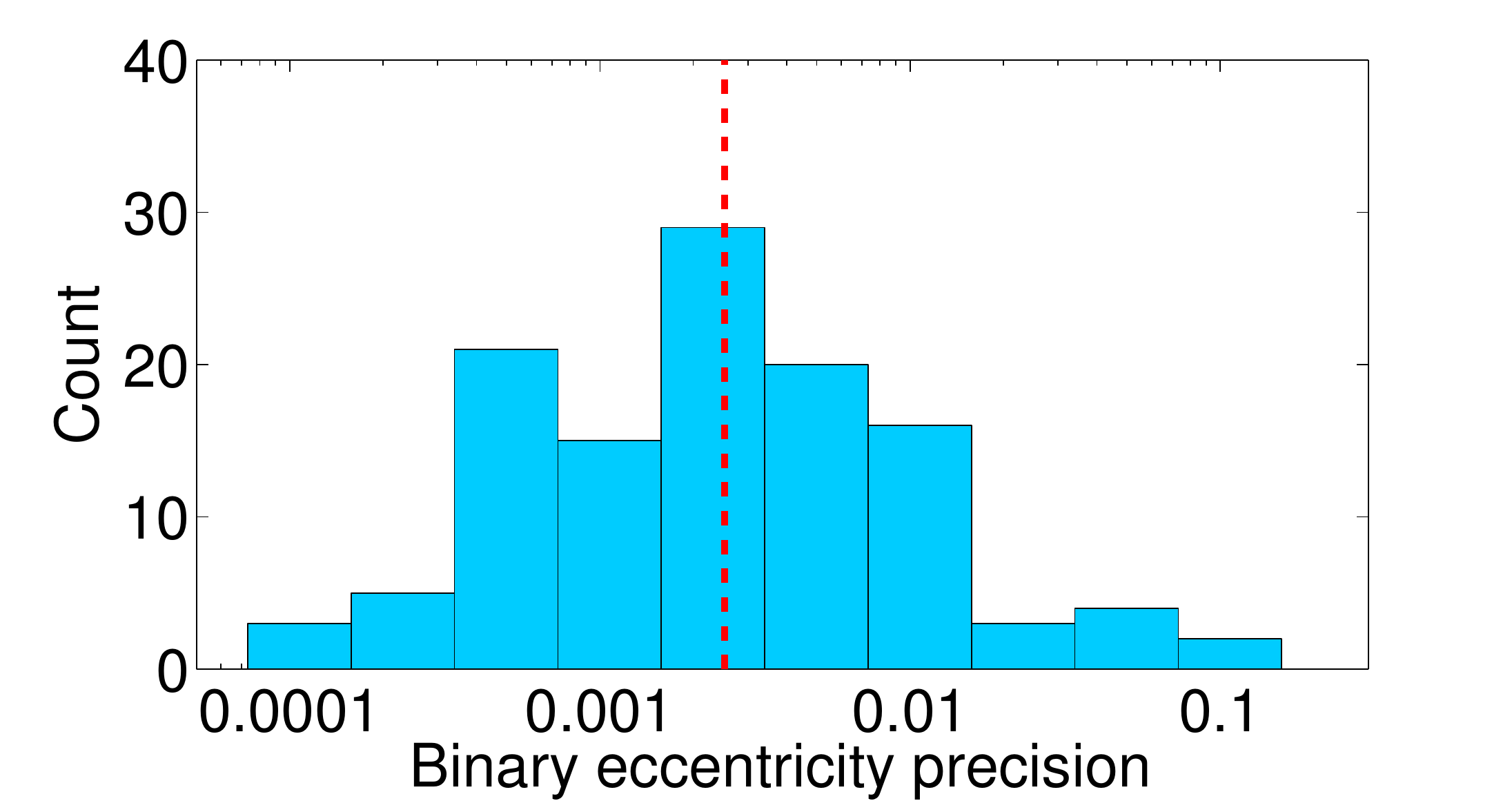}
\end{subfigure}
\begin{subfigure}[b]{0.49\textwidth}
\includegraphics[width=\textwidth,trim={0 0 0 0},clip]{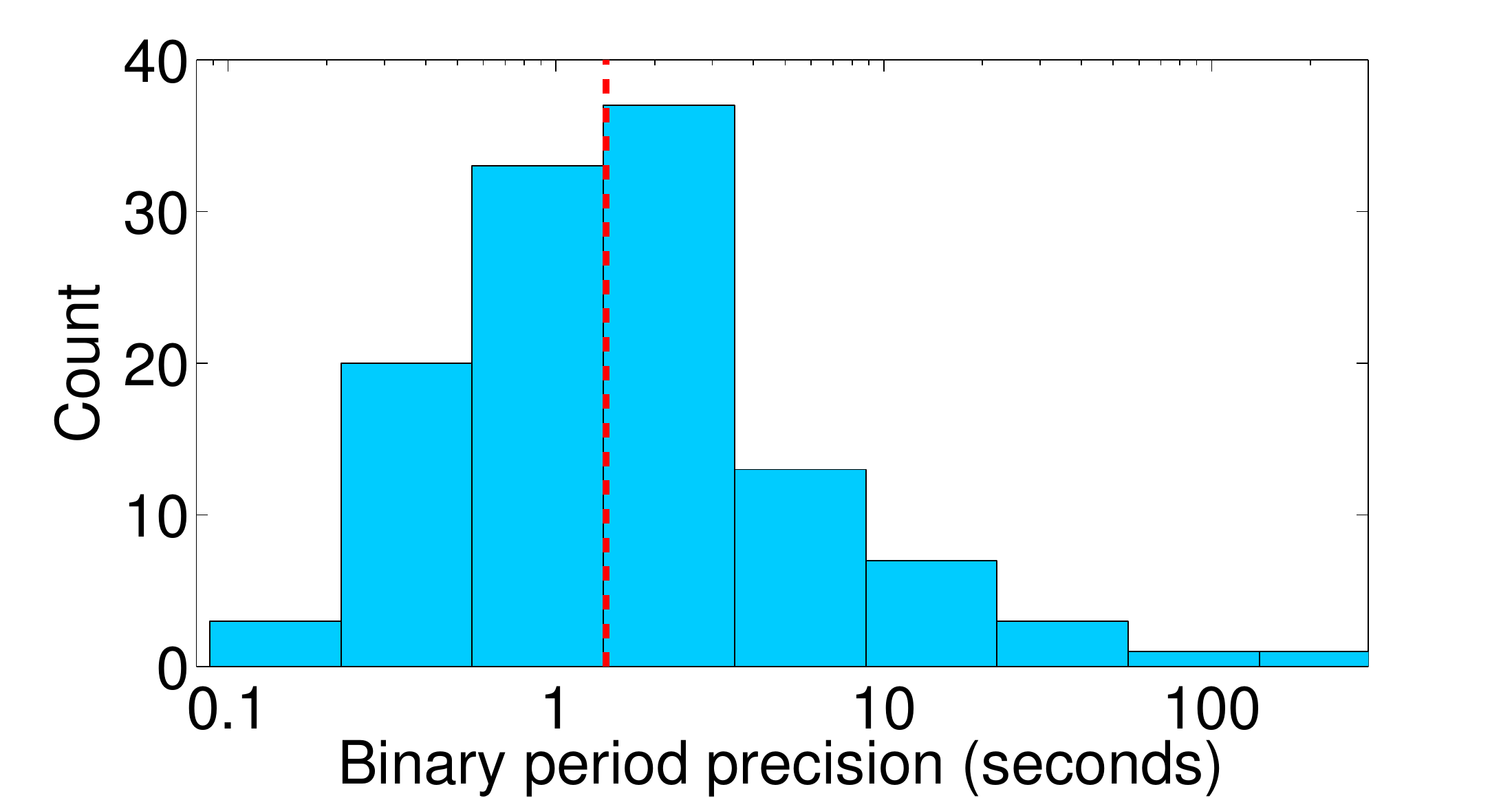}
\end{subfigure}
\caption{ Top: histogram of the precision of the eccentricity for all of our 118 binaries. For models where an eccentricity is favoured it is the $1\sigma$ error bar. For models where a circular solution is favoured it is the upper limit, which is equal to the fitted eccentricity value plus the $1\sigma$ uncertainty. The vertical red dashed line is the median precision of 0.0025. Bottom: histogram of the precision of the binary period, in seconds. The red dashed line is the median precision of 1.4 seconds. We emphasise that the period precision is obtained purely by the radial velocity fit, not from the eclipse timing.}
\label{fig:precision_histogram}
\end{center}
\end{figure}

In Figure~\ref{fig:eccentricity_vs_a_and_P} we show the eccentricity of our systems as a function of the semi-major axis (top two plots) and period (bottom two plots). For the unzoomed $P$ vs $e$ plot we show a fit to the data using the Meibom--Mathieu function \citep{Meibom:2005kx}, which is presented in Sect.~\ref{sec:tides}. For both the semi-major axis and period we show zoomed versions of the plots for eccentricities between 0 and 0.022. There are many binaries for which we can constrain their orbits to being circular within these small bounds of eccentricity. Furthermore, there are some binaries for which we actually measure eccentricities that are small but significantly non-zero. This high precision is by virtue of using the CORALIE instrument with planet-finding precision to observe  much larger amplitude binaries. The smallest significantly non-zero eccentricity measured is $e=0.00108 \pm 0.00017$ for J0353+05\footnote{For J0540-17 our BIC selection algorithm favoured an eccentric solution over a circular one, however the calculated eccentricity is $e=0.00025 \pm 0.00055$ is smaller than that for J0353+05 but compatible with 0 within $1\sigma$. This is a one off case.}. The most precisely measured eccentricity is $e=0.051004 \pm 0.000086$ for J0042-17. These results are expanded upon in Figure~\ref{fig:precision_histogram}a which is a histogram of the eccentricity precision obtained in the EBLM program. For half of our targets we can constrain eccentricities to a precision of 0.0025. For 29\% we obtain a precision better than 0.001 and for 14\% a precision better than 0.0005.

The periods of our binaries are measured to an even better precision. In Figure~\ref{fig:precision_histogram}b is a histogram of the precision obtained on the binary period, shown in a scale of seconds. Half of the sample have a period measured to better than 1.4 seconds. As a percentage error, 50\% of our targets have their period measured to better than 0.00031\%. with the worst being 0.047\% for J0629-67. Our highly precise periods are a result of both the high resolution CORALIE instrument and the fact that all orbits have been observed for a timespan of at least $16P_{\rm bin}$ and 86\% are observed over a timespan of more than $100P_{\rm bin}$.

The longest period binary for which we measure a zero eccentricity is J1008-29 with a 10.4 day period and $e<0.0016$.

The implications of our measured eccentricities as a function of period and semi-major axis are discussed in Sect.~\ref{sec:tides}.

\subsection{Triple star systems}\label{subsec:triples}

\begin{figure}
\centering
\includegraphics[width=0.5\textwidth]{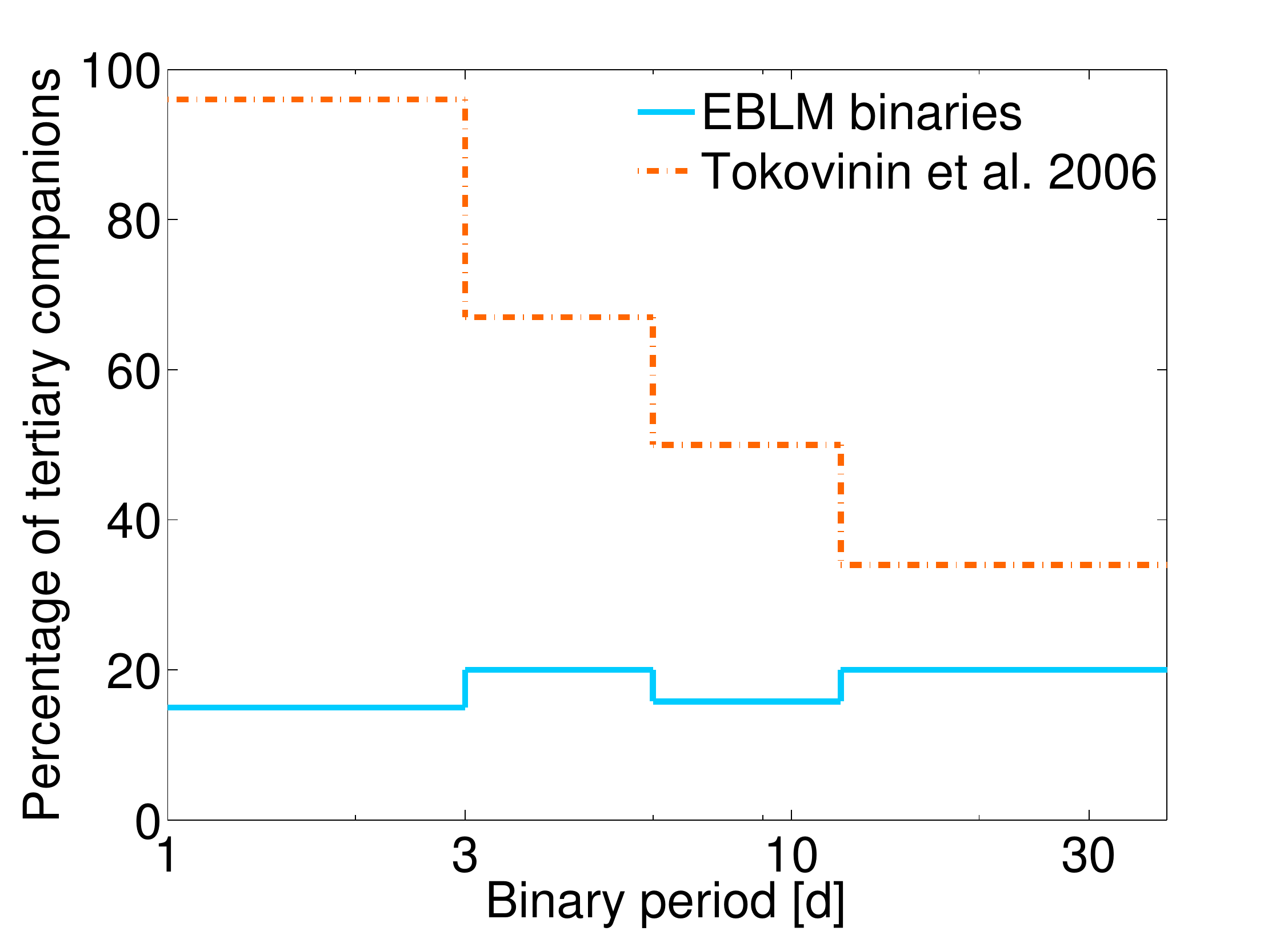}
\caption{Percentage of our EBLM binaries with a tertiary companion (blue solid line). Binaries are said to have a tertiary companion if the best fitting radial velocity model is not a single Keplerian. For comparison we show the results of the \citet{Tokovinin:2006la} imaging survey of close spectroscopic binaries (orange dashed line).}
\label{fig:triple_percentage}
\end{figure}

According to Table~\ref{tab:BIC} there are 21 systems fitted with a model other than a single Keplerian (circular or eccentric). All solutions fitted with a drift or second Keplerian are indicative of a third body, most likely a tertiary star but possibly a circumbinary brown dwarf or massive planet. Our overall tertiary rate is $21/118 = 17.8\%$. In Fig.~\ref{fig:triple_percentage} we plot the percentage of systems with indications of a close tertiary companion as a function of inner binary period (blue solid line). Bin edges are chosen to match the study of \citet{Tokovinin:2006la}: 3, 6, 9, 12, 40. This plot indicates a roughly flat rate of triples as a function of $P$. This contrasts with the results of \citet{Tokovinin:2006la} (orange dash-dotted line) in two ways. First, our results are significantly lower at all binary periods. The cause of this is however simple. Our only indicator of a third star is an additional radial-velocity signal, and this is only sensitive to close triples (estimate period range) and can miss tertiary stars with inclinations near zero. By contrast, \citet{Tokovinin:2006la} was an imaging survey of known spectroscopic binaries, so whilst it may miss some very tight triple systems it is sensitive to a much larger range. 

Second, one of the most important results of \citet{Tokovinin:2006la} was the sharp dependence of tertiary fraction on binary period. This has been interpreted as evidence for the formation of close binaries via Kozai-Lidov cycles followed by tidal circularisation \citep{Lidov:1962kx,Kozai:1962qf,Mazeh:1979eu,Fabrycky:2007pd} and this is different to the flat distribution seen in our raw triple fraction. If the distributions of tertiary periods and masses are uncorrelated with the inner binary period, then on average the slope of the radial velocity drift would be independent of the inner binary period.  However, our detectability of this radial-velocity drift is not uniform with binary period in our sample. As shown in the top of Fig.~\ref{fig:timespan}, the observing time spent on a given target is not dependent on the binary period, so this does not introduce a bias. However, a significant bias is the trend of decreasing precision with closer binaries. This is due to tidal locking leading to broadened spectral lines, as discussed in Sect.~\ref{subsec:sample} and evidenced in Fig.~\ref{fig:precision}. For $P_{\rm bin}<3$ days there are 20 binaries and the median precision obtained is 477 m/s. Between 3 and 6 days the median precision improves to 118 m/s. For 6 to 12 day binaries there is a further improvement to 56 m/s and for our 10 binaries with a period longer than 12 days we have an excellent median precision of 24 m/s. This strong bias hurts our ability to detect tertiary companions to very close binaries. This was not shared with \citet{Tokovinin:2006la}, who used imaging.


Of the 21 binaries identified as having a tertiary companion in four cases our observations allow a characterisation of the tertiary orbit: J0543-57, J1146-42, J2011-71 and J2046-40. These four triples have outer periods of 3062, 260, 663 and 5584 days, respectively. All orbital parameters are provided in Table~\ref{tab:triple} in Appendix~\ref{app:triple}. In this appendix we also provide orbits of both the inner binary and outer tertiary and a top-down view. J1146-42 in particular, with three stars within $\sim 1$ AU { (modulo $\sin i_{\rm C}$)}, is a rare tight triple star system. We note that for the best fitting model $\chi_{\rm red}^2=15.06$ which is actually the worst in our sample, which may seem surprising given the precision obtained is so high (median 18~m~s$^{-1}$) owing to its brightness (Vmag$=10.3$) and long inner period ($P=10.47$ days). We suggest that the cause of these large residuals is a Newtonian perturbation between the two orbits causing them to become non-Keplerian, which is not accounted for within {\sc Yorbit}. This arises because the inner and outer orbits are close: $a_{\rm out}/a_{\rm in}= 9.1$. Our observing timespan of 3.34 years and high precision make us sensitive to these perturbations. It is a future task to analyse the orbital dynamics of these close triple systems and try to exploit them to calculate additional parameters in the system, such as the mutual inclination between the two orbits \citep[e.g.][]{Correia:2010vn}. 

It is also likely that the ongoing {\it GAIA} astrometric survey will provide additional orbital constraints on these triple systems.

\subsection{Active systems}\label{subsec:activity}

Throughout the history of radial-velocity surveys for extra-solar planets, stellar activity has often contrived to confuse observers by creating spurious radial velocity variations that may be mistaken as planets \citep[e.g.][]{Queloz:2001lr}. Fortunately in a survey of binaries, the amplitude of the Keplerian signal is tens of km~s$^{-1}$, which is higher by orders of magnitude compared to stellar activity. From this perspective there is therefore no doubt about the existence of our binaries. What stellar activity may do, however, is inhibit our ability to detect smaller amplitude effects such as radial-velocity drifts indicative of a third star, as these may be of a low amplitude of only tens of metres per second. 

In exoplanet studies, a classic diagnostic for stellar activity is the span of the bisector slope \citep{Queloz:2001lr,Figueira:2013lr}, which we use here as well. An anti-correlation between a motion in radial-velocity and in the slope of the bisector indicates a distortion of the absorption lines, caused by stellar activity.

\begin{figure*}
\captionsetup[subfigure]{}\begin{center}
\begin{subfigure}[b]{0.33\textwidth}
\caption{J0021-16; k1}\label{fig:correlation_a}
\includegraphics[width=\textwidth,trim={0 0 0 0},clip]{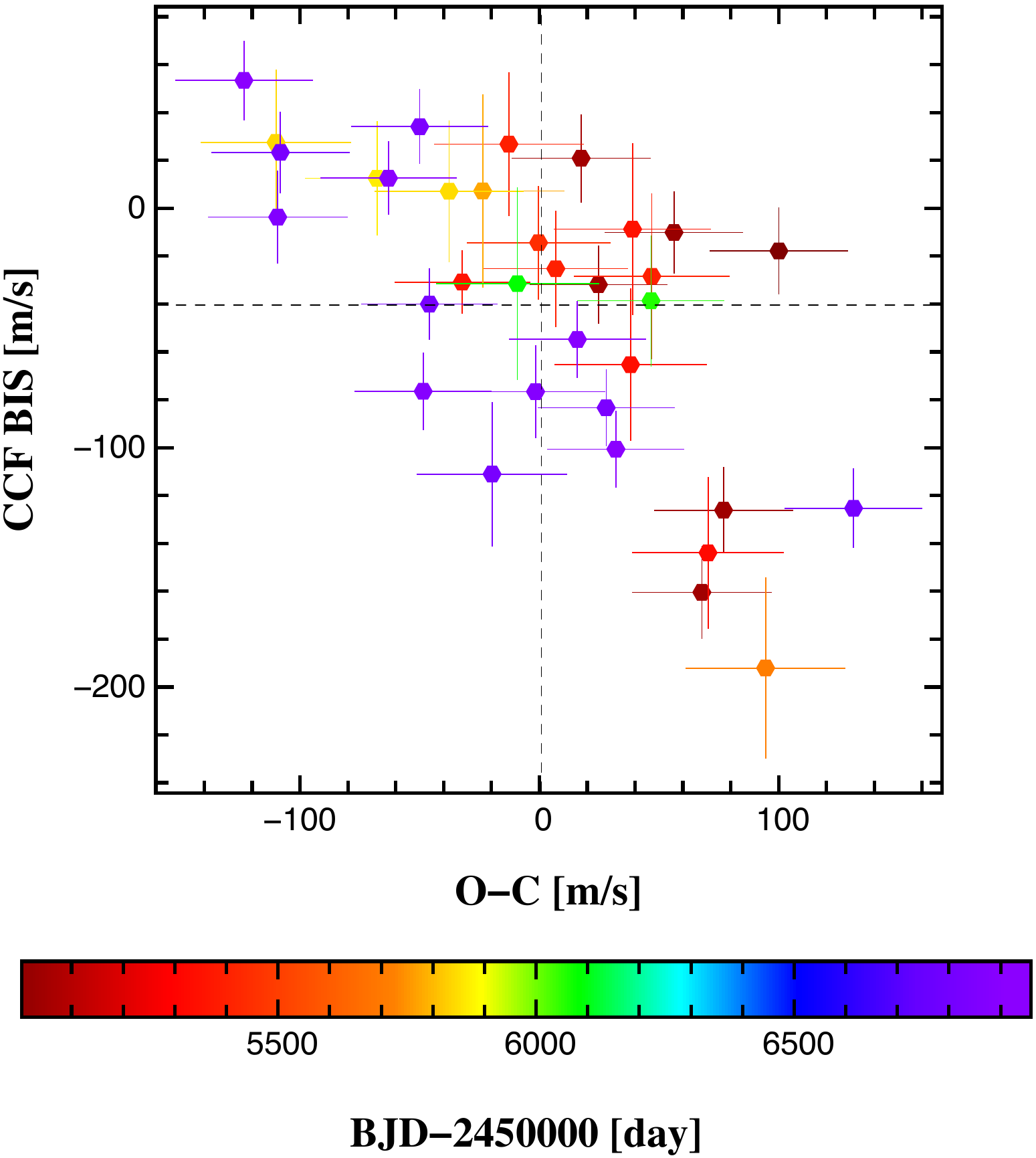}
\end{subfigure}
\begin{subfigure}[b]{0.33\textwidth}
\caption{J0021-16; k1d1}\label{fig:correlation_b}
\includegraphics[width=\textwidth,trim={0 0 0 0},clip]{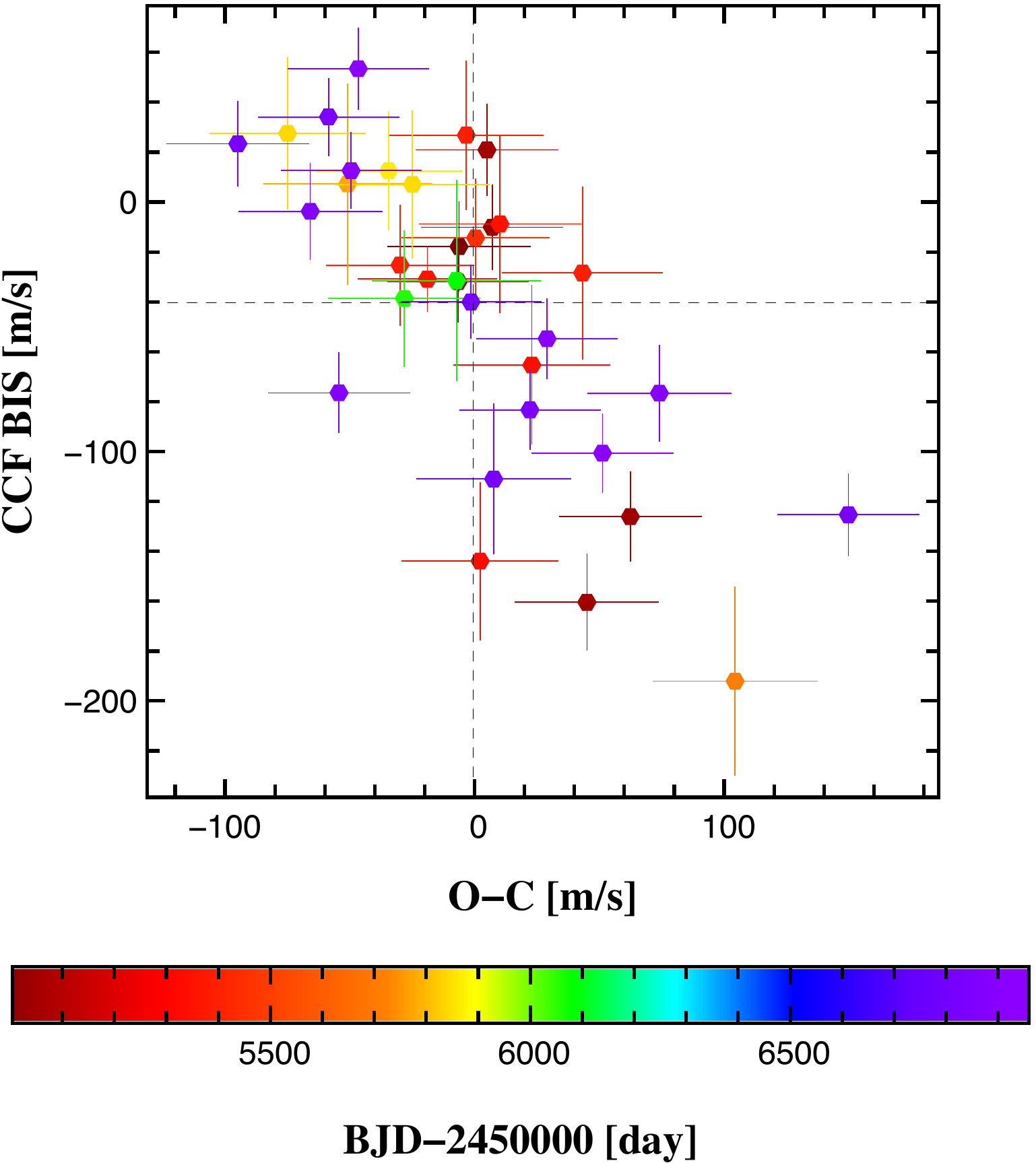}
\end{subfigure}
\begin{subfigure}[b]{0.33\textwidth}
\caption{J2025-45; k1}\label{fig:correlation_c}
\includegraphics[width=\textwidth,trim={0 0 0 0},clip]{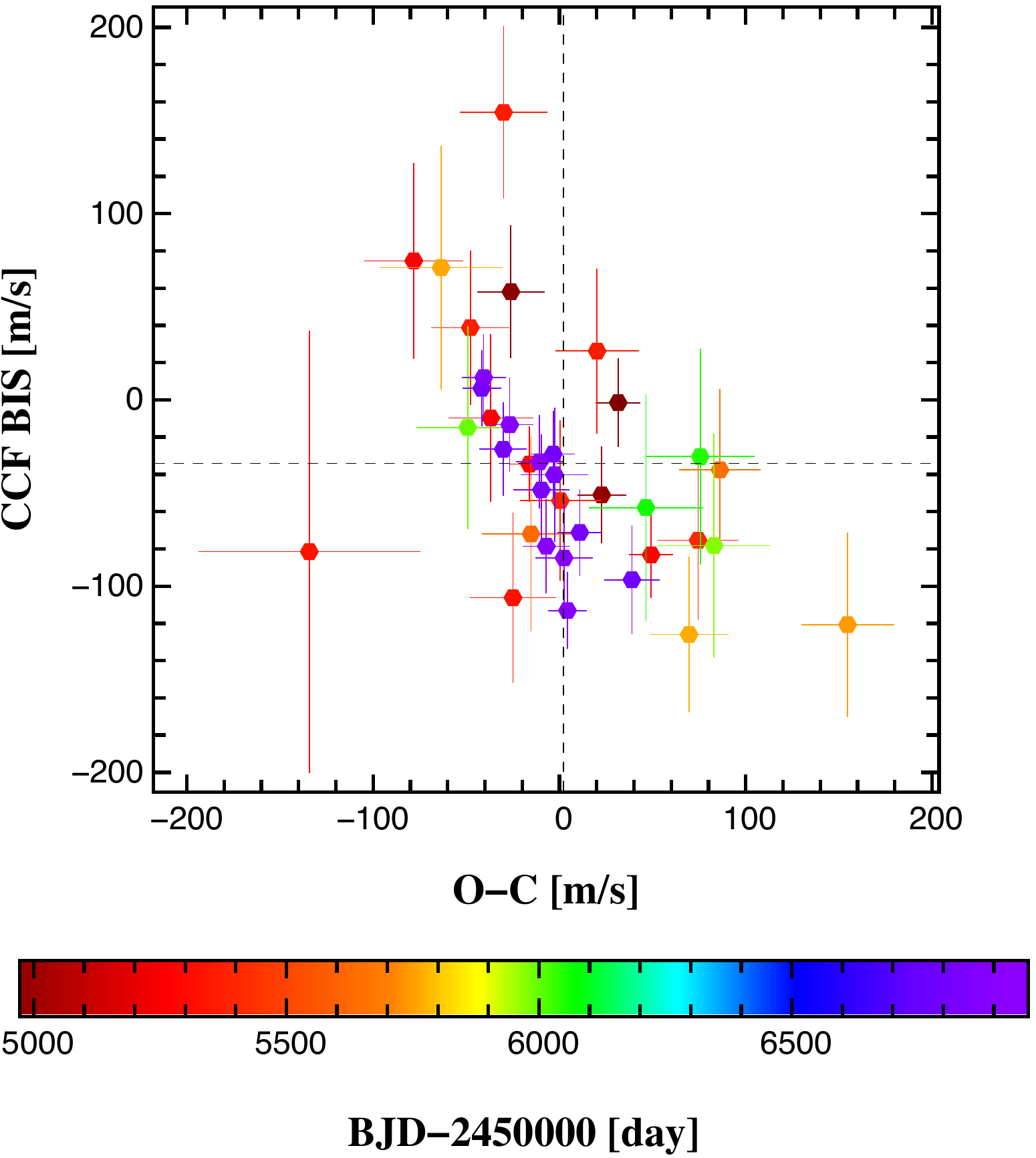}
\end{subfigure}
\caption{Correlation between the bisector and residuals to an eccentric single Keplerian fit for two of our targets showing signs of stellar activity: J0021-16 and J2025-45. This negative linear trend is an indicator of stellar activity.}
\label{fig:correlation}

\end{center}
\end{figure*}

For two of our binaries we see clear signs of stellar activity: J0021-16 and J2025-45. In Fig.~\ref{fig:correlation_a} we plot for J0021-16 the residuals from a single Keplerian fit and in Fig.~\ref{fig:correlation_b} for a single Keplerian plus linear drift. When only a single Keplerian is fitted there is a clear linear negative trend between the residuals and the bisector. However, if we look in more detail we see that the more recent points, denoted in purple, have a systematic shift in bisector to the left in comparison with the older points. This indicates that whilst there is stellar activity present throughout all of the observations, there is an additional source of the residuals. When a linear drift is added, we see in Fig.~\ref{fig:correlation_b} that the latest points in purple now overlap the other points. We conclude that this system has both stellar activity and a tertiary stellar companion inducing a drift. This illustrates the advantage of our long observing baseline, which was 5.34 yr for this target. The case is different for J2025-45, for which there is no discernible difference in the bisector-residuals correlation between the old and new observations. This is not due to a lack of time spent on the target, as the observations span a total of 5.45 yr. We assign a single eccentric Keplerian fit to this target.

%

\section{A discussion on tidal evolution}\label{sec:tides}

One of the scientific advantages of having very precise eccentricities and periods is to allow investigations into tidal interactions. A future work is planned to exploit the results of the EBLM survey in more details and to better our understanding tidal evolution in close binaries. For now, we discuss some of the first order implications that our results may have.

Tidal interactions between two close stars have several effects \citep{Zahn:1975fp,Zahn:1977yq}:

\begin{itemize}
\item Synchronisation of the rotation and orbital periods for circular orbits (pseudo-synchronisation for eccentric orbits \citep{Hut:1981kx}; 
\item Alignment of the orbital and spin axes of the stars
\item Circularisation of the orbit
\end{itemize}

In the plot of FWHM and precision as a function of the binary period (Figure~\ref{fig:precision}) we saw evidence for { tidal} synchronisation, which manifests itself as { spectral lines being more broadened with a reduced orbital period}. Measurements of the Rossiter--McLaughlin effect, which probe the projected obliquity of the orbit, will serve to help better understand the strength of tidal realignment, and confront theoretical expectations \citep[e.g.][]{Anderson:2016lr}.

These are to be presented in a future paper. Finally, the plots of binary eccentricity as a function of semi-major axis and period in Figure~\ref{fig:eccentricity_vs_a_and_P} allow us to probe the circularisation of the orbit. By eye, it is evident that there is a trend of increased eccentricity with semi-major axis and period, which is expected since tides are mostly effective over very short distances. Binaries are not expected to form on primordially circular orbits. Rather, they circularise over time \citep[e.g.][]{Mazeh:1979eu,Fabrycky:2007pd,Bate:2012rt}. A consequence of this is that one may define a cut-off period, $P_{\rm cut}$, above which orbits are found eccentric. \citet{Meibom:2005kx} provided a means of measuring $P_{\rm cut}$ as follows:

\rowcolors{2}{white}{white}
\begin{equation}
\label{eq:mm_fit}
e(P) = \twopartdef{0.0}{P \leq P_{\rm cut}}{\alpha\left(1-e^{\beta(P_{\rm cut}-P)}\right)^{\gamma}}{P>P_{\rm cut}},
\end{equation}
where the constants were calculated to be $\alpha=0.35$, $\beta=0.14$ and $\gamma=1.0$. The value of $\alpha$ is defined based on the mean eccentricity of all field binaries of periods greater than 50 days being 0.35, whilst $\beta$ and $\gamma$ were shown to optimise the fit. In Figure~\ref{fig:eccentricity_vs_a_and_P} we include two versions of our fitted Eq.~\ref{eq:mm_fit}. First, we make a fit to all of the data, and calculate $P_{\rm cut}=8.9$ days. Second, we make a fit to only the binaries where $M_{\rm A}<1.3M_{\odot}$ and there is no sign of a tertiary companion. This second adjustment yields a more accurate $P_{\rm cut}=7.0$ days. We base our preference on the following. Heavier stars have a radiative outer envelope, rather than a convective one \citep{Pinsonneault:2001kx}. Tidal dissipation in radiative envelopes is less efficient, which causes tidal circularisation to be slower  \citep[similar arguments have been invoked in the exoplanet literature;][]{Albrecht:2012lp,Dawson:2014kx}. Additionally, outer tertiary companions may induce some eccentricity in the inner binary via secular perturbations.

Our $P_{\rm cut}$ result comes in conflict with other estimates discussed in the literature and compiled in \citet{Meibom:2005kx} with an update in \citet{Milliman:2014fj}. Results obtained in the field \citep{Duquennoy:1991kx}, the halo \citep{Latham:2002vn}, in M\,67 \citep{Mathieu:1990fk} and NGC\,188 \citep{Mathieu:2004yq}, all with ages $> 1$ Gyr, are found with $P_{\rm cut} > 10$ days. The only exception is NGC\,6819 \citep{Milliman:2014fj} whose value $P_{\rm cut}=6.2\pm1.1$ days is consistent with ours. This is also consistent with results on young open clusters ($< 100$ Myr), where $P_{\rm cut}$ is found around 8 days \citep{Melo:2001yu}. The reason for disagreement between our value and the bulk of other results on old populations is not presently known. A contributing factor may be that our survey is, by design, biased towards small mass ratio binaries, and the circularisation timescale is dependent on the mass ratio \citep{Zahn:1977yq,Zahn:1978db}. Past surveys may also suffer from small number statistics and a poorer  precision on eccentricities than what we can produce nowadays. { This preliminary result is consistent with there only being marginal tidal evolution during} the Main Sequence. We have another 100+ binary systems under observation at the moment and will update our $P_{\rm cut}$ and analysis once observations on those are completed.\\

\section{Conclusion}\label{sec:conclusion}

We present the spectroscopic orbits of 118 stellar systems, all eclipsing single-line binaries. We produce that sample in order to map out the { sky} position of eclipsing systems mimicking transiting hot-Jupiters. This will be of great help in the advent of large scale exoplanet surveys such as {\it TESS} and {\it PLATO}. In addition, this sample can be used as a comparison to hot-Jupiters for a host of topics that are detailed in the introduction. 

Our release of these systems opens multiple opportunity for further research, for instance to detect tertiary companions with direct imaging,  astrometry or eclipse timing variations. As high-resolution, near-infrared spectrographs are coming online, it will become feasible to transform our single-line binaries into double-line systems and derive accurate masses and radii \citep{Torres:2002fj,Brogi:2012fk,Rodler:2012qy}. As part of the efforts of our team, we will double the current sample of eclipsing low-mass binaries (we are currently acquiring data to reach a minimum of 13 radial-velocity measurements), measure some primary eclipses of secondaries with mass $< 0.2 M_\odot$, prepare a publication on the Rossiter--McLaughlin effect of 20 of our binaries. .

We derive orbital periods with a precision of order one second and compute eccentricities well below 1\%. This is thanks to { longterm observations} using a stable, high precision spectrograph usually employed in the discovery of exoplanets. We use our results to carry-out a preliminary investigation of the strength of tidal forces in stars. To a first order, we find that binaries in the field have a similar eccentricity distribution to pre main-sequence binaries, { This is consistent with there being} marginal tidal evolution over a main sequence lifetime, { although further investigation is needed to make a definitive statement.}

{ Ordering our sample as a function of mass, and adding the results, with the same instrument, from WASP and CORALIE on exoplanets, we construct a preliminary mass spectrum. Its appears to show a deficit of planets with masses $> 3 M_{\rm Jup}$ compared to earlier results considering wider orbital separations.}

{ For 21 of our systems there is a significant indication of an outer tertiary companion, most likely stellar in nature and long period.} For four systems we { are actually able to characterise the tertiary orbit, including one system with three stars packed within 1 astronomical unit. } We will also intensify our monitoring of a subset of our systems in search of circumbinary gas-giants, { in a connected program known as BEBOP}


\paragraph{\textbf{Nota Bene}}
We used the Barycentric Julian Dates in our analysis. Our results are based on the equatorial solar and jovian radii and masses taken from Allen's Astrophysical Quantities \citep{Cox:2000vn}

\begin{acknowledgements} 
We benefitted from enlightening discussions with Dan Bayliss, Corinne Charbonnel, Florian Gallet, Dave Latham, Rosemary Mardling, Bob Mathieu, John Papaloizou, Andrei Tokovinin and Yanqin Wu. We also thank Dan Fabrycky for providing comments on the manuscript. The authors would like to attract attention on the help and kind attention of the ESO staff at La Silla  and on the dedication of the many technicians and observers from the University of Geneva, to upkeep the telescope and acquire the data that we present here. We would also like to acknowledge that the {\it Euler} Swiss Telescope at La Silla is a project funded by the Swiss National Science Foundation (SNSF).
Over the time required to collect and analyse the data, AHMJT has received funding from the SNSF, the University of Toronto and now the University of Cambridge. DVM is supported by the SNSF.

This publication makes use of data products from two projects, which were obtained through the  \href{http://simbad.u-strasbg.fr/simbad/}{Simbad} and \href{http://vizier.u-strasbg.fr/viz-bin/VizieR}{VizieR} services hosted at the \href{http://cds.u-strasbg.fr}{CDS-Strasbourg}:
\begin{itemize}
\item The Two Micron All Sky Survey (2MASS), which is a joint project of the University of Massachusetts and the Infrared Processing and Analysis Center/California Institute of Technology, funded by the National Aeronautics and Space Administration and the National Science Foundation \citep{Skrutskie:2006kx}.
\item The Naval Observatory Merged Astrometric Dataset (NOMAD), which is project of the US Naval Observatory \citep{Monet:2003fk}.
\item The Tycho2 catalog \citep{Hog:2000uq}.
\item The AAVSO Photometric All-Sky Survey (APASS), funded by the Robert Martin Ayers Sciences Fund \citep{Henden:2015ad}

\end{itemize}
\end{acknowledgements}

\bibliographystyle{aa}
\bibliography{1Mybib.bib}

\rowcolors{2}{gray!25}{white}

\appendix

\section{Model Selection using the Bayesian Information Criterion}\label{app:bic}

\onecolumn
\begin{landscape}
\tiny
\begin{longtable}{lccc|cc|cc|cc|c|cc|cc|c}
\caption{Bayesian Information Criterion (BIC) with selected model in bold}\label{tab:BIC}\\
\hline 
\multicolumn{4}{c|}{System} & \multicolumn{7}{c|}{Base Models} &  \multicolumn{5}{c}{Complex Models} \\
\hline
\multicolumn{4}{c|}{} & \multicolumn{2}{c|}{k1} & \multicolumn{2}{c|}{k1d1} &  \multicolumn{2}{c|}{k1d2} & &\multicolumn{2}{c|}{k1d3} & \multicolumn{2}{c}{k2} &\\
name & num. & chosen & flag & circ & ecc & circ & ecc & circ & ecc & $\chi_{\rm red}^2$ & circ & ecc  & circ & ecc & $\chi_{\rm red}^2$  \\
 & obs. & model && 4 params. & 6 params. & 5 params. & 7 params. & 6 params. & 8 params. & & 7 params. & 9 params.  & 10 params. & 12 params. &  \\
\hline
\endfirsthead
\hline
\multicolumn{4}{c|}{System} & \multicolumn{7}{c|}{Base Models} &  \multicolumn{5}{c}{Complex Models} \\
\hline
\multicolumn{4}{c|}{} & \multicolumn{2}{c|}{k1} & \multicolumn{2}{c|}{k1d1} &  \multicolumn{2}{c|}{k1d2} & &\multicolumn{2}{c|}{k1d3} & \multicolumn{2}{c|}{k2} &\\
name & num. & chosen & flag & circ & ecc & circ & ecc & circ & ecc & $\chi_{\rm red}^2$ & circ & ecc  & circ & ecc & $\chi_{\rm red}^2$  \\
 & obs. & model && 4 params. & 6 params. & 5 params. & 7 params. & 6 params. & 8 params. & & 7 params. & 9 params.  & 10 params. & 12 params. &  \\
\hline
\endhead
\hline
\multicolumn{14}{l}{Table continues next page...}\\
\hline
\endfoot
\hline
\endlastfoot 
EBLM J0008+02 & 25 & k1d2 (ecc) & drift &455341 &9907 &423270 &229 &432023 & {55} & 1.71& -- & -- & -- & -- & -- \\
EBLM J0017-38 & 13 & k1 (circ) &  & {22} &27 &24 &29 &26 &30 & 1.33& -- & -- & -- & -- & -- \\
EBLM J0021-16 & 34 & k1d1 (circ) & active &221 &178 & {119} &122 &122 &121 & 3.49& -- & -- & -- & -- & -- \\
EBLM J0027-41 & 14 & k1 (ecc) &  &310 & {20} &247 &23 &240 &25 & 0.54& -- & -- & -- & -- & -- \\
EBLM J0035-69 & 21 & k1 (ecc) &  &41514 & {40} &39897 &41 &41171 &42 & 1.42& -- & -- & -- & -- & -- \\
EBLM J0040+01 & 20 & k1 (ecc) &  &13223 & {37} &13051 &39 &12830 &46 & 1.34& -- & -- & -- & -- & -- \\
EBLM J0042-17 & 17 & k1 (ecc) &  &541568 & {34} &522070 &37 &598991 &39 & 1.57& -- & -- & -- & -- & -- \\
EBLM J0048-66 & 18 & k1 (ecc) &  &2897 & {22} &5777 &25 &3767 &26 & 0.42& -- & -- & -- & -- & -- \\
EBLM J0057-19 & 18 & k1 (circ) &  & {24} &23 &21 &23 &26 &26 & 0.9& -- & -- & -- & -- & -- \\
EBLM J0104-38 & 16 & k1d1 (ecc) & drift &92 &41 &52 & {28} &58 &29 & 0.92& -- & -- & -- & -- & -- \\
EBLM J0109-67 & 21 & k1 (ecc) &  &523 & {34} &531 &37 &551 &40 & 1.07& -- & -- & -- & -- & -- \\
EBLM J0218-31 & 45 & k1d2 (circ) & drift &1886 &1789 &276 &275 & {80} &86 & 1.46& -- & -- & -- & -- & -- \\
EBLM J0228+05 & 15 & k1 (circ) &  & {16} &21 &19 &24 &24 &27 & 0.48& -- & -- & -- & -- & -- \\
EBLM J0239-20 & 21 & k1d1 (circ) & drift &45 &48 & {35} &41 &37 &44 & 1.23& -- & -- & -- & -- & -- \\
EBLM J0247-51 & 19 & k1 (circ) &  & {34} &38 &36 &41 &37 &42 & 1.48& -- & -- & -- & -- & -- \\
EBLM J0310-31 & 15 & k1 (ecc) &  &9932207 & {24} &9939367 &27 &9756689 &27 & 0.9& -- & -- & -- & -- & -- \\
EBLM J0315-24 & 21 & k1 (circ) &  & {32} &37 &34 &40 &37 &43 & 1.15& -- & -- & -- & -- & -- \\
EBLM J0326-09 & 14 & k1 (circ) &  & {21} &23 &22 &25 &24 &27 & 1.02& -- & -- & -- & -- & -- \\
EBLM J0339+03 & 15 & k1 (circ) &  & {26} &30 &29 &35 &27 &28 & 1.37& -- & -- & -- & -- & -- \\
EBLM J0351-07 & 21 & k1 (ecc) &  &1488 & {35} &1437 &38 &1945 &41 & 1.12& -- & -- & -- & -- & -- \\
EBLM J0353+05 & 51 & k1d3 (ecc) & drift &52221 &50336 &1593 &2327 &145 &127 & 2.23 &94 & {69} &102 &72 & 0.81\\
EBLM J0353-16 & 29 & k1 (ecc) &  &621 & {59} &590 &63 &746 &67 & 1.69& -- & -- & -- & -- & -- \\
EBLM J0400-51 & 13 & k1 (circ) &  & {15} &17 &17 &20 &22 &22 & 0.52& -- & -- & -- & -- & -- \\
EBLM J0425-46 & 14 & k1 (ecc) &  &20607 & {18} &20581 &21 &21047 &23 & 0.27& -- & -- & -- & -- & -- \\
EBLM J0432-33 & 21 & k1 (circ) &  & {25} &28 &28 &31 &36 &34 & 0.74& -- & -- & -- & -- & -- \\
EBLM J0440-48 & 21 & k1 (circ) &  & {25} &29 &21 &26 &23 &28 & 0.77& -- & -- & -- & -- & -- \\
EBLM J0443-06 & 20 & k1 (ecc) &  &383 & {20} &386 &23 &393 &26 & 0.15& -- & -- & -- & -- & -- \\
EBLM J0454-09 & 19 & k1 (circ) &  & {33} &37 &36 &40 &39 &43 & 1.4& -- & -- & -- & -- & -- \\
EBLM J0500-46 & 13 & k1 (ecc) &  &12029 & {23} &12138 &37 &13398 &48 & 1.02& -- & -- & -- & -- & -- \\
EBLM J0502-38 & 16 & k1 (circ) &  & {17} &21 &20 &23 &23 &26 & 0.5& -- & -- & -- & -- & -- \\
EBLM J0504-09 & 22 & k1 (circ) &  & {49} &52 &52 &55 &55 &57 & 2.05 &58 &54 &42 &48 & 2.42\\
EBLM J0518-39 & 21 & k1 (circ) &  & {34} &35 &37 &38 &40 &42 & 1.28& -- & -- & -- & -- & -- \\
EBLM J0520-06 & 14 & k1 (circ) &  & {14} &17 &17 &20 &20 &22 & 0.37& -- & -- & -- & -- & -- \\
EBLM J0525-55 & 14 & k1 (circ) &  & {40} &32 &63 &40 &126 &35 & 2.93 &210 &37 &35 &32 & 2.72\\
EBLM J0526+04 & 14 & k1 (circ) &  & {25} &27 &31 &30 &28 &30 & 1.48& -- & -- & -- & -- & -- \\
EBLM J0526-34 & 21 & k1 (ecc) &  &83506 & {37} &81877 &39 &81525 &40 & 1.26& -- & -- & -- & -- & -- \\
EBLM J0540-17 & 18 & k1d3 (ecc) & drift &3785 &3391 &329 &336 &63 &57 & 3.79 &140 & {38} &38 &43 & 1.34\\
EBLM J0543-57 & 35 & k2 (circ) & triple &205949 &200239 &29371 &27488 &1537 &1178 & 42.57 &514 &211 & {59} &62 & 0.95\\
EBLM J0546-18 & 21 & k1 (circ) &  & {28} &30 &25 &29 &29 &32 & 0.93& -- & -- & -- & -- & -- \\
EBLM J0608-59 & 21 & k1 (ecc) &  &166959 & {28} &146603 &31 &205515 &35 & 0.68& -- & -- & -- & -- & -- \\
EBLM J0610-52 & 19 & k1 (circ) &  & {14} &19 &16 &22 &19 &25 & 0.13& -- & -- & -- & -- & -- \\
EBLM J0621-46 & 19 & k1 (circ) &  & {14} &19 &17 &22 &19 &24 & 0.13& -- & -- & -- & -- & -- \\
EBLM J0621-50 & 25 & k1 (circ) &  & {45} &39 &48 &41 &51 &44 & 1.52& -- & -- & -- & -- & -- \\
EBLM J0623-27 & 14 & k1 (ecc) &  &2782 & {21} &2399 &24 &3847 &26 & 0.66& -- & -- & -- & -- & -- \\
EBLM J0625-43 & 21 & k1 (circ) &  & {28} &30 &31 &33 &34 &34 & 0.93& -- & -- & -- & -- & -- \\
EBLM J0627-67 & 24 & k1 (ecc) &  &21450 & {26} &21452 &29 &21561 &32 & 0.38& -- & -- & -- & -- & -- \\
EBLM J0627-59 & 13 & k1 (circ) &  & {15} &19 &20 &19 &36 &21 & 0.56& -- & -- & -- & -- & -- \\
EBLM J0629-67 & 15 & k1d3 (ecc) & drift &3564 &1508 &188 &150 &286 &71 & 7.01 &233 & {62} &277 &54 & 6.35\\
EBLM J0642-60 & 16 & k1 (circ) &  & {25} &26 &28 &28 &31 &31 & 1.2& -- & -- & -- & -- & -- \\
EBLM J0645-61 & 36 & k1 (ecc) &  &2249 & {28} &2241 &32 &2550 &35 & 0.22& -- & -- & -- & -- & -- \\
EBLM J0645-26 & 22 & k1 (ecc) &  &3296 & {23} &3167 &26 &3525 &28 & 0.26& -- & -- & -- & -- & -- \\
EBLM J0649-27 & 20 & k1 (circ) &  & {32} &27 &34 &30 &38 &33 & 1.24& -- & -- & -- & -- & -- \\
EBLM J0650-34 & 13 & k1 (circ) &  & {12} &15 &15 &18 &18 &21 & 0.19& -- & -- & -- & -- & -- \\
EBLM J0659-61 & 19 & k1d2 (ecc) & drift &63 &65 &72 &41 &75 & {35} & 1.01& -- & -- & -- & -- & -- \\
EBLM J0700-30 & 13 & k1 (circ) &  & {17} &21 &26 &31 &36 &25 & 0.78& -- & -- & -- & -- & -- \\
EBLM J0709-52 & 16 & k1 (ecc) &  &3943 & {18} &3958 &21 &4061 &23 & 0.18& -- & -- & -- & -- & -- \\
EBLM J0801+02 & 13 & k1 (circ) &  & {21} &25 &23 &27 &25 &30 & 1.19& -- & -- & -- & -- & -- \\
EBLM J0851+05 & 16 & k1 (circ) &  & {17} &22 &19 &24 &21 &26 & 0.46& -- & -- & -- & -- & -- \\
EBLM J0855+04 & 22 & k1 (ecc) &  &545 & {28} &547 &30 &553 &33 & 0.6& -- & -- & -- & -- & -- \\
EBLM J0941-31 & 21 & k1 (ecc) &  &24573 & {47} &23997 &52 &32212 &47 & 1.89& -- & -- & -- & -- & -- \\
EBLM J0948-08 & 26 & k1d2 (ecc) & drift &120780 &2705 &92104 &154 &95172 & {49} & 1.3& -- & -- & -- & -- & -- \\
EBLM J0954-23 & 21 & k1 (ecc) &  &648 & {28} &621 &31 &768 &34 & 0.67& -- & -- & -- & -- & -- \\
EBLM J0954-45 & 23 & k1 (ecc) &  &184316 & {33} &179057 &36 &201067 &39 & 0.86& -- & -- & -- & -- & -- \\
EBLM J0955-39 & 23 & k1 (circ) &  & {34} &38 &36 &44 &34 &40 & 1.11& -- & -- & -- & -- & -- \\
EBLM J1007-40 & 21 & k1 (circ) &  & {17} &23 &19 &25 &22 &28 & 0.28& -- & -- & -- & -- & -- \\
EBLM J1008-29 & 13 & k1 (circ) &  & {12} &16 &14 &19 &17 &21 & 0.15& -- & -- & -- & -- & -- \\
EBLM J1013+01 & 21 & k1 (circ) &  & {34} &38 &37 &40 &41 &43 & 1.31& -- & -- & -- & -- & -- \\
EBLM J1014-07 & 24 & k1d1 (ecc) & drift &66406 &316 &66383 & {34} &67380 &36 & 0.71& -- & -- & -- & -- & -- \\
EBLM J1023-43 & 16 & k1 (circ) &  & {14} &19 &16 &21 &20 &24 & 0.25& -- & -- & -- & -- & -- \\
EBLM J1034-29 & 24 & k1 (circ) &  & {27} &32 &29 &34 &33 &35 & 0.73& -- & -- & -- & -- & -- \\
EBLM J1037-25 & 20 & k1 (ecc) &  &29835 & {33} &29440 &36 &32168 &36 & 1.05& -- & -- & -- & -- & -- \\
EBLM J1037-45 & 13 & k1 (circ) &  & {11} &16 &13 &18 &17 &21 & 0.07& -- & -- & -- & -- & -- \\
EBLM J1038-37 & 13 & k1d3 (ecc) & drift &2998 &2583 &420 &374 &68 &42 & 4.29 &42 & {27} &28 &32 & 1.05\\
EBLM J1104-43 & 18 & k1 (circ) &  & {14} &19 &16 &22 &20 &25 & 0.15& -- & -- & -- & -- & -- \\
EBLM J1105-13 & 17 & k1 (circ) &  & {23} &26 &25 &29 &26 &31 & 0.87& -- & -- & -- & -- & -- \\
EBLM J1116-32 & 22 & k1 (circ) &  & {21} &26 &24 &29 &28 &31 & 0.46& -- & -- & -- & -- & -- \\
EBLM J1116-01 & 14 & k1 (circ) &  & {13} &18 &15 &20 &18 &22 & 0.28& -- & -- & -- & -- & -- \\
EBLM J1141-37 & 21 & k1 (circ) &  & {27} &31 &29 &33 &32 &36 & 0.87& -- & -- & -- & -- & -- \\
EBLM J1146-42 & 13 & k2 (ecc) & triple &674940 &472743 &703101 &484807 &704196 &352448 & 52039.93 &175583 &100473 &156145 & {46} & 15.06\\
EBLM J1201-36 & 15 & k1 (ecc) &  &27031 & {22} &27034 &30 &27088 &28 & 0.67& -- & -- & -- & -- & -- \\
EBLM J1208-29 & 20 & k1 (ecc) &  &81 & {21} &81 &23 &87 &25 & 0.22& -- & -- & -- & -- & -- \\
EBLM J1219-39 & 22 & k1 (ecc) &  &32382 & {45} &29437 &66 &39664 &71 & 1.66& -- & -- & -- & -- & -- \\
EBLM J1301-37 & 13 & k1 (ecc) &  &1278 & {21} &1292 &22 &1440 &24 & 0.87& -- & -- & -- & -- & -- \\
EBLM J1305-31 & 17 & k1 (ecc) &  &6128 & {29} &3074 &42 &10852 &41 & 1.12& -- & -- & -- & -- & -- \\
EBLM J1420-07 & 20 & k1 (ecc) &  &595 & {21} &587 &24 &803 &27 & 0.23& -- & -- & -- & -- & -- \\
EBLM J1431-11 & 19 & k1 (circ) &  & {19} &24 &23 &27 &28 &30 & 0.48& -- & -- & -- & -- & -- \\
EBLM J1433-43 & 16 & k1 (circ) &  & {18} &23 &21 &26 &23 &27 & 0.6& -- & -- & -- & -- & -- \\
EBLM J1436-13 & 22 & k1 (circ) &  & {32} &38 &35 &41 &35 &41 & 1.09& -- & -- & -- & -- & -- \\
EBLM J1500-33 & 25 & k1 (ecc) &  &484 & {34} &487 &36 &487 &38 & 0.76& -- & -- & -- & -- & -- \\
EBLM J1509-10 & 20 & k1 (circ) &  & {29} &29 &33 &32 &48 &35 & 1.05& -- & -- & -- & -- & -- \\
EBLM J1525-36 & 22 & k1 (circ) &  & {28} &34 &31 &37 &45 &40 & 0.88& -- & -- & -- & -- & -- \\
EBLM J1559-05 & 18 & k1 (circ) &  & {24} &29 &22 &27 &24 &29 & 0.88& -- & -- & -- & -- & -- \\
EBLM J1630+10 & 20 & k1d1 (ecc) & drift &105475 &37 &101974 & {25} &102677 &27 & 0.29& -- & -- & -- & -- & -- \\
EBLM J1928-38 & 17 & k1 (ecc) &  &32439 & {29} &32265 &44 &32669 &55 & 1.05& -- & -- & -- & -- & -- \\
EBLM J1934-42 & 14 & k1 (circ) &  & {20} &24 &374 &48 &1283 &127 & 0.95& -- & -- & -- & -- & -- \\
EBLM J1944-20 & 13 & k1 (circ) &  & {11} &16 &13 &18 &16 &21 & 0.04& -- & -- & -- & -- & -- \\
EBLM J1947-23 & 16 & k1d3 (circ) & drift &48871 &617 &15503 &610 &13903 &76 & 6.77 & {47} &51 &76 &83 & 3.07\\
EBLM J2011-71 & 23 & k2 (ecc) & triple &1663926 &1446864 &1597818 &1471188 &348043 &258310 & 15317.89 &465403 &237489 &199164 & {58} & 1.88\\
EBLM J2025-45 & 36 & k1 (ecc) & active &476726 & {234} &471748 &397 &487099 &465 & 7.09& -- & -- & -- & -- & -- \\
EBLM J2027+03 & 15 & k1 (circ) &  & {19} &24 &21 &26 &25 &29 & 0.75& -- & -- & -- & -- & -- \\
EBLM J2040-41 & 16 & k1 (ecc) &  &58162 & {21} &57381 &25 &56470 &30 & 0.47& -- & -- & -- & -- & -- \\
EBLM J2043-18 & 15 & k1 (circ) &  & {17} &20 &19 &24 &19 &24 & 0.53& -- & -- & -- & -- & -- \\
EBLM J2046-40 & 29 & k2 (ecc) & triple &420157 &29902 &336728 &93888 &361383 &13878 & 659.56 &326950 &180 &242747 & {49} & 0.49\\
EBLM J2046+06 & 14 & k1 (ecc) &  &506924 & {26} &490452 &28 &478984 &31 & 1.32& -- & -- & -- & -- & -- \\
EBLM J2101-45 & 20 & k1 (ecc) &  &43664 & {28} &42704 &30 &47989 &31 & 0.72& -- & -- & -- & -- & -- \\
EBLM J2104-46 & 20 & k1d3 (ecc) & drift &16859 &16233 &2185 &1537 &226 &80 & 4.66 &204 & {39} &116 &43 & 1.1\\
EBLM J2107-39 & 20 & k1 (circ) &  & {36} &34 &51 &30 &53 &32 & 1.47& -- & -- & -- & -- & -- \\
EBLM J2122-32 & 13 & k1 (ecc) &  &2810119 & {32} &2783724 &41 &1181333 &83601 & 2.41 &2786465 &42 &1546605 &32 & 4.67\\
EBLM J2153-55 & 16 & k1 (circ) &  & {20} &25 &22 &26 &24 &28 & 0.75& -- & -- & -- & -- & -- \\
EBLM J2207-41 & 13 & k1 (ecc) &  &3936 & {25} &3783 &28 &3605 &25 & 1.37& -- & -- & -- & -- & -- \\
EBLM J2210-48 & 15 & k1d1 (circ) & drift &41 &37 & {30} &35 &28 &29 & 1.67& -- & -- & -- & -- & -- \\
EBLM J2217-04 & 15 & k1 (ecc) &  &528 & {19} &528 &21 &534 &24 & 0.28& -- & -- & -- & -- & -- \\
EBLM J2232-31 & 13 & k1d1 (circ) & drift &29 &30 & {20} &23 &22 &25 & 0.88& -- & -- & -- & -- & -- \\
EBLM J2236-36 & 18 & k1 (circ) &  & {17} &23 &20 &26 &23 &28 & 0.41& -- & -- & -- & -- & -- \\
EBLM J2308-46 & 19 & k1 (circ) &  & {19} &22 &21 &24 &24 &27 & 0.45& -- & -- & -- & -- & -- \\
EBLM J2330-61 & 16 & k1 (circ) &  & {19} &24 &22 &26 &31 &29 & 0.7& -- & -- & -- & -- & -- \\
EBLM J2349-32 & 20 & k1 (circ) &  & {18} &24 &19 &27 &22 &27 & 0.36& -- & -- & -- & -- & -- \\
EBLM J2353-10 & 15 & k1 (circ) &  & {16} &20 &18 &23 &34 &26 & 0.5& -- & -- & -- & -- & -- \\

\end{longtable}
\end{landscape}

\section{Binary orbital parameters}\label{app:params}

\rowcolors{2}{gray!25}{white}

\onecolumn

\begin{landscape}
\tiny
\begin{longtable}{cccccccccccccc}
\caption{Orbital parameters from the selected models.}\label{tab:params}\\
\hline
name & $P$ & $a$ & $K$ & $e$ & $\omega$ & $T_{\rm peri}$ & $f(m)$ & $m_{\rm A}$ & $m_{\rm B}$ & lin & quad  & cubic  \\
 & [day] & [AU] & [km/s] &  & [deg] & [BJD-2,455,000] & [$10^{-3} M_{\odot}$] & [$M_{\odot}$] & [$M_{\odot}$] & [m/s/yr] & [m/s$^2$/yr] & [m/s$^3$/yr]  \\
\hline
\endfirsthead
\hline 
name & $P$ & $a$ & $K$ & $e$ & $\omega$ & $T_{\rm peri}$ & $f(m)$ & $m_{\rm A}$ & $m_{\rm B}$ & lin & quad  & cubic  \\
 & [day] & [AU] & [km/s] &  & [deg] & [BJD-2,455,000] & [$10^{-3} M_{\odot}$] & [$M_{\odot}$] & [$M_{\odot}$] & [m/s/yr] & [m/s$^2$/yr] & [m/s$^3$/yr]  \\
\hline
\endhead
\hline
\multicolumn{14}{c}{Table continues next page...}\\
\hline
\endfoot
\hline
\endlastfoot 
EBLM J0008+02&4.7222907(63)&0.0668(17)&16.2711(88)&0.24476(44)&-51.10(15)&1707.4700(19)&1.9212(65)&1.60(11)&0.183(28)&-377.092(25)&71.395(19)&$--$\\
EBLM J0017-38&6.34008(13)&0.0747(19)&17.08(11)&$<$0.015&$--$&1790.1599(67)&3.271(64)&1.200(80)&0.184(24)&$<$203&$--$&$--$\\
EBLM J0021-16&5.9672751(63)&0.0703(17)&19.060(13)&$<$0.0016&$--$&1067.82406(80)&4.2810(88)&1.110(70)&0.194(22)&-23.3(4.8)&$--$&$--$\\
EBLM J0027-41&4.9279889(71)&0.0677(18)&46.223(41)&0.0243(11)&65.3(2.4)&1311.199(32)&50.38(30)&1.180(70)&0.528(65)&$<$22.5&$--$&$--$\\
EBLM J0035-69&8.414620(24)&0.0899(20)&17.335(24)&0.2465(14)&-12.79(45)&1230.4976(95)&4.134(39)&1.170(70)&0.198(20)&$<$33&$--$&$--$\\
EBLM J0040+01&7.2348410(79)&0.0710(18)&11.8654(66)&0.06959(66)&5.72(43)&1399.4294(85)&1.2431(47)&0.810(60)&0.102(10)&$<$3.8&$--$&$--$\\
EBLM J0042-17&10.3475294(22)&0.1118(27)&43.3961(37)&0.051004(86)&-71.60(10)&772.6373(30)&87.276(46)&1.100(70)&0.642(56)&$<$1.4&$--$&$--$\\
EBLM J0048-66&6.649275(17)&0.0825(23)&35.631(25)&0.06876(80)&-62.93(65)&1313.619(12)&30.94(15)&1.250(90)&0.447(55)&$<$55&$--$&$--$\\
EBLM J0057-19&4.300510(15)&0.0554(15)&15.523(25)&$<$0.0062&$--$&631.2020(14)&1.6668(80)&1.090(80)&0.136(19)&$<$45&$--$&$--$\\
EBLM J0104-38&8.256058(12)&0.0945(27)&20.627(11)&0.00285(50)&-112(13)&827.93(30)&7.507(24)&1.38(11)&0.274(33)&16.5(5.8)&$--$&$--$\\
EBLM J0109-67&9.029996(46)&0.1044(28)&46.608(26)&0.01629(76)&6.2(2.5)&1015.478(62)&94.69(38)&1.170(80)&0.689(68)&$<$72.9&$--$&$--$\\
EBLM J0218-31&8.8841033(48)&0.0967(25)&27.7851(58)&$<$0.00037&$--$&1126.45624(33)&19.745(12)&1.170(80)&0.359(36)&-72.468400(35)&12.5(1.2)&$--$\\
EBLM J0228+05&6.634727(18)&0.0826(22)&14.2270(67)&$<$0.0013&$--$&1786.0418(11)&1.9795(28)&1.53(11)&0.180(23)&$<$16&$--$&$--$\\
EBLM J0239-20&2.7786835(54)&0.0425(12)&21.316(36)&$<$0.0032&$--$&459.01896(92)&2.788(14)&1.160(80)&0.170(29)&75(23)&$--$&$--$\\
EBLM J0247-51&4.007851(38)&0.0564(19)&23.683(50)&$<$0.0061&$--$&1321.1096(14)&5.516(35)&1.26(11)&0.231(38)&$<$110&$--$&$--$\\
EBLM J0310-31&12.642818(22)&0.1260(36)&27.8645(29)&0.308466(98)&-174.215(27)&1670.3636(10)&24.393(18)&1.26(10)&0.408(41)&$<$30.2&$--$&$--$\\
EBLM J0315-24&3.190524(13)&0.0493(14)&27.692(59)&$<$0.0036&$--$&687.8683(13)&7.020(45)&1.310(90)&0.258(45)&$<$61&$--$&$--$\\
EBLM J0326-09&2.400396(38)&0.0388(11)&25.11(35)&$<$0.041&$--$&1608.2352(60)&3.94(17)&1.160(80)&0.193(38)&$<$591&$--$&$--$\\
EBLM J0339+03&3.5806689(86)&0.0533(26)&24.877(49)&$<$0.0036&$--$&1042.75845(87)&5.712(34)&1.33(18)&0.242(51)&$<$39&$--$&$--$\\
EBLM J0351-07&4.080905(20)&0.0576(19)&24.331(27)&0.0494(14)&98.2(1.5)&642.827(17)&6.068(46)&1.29(11)&0.242(40)&$<$41&$--$&$--$\\
EBLM J0353+05&6.8620266(21)&0.0785(30)&16.3058(26)&0.00108(17)&-107.3(8.2)&1111.61(16)&3.0823(31)&1.19(13)&0.179(26)&236(65)&-30.07(42)&2.7849(28)\\
EBLM J0353-16&11.761212(13)&0.1130(27)&16.6432(42)&0.00524(29)&40.9(2.8)&869.216(92)&5.6176(92)&1.170(80)&0.222(21)&$<$1.8&$--$&$--$\\
EBLM J0400-51&2.692078(48)&0.0436(14)&30.73(13)&$<$0.019&$--$&1592.7123(26)&8.09(10)&1.26(10)&0.266(52)&$<$570&$--$&$--$\\
EBLM J0425-46&16.587865(39)&0.1553(37)&35.1977(72)&0.04772(17)&16.15(32)&1713.054(15)&74.688(86)&1.190(80)&0.627(51)&$<$18&$--$&$--$\\
EBLM J0432-33&5.305486(23)&0.0693(18)&23.424(16)&$<$0.0031&$--$&986.5030(11)&7.065(15)&1.320(90)&0.260(35)&$<$51&$--$&$--$\\
EBLM J0440-48&2.543040(47)&0.0416(13)&22.82(12)&$<$0.015&$--$&1006.1964(24)&3.132(51)&1.29(10)&0.190(39)&$<$660&$--$&$--$\\
EBLM J0443-06&3.1119219(43)&0.0523(21)&53.665(61)&0.0590(12)&27.4(1.2)&992.238(11)&49.57(36)&1.39(13)&0.58(11)&$<$107&$--$&$--$\\
EBLM J0454-09&5.0134508(65)&0.0709(28)&57.640(26)&$<$0.0018&$--$&415.78686(44)&99.48(13)&1.18(12)&0.71(10)&$<$23&$--$&$--$\\
EBLM J0500-46&8.28437(11)&0.0890(22)&15.914(34)&0.2346(28)&-8.23(38)&2210.0597(68)&3.178(55)&1.190(80)&0.182(21)&$<$270&$--$&$--$\\
EBLM J0502-38&3.256303(20)&0.0499(16)&32.68(11)&$<$0.01&$--$&1172.0773(18)&11.78(12)&1.26(10)&0.307(54)&$<$154&$--$&$--$\\
EBLM J0504-09&2.6989669(77)&0.0457(14)&28.215(96)&$<$0.0098&$--$&1422.2567(19)&6.281(64)&1.48(11)&0.268(54)&$<$66&$--$&$--$\\
EBLM J0518-39&3.6497977(70)&0.0553(15)&21.438(23)&$<$0.0032&$--$&579.97039(75)&3.726(12)&1.47(10)&0.220(37)&$<$20&$--$&$--$\\
EBLM J0520-06&2.131514(38)&0.0386(17)&58.84(65)&$<$0.047&$--$&796.4118(53)&45.0(1.5)&1.18(11)&0.50(11)&$<$780&$--$&$--$\\
EBLM J0525-55&4.800097(55)&0.0649(21)&25.157(43)&$<$0.0059&$--$&2061.7363(21)&7.919(41)&1.31(11)&0.270(41)&$<$180&$--$&$--$\\
EBLM J0526+04&4.0310137(73)&0.0550(17)&21.032(35)&$<$0.0048&$--$&1327.63931(91)&3.885(19)&1.17(10)&0.193(30)&$<$53.3&$--$&$--$\\
EBLM J0526-34&10.1909001(65)&0.1095(32)&23.598(13)&0.12609(44)&-163.78(18)&700.0666(50)&13.546(43)&1.35(11)&0.338(38)&$<$9.1&$--$&$--$\\
EBLM J0540-17&6.004884(16)&0.0718(20)&16.199(10)&0.00029(57)&-164(10)&1851.0(1.7)&2.6444(96)&1.200(90)&0.171(22)&709.3(1.2)&308.6186(90)&-106.37420(10)\\
EBLM J0543-57&4.4638343(29)&0.0592(18)&16.6460(60)&$<$0.0018&$--$&903.33248(40)&2.1332(23)&1.23(10)&0.160(25)&$--$&$--$&$--$\\
EBLM J0546-18&3.191910(21)&0.0479(14)&26.15(10)&$<$0.015&$--$&612.0106(24)&5.912(69)&1.210(90)&0.231(40)&$<$240&$--$&$--$\\
EBLM J0608-59&14.608515(32)&0.1346(32)&21.6144(77)&0.15606(32)&117.48(13)&562.0090(50)&14.729(32)&1.200(80)&0.325(29)&$<$5&$--$&$--$\\
EBLM J0610-52&2.4169923(76)&0.0449(17)&58.41(15)&$<$0.0036&$--$&652.98202(93)&49.90(38)&1.47(11)&0.60(12)&$<$288&$--$&$--$\\
EBLM J0621-46&1.550830(19)&0.0294(11)&32.25(36)&$<$0.049&$--$&701.2661(37)&5.39(18)&1.19(10)&0.221(58)&$<$852&$--$&$--$\\
EBLM J0621-50&4.9638415(31)&0.0673(22)&37.537(26)&$<$0.0026&$--$&837.75550(55)&27.201(57)&1.23(10)&0.420(60)&$<$10.9&$--$&$--$\\
EBLM J0623-27&5.777931(20)&0.0719(19)&28.051(19)&0.0571(12)&33.94(93)&2042.051(15)&13.150(76)&1.180(80)&0.308(38)&$<$149&$--$&$--$\\
EBLM J0625-43&3.968989(11)&0.0555(16)&30.570(65)&$<$0.0075&$--$&705.5561(12)&11.748(75)&1.160(80)&0.291(42)&$<$45&$--$&$--$\\
EBLM J0627-67&9.468894(20)&0.1018(30)&17.277(10)&0.15879(65)&84.57(24)&771.8274(62)&4.869(20)&1.34(11)&0.229(27)&$<$9.1&$--$&$--$\\
EBLM J0627-59&5.72958(19)&0.0725(20)&34.564(64)&$<$0.0045&$--$&2039.5308(23)&24.51(14)&1.160(80)&0.389(47)&$<$520&$--$&$--$\\
EBLM J0629-67&18.2903(86)&0.1678(51)&23.07(46)&0.116(25)&140.0(8.4)&1136.45(42)&22.8(3.3)&1.45(11)&0.432(60)&-3908.879900(10)&-9021.84(66)&-108178.741(13)\\
EBLM J0642-60&5.011537(39)&0.0715(26)&22.515(36)&$<$0.0055&$--$&1103.0059(16)&5.927(29)&1.66(16)&0.282(48)&$<$108&$--$&$--$\\
EBLM J0645-61&4.453665(18)&0.0618(20)&13.431(26)&0.2040(19)&21.99(54)&896.1007(60)&1.049(14)&1.45(13)&0.138(24)&$<$30&$--$&$--$\\
EBLM J0645-26&7.564792(22)&0.0889(27)&30.425(20)&0.07786(73)&119.53(55)&669.406(12)&21.875(94)&1.25(10)&0.389(48)&$<$49&$--$&$--$\\
EBLM J0649-27&4.3080484(29)&0.0633(22)&32.522(16)&$<$0.0023&$--$&802.73526(41)&15.354(23)&1.45(13)&0.371(61)&$<$15&$--$&$--$\\
EBLM J0650-34&8.95770(73)&0.1072(35)&19.57(77)&$<$0.12&$--$&1099.391(53)&6.95(83)&1.74(15)&0.308(52)&$<$620&$--$&$--$\\
EBLM J0659-61&4.235638(12)&0.0601(19)&43.638(32)&0.00364(93)&-168(14)&1711.52(17)&36.47(18)&1.160(90)&0.457(65)&-265.62(21)&126.33(18)&$--$\\
EBLM J0700-30&6.545624(96)&0.0835(28)&36.782(39)&$<$0.0034&$--$&1974.3493(22)&33.75(11)&1.33(12)&0.480(65)&$<$360&$--$&$--$\\
EBLM J0709-52&9.10801(43)&0.0971(34)&30.266(61)&0.3404(17)&-44.34(43)&1689.0658(87)&21.75(30)&1.11(11)&0.361(44)&$<$273&$--$&$--$\\
EBLM J0801+02&3.348823(16)&0.0520(15)&20.373(56)&$<$0.0089&$--$&1711.6792(20)&2.934(24)&1.47(11)&0.202(37)&$<$62&$--$&$--$\\
EBLM J0851+05&2.553695(39)&0.0407(12)&20.11(31)&$<$0.048&$--$&929.9936(60)&2.152(99)&1.220(90)&0.160(33)&$<$590&$--$&$--$\\
EBLM J0855+04&2.2269646(58)&0.0399(11)&21.136(51)&0.0645(22)&-165.0(2.1)&843.480(13)&2.165(31)&1.52(10)&0.185(41)&$<$105.8&$--$&$--$\\
EBLM J0941-31&5.545628(18)&0.0687(21)&21.312(36)&0.2006(17)&5.02(52)&340.5756(79)&5.230(60)&1.19(10)&0.218(31)&$<$32.4&$--$&$--$\\
EBLM J0948-08&5.3798003(13)&0.0768(26)&50.3024(87)&0.04918(20)&4.41(20)&872.2636(30)&70.690(82)&1.41(12)&0.675(96)&-202.858(23)&18.214(12)&$--$\\
EBLM J0954-23&7.574635(26)&0.0873(23)&8.677(12)&0.0428(14)&-107.6(1.6)&1360.974(34)&0.5112(42)&1.44(11)&0.107(14)&$<$21.4&$--$&$--$\\
EBLM J0954-45&8.0726432(86)&0.1009(40)&27.886(17)&0.29592(59)&63.27(12)&741.3379(26)&15.809(69)&1.69(19)&0.412(62)&$<$13&$--$&$--$\\
EBLM J0955-39&5.313599(12)&0.0675(26)&21.446(34)&$<$0.0042&$--$&458.3077(14)&5.430(26)&1.23(13)&0.226(36)&$<$29&$--$&$--$\\
EBLM J1007-40&3.9360378(68)&0.0604(23)&33.208(27)&$<$0.0016&$--$&599.73222(60)&14.934(36)&1.52(15)&0.377(68)&$<$53&$--$&$--$\\
EBLM J1008-29&10.400866(55)&0.1190(31)&22.026(11)&$<$0.0016&$--$&1861.9844(11)&11.516(17)&1.71(12)&0.368(41)&$<$12.8&$--$&$--$\\
EBLM J1013+01&2.8922811(87)&0.0414(13)&23.193(80)&$<$0.0089&$--$&729.4973(16)&3.739(39)&0.960(80)&0.168(27)&$<$72&$--$&$--$\\
EBLM J1014-07&4.5574702(33)&0.0621(21)&23.696(12)&0.20558(62)&-44.72(27)&822.2464(29)&5.888(23)&1.30(12)&0.241(39)&115.5(5.8)&$--$&$--$\\
EBLM J1023-43&3.684071(25)&0.0628(23)&64.84(30)&$<$0.015&$--$&716.1783(27)&104.0(1.4)&1.58(12)&0.85(14)&$<$740&$--$&$--$\\
EBLM J1034-29&2.1742624(40)&0.0383(12)&18.229(52)&$<$0.007&$--$&841.46294(86)&1.365(12)&1.44(11)&0.151(34)&$<$64.8&$--$&$--$\\
EBLM J1037-25&4.9365623(34)&0.0652(20)&24.797(15)&0.12075(68)&-74.07(33)&767.0382(42)&7.629(31)&1.26(10)&0.260(38)&$<$7.9&$--$&$--$\\
EBLM J1037-45&1.593894(16)&0.0311(14)&37.5(1.1)&$<$0.089&$--$&700.9886(63)&8.68(75)&1.30(13)&0.279(82)&$<$670&$--$&$--$\\
EBLM J1038-37&5.021663(30)&0.0633(16)&17.670(37)&0.0024(28)&119(84)&1380.5(1.2)&2.870(42)&1.170(80)&0.173(23)&1642.5(2.9)&-451.588(20)&64.69890(30)\\
EBLM J1104-43&1.7615796(48)&0.0349(15)&46.71(14)&$<$0.0063&$--$&744.14989(89)&18.60(17)&1.43(13)&0.40(10)&$<$205&$--$&$--$\\
EBLM J1105-13&3.934256(23)&0.0556(17)&15.493(43)&$<$0.0096&$--$&578.2029(20)&1.516(13)&1.33(11)&0.149(25)&$<$39&$--$&$--$\\
EBLM J1116-32&4.7456177(36)&0.0667(20)&51.249(25)&$<$0.0011&$--$&545.79890(38)&66.184(96)&1.170(80)&0.590(75)&$<$30.3&$--$&$--$\\
EBLM J1116-01&7.375828(66)&0.0837(25)&17.881(22)&$<$0.0045&$--$&881.8005(22)&4.369(16)&1.23(10)&0.208(26)&$<$69&$--$&$--$\\
EBLM J1141-37&5.1476797(45)&0.0679(22)&32.284(19)&$<$0.0017&$--$&907.49336(56)&17.946(32)&1.22(10)&0.354(50)&$<$15.36&$--$&$--$\\
EBLM J1146-42&10.46644(16)&0.1158(43)&34.418(68)&0.0598(28)&96.1(3.4)&1453.898(98)&43.98(65)&1.35(14)&0.539(69)&$--$&$--$&$--$\\
EBLM J1201-36&9.113113(23)&0.0930(27)&8.7366(71)&0.15350(83)&-158.43(39)&1595.8848(94)&0.6075(33)&1.19(10)&0.101(12)&$<$9.37&$--$&$--$\\
EBLM J1208-29&2.676017(59)&0.0463(14)&25.72(22)&0.195(11)&-60.5(3.0)&741.907(19)&4.45(31)&1.60(11)&0.248(56)&$<$460&$--$&$--$\\
EBLM J1219-39&6.7599941(48)&0.0711(18)&10.8285(38)&0.05594(39)&20.90(32)&740.8590(60)&0.8851(20)&0.950(70)&0.100(11)&$<$7.4&$--$&$--$\\
EBLM J1301-37&6.549848(89)&0.0796(25)&23.14(21)&0.3147(67)&141.6(1.8)&1083.297(28)&7.19(41)&1.31(11)&0.261(40)&$<$350&$--$&$--$\\
EBLM J1305-31&10.619149(15)&0.1055(27)&22.398(11)&0.03694(53)&-154.2(1.2)&1351.878(34)&12.337(39)&1.100(80)&0.287(28)&$<$26&$--$&$--$\\
EBLM J1420-07&2.7038926(69)&0.0456(14)&24.121(49)&0.1263(24)&177.23(86)&749.7115(65)&3.838(56)&1.50(11)&0.225(46)&$<$58&$--$&$--$\\
EBLM J1431-11&4.450132(31)&0.0572(16)&12.990(32)&$<$0.006&$--$&1037.0491(19)&1.0107(75)&1.140(90)&0.117(17)&$<$53&$--$&$--$\\
EBLM J1433-43&3.082484(11)&0.0498(21)&40.009(57)&$<$0.0033&$--$&1470.70746(86)&20.454(87)&1.34(14)&0.395(78)&$<$170&$--$&$--$\\
EBLM J1436-13&3.997529(20)&0.0586(19)&46.406(99)&$<$0.0033&$--$&651.8492(15)&41.39(26)&1.190(90)&0.489(73)&$<$125&$--$&$--$\\
EBLM J1500-33&3.7381773(83)&0.0548(21)&34.884(72)&0.0452(19)&-0.8(2.6)&583.549(27)&16.39(20)&1.23(12)&0.344(61)&$<$79&$--$&$--$\\
EBLM J1509-10&6.867840(21)&0.0874(30)&49.095(32)&$<$0.003&$--$&480.99778(82)&84.20(17)&1.22(11)&0.670(86)&$<$38&$--$&$--$\\
EBLM J1525-36&9.008921(33)&0.0943(28)&17.115(20)&$<$0.0027&$--$&506.4452(19)&4.679(16)&1.17(10)&0.207(24)&$<$23&$--$&$--$\\
EBLM J1559-05&3.760074(16)&0.0523(23)&18.063(42)&$<$0.0048&$--$&1172.7600(12)&2.296(16)&1.19(15)&0.161(31)&$<$110&$--$&$--$\\
EBLM J1630+10&10.963789(13)&0.1056(28)&19.4083(54)&0.18183(47)&111.61(16)&692.7400(50)&7.896(20)&1.070(80)&0.238(22)&20.9(2.9)&$--$&$--$\\
EBLM J1928-38&23.32270(15)&0.1720(47)&17.2679(56)&0.07370(40)&137.18(22)&1789.254(14)&12.341(28)&0.980(80)&0.268(21)&$<$38&$--$&$--$\\
EBLM J1934-42&6.352376(99)&0.0703(18)&18.623(16)&$<$0.0046&$--$&1885.8421(18)&4.251(11)&0.970(70)&0.178(19)&$<$400&$--$&$--$\\
EBLM J1944-20&2.7408047(89)&0.0478(18)&49.51(15)&$<$0.009&$--$&1042.7228(12)&34.47(31)&1.43(12)&0.505(99)&$<$230&$--$&$--$\\
EBLM J1947-23&1.919555(18)&0.0387(12)&24.17(22)&$<$0.014&$--$&1189.7009(22)&2.807(76)&1.86(14)&0.231(60)&4484.775800(50)&-507(32)&-406.23(21)\\
EBLM J2011-71&5.8727000(59)&0.0760(26)&23.6638(22)&0.03099(15)&-106.45(24)&1781.7851(40)&8.0513(61)&1.41(13)&0.285(42)&$--$&$--$&$--$\\
EBLM J2025-45&6.1919863(33)&0.0697(16)&22.856(10)&0.12642(49)&-77.47(21)&1255.0087(33)&7.477(23)&0.960(60)&0.218(22)&$<$16&$--$&$--$\\
EBLM J2027+03&3.8397112(87)&0.0575(19)&27.568(51)&$<$0.003&$--$&1611.99856(93)&8.335(47)&1.43(12)&0.291(50)&$<$59&$--$&$--$\\
EBLM J2040-41&14.456245(86)&0.1266(31)&12.4673(66)&0.22684(64)&-36.90(21)&1711.9462(91)&2.681(11)&1.130(80)&0.165(15)&$<$47.2&$--$&$--$\\
EBLM J2043-18&6.911406(87)&0.0809(23)&23.315(35)&$<$0.01&$--$&1447.0988(34)&9.075(41)&1.210(90)&0.271(33)&$<$52&$--$&$--$\\
EBLM J2046-40&37.01426(33)&0.2350(58)&11.986(12)&0.47316(56)&155.771(61)&1276.0866(48)&4.515(28)&1.070(80)&0.193(14)&$--$&$--$&$--$\\
EBLM J2046+06&10.107806(15)&0.1041(31)&15.5493(89)&0.34375(71)&108.84(12)&846.2647(37)&3.260(16)&1.28(11)&0.192(23)&$<$5.3833&$--$&$--$\\
EBLM J2101-45&25.57688(10)&0.2072(54)&25.5082(73)&0.09082(38)&19.90(17)&1514.890(13)&43.441(91)&1.29(10)&0.523(43)&$<$12&$--$&$--$\\
EBLM J2104-46&4.3573411(74)&0.0572(16)&35.568(34)&0.00771(87)&75.2(6.3)&400.062(77)&20.31(11)&0.990(70)&0.328(41)&-2513.737000(64)&237.3(4.2)&25.796(28)\\
EBLM J2107-39&3.961800(15)&0.0555(15)&26.516(80)&$<$0.012&$--$&557.6040(23)&7.653(70)&1.200(80)&0.253(38)&$<$130&$--$&$--$\\
EBLM J2122-32&18.42143(26)&0.1655(52)&35.539(13)&0.40518(54)&-135.317(91)&1995.5982(56)&65.47(25)&1.19(11)&0.593(57)&$<$257&$--$&$--$\\
EBLM J2153-55&8.544828(57)&0.0905(22)&26.852(25)&$<$0.004&$--$&1254.2567(34)&17.140(48)&1.040(70)&0.316(30)&$<$37&$--$&$--$\\
EBLM J2207-41&14.77480(22)&0.1296(33)&8.6853(51)&0.0668(13)&118.64(60)&1874.449(25)&0.9962(60)&1.210(90)&0.121(12)&$<$6.4&$--$&$--$\\
EBLM J2210-48&2.8200982(39)&0.0469(16)&37.869(30)&$<$0.0032&$--$&1735.57409(57)&15.868(38)&1.37(11)&0.362(69)&-54(19)&$--$&$--$\\
EBLM J2217-04&8.155259(11)&0.0833(25)&19.9668(92)&0.0480(12)&47.70(82)&867.215(19)&6.703(34)&0.950(80)&0.208(22)&$<$18.4&$--$&$--$\\
EBLM J2232-31&3.141524(23)&0.0477(15)&24.18(10)&$<$0.019&$--$&1749.9377(28)&4.600(57)&1.25(10)&0.215(39)&-271(74)&$--$&$--$\\
EBLM J2236-36&3.0671665(39)&0.0474(14)&29.797(30)&$<$0.0025&$--$&587.28178(59)&8.407(25)&1.240(90)&0.267(47)&$<$35&$--$&$--$\\
EBLM J2308-46&2.1992157(79)&0.0371(12)&23.555(98)&$<$0.014&$--$&712.1248(18)&2.978(37)&1.23(10)&0.181(40)&$<$100&$--$&$--$\\
EBLM J2330-61&7.457233(28)&0.0910(25)&45.216(41)&$<$0.0027&$--$&1306.4390(17)&71.43(19)&1.190(80)&0.615(67)&$<$56&$--$&$--$\\
EBLM J2349-32&3.5496719(87)&0.0508(14)&21.918(24)&$<$0.0023&$--$&531.94656(85)&3.873(13)&1.190(80)&0.195(31)&$<$37&$--$&$--$\\
EBLM J2353-10&4.534528(18)&0.0616(18)&22.181(30)&$<$0.0041&$--$&1631.0558(11)&5.127(21)&1.29(10)&0.228(35)&$<$51.1&$--$&$--$\\

\end{longtable}
\end{landscape}

\section{Parameters for the primary and secondary stars}

\rowcolors{2}{gray!25}{white}

\onecolumn

\begin{landscape}
\tiny
\begin{longtable}{ccccccccc}
\caption{Observational and calculated parameters of the primary stars.}\label{tab:primary}\\
\hline
name & $m_{\rm A}$ & $R_{\rm A}$ & $T_{\rm eff}$ & $E({\rm B}-{\rm V})$ & Vmag & Rmag & Jmag & Spectral \\
& $M_{\odot}$ & $R_{\odot}$ & [K] & & & & & Type \\
\hline
\endfirsthead
\hline 
name & $m_{\rm A}$ & $R_{\rm A}$ & $T_{\rm eff}$ & $E({\rm B}-{\rm V})$ & Vmag & Rmag & Jmag & Spectral \\
& $M_{\odot}$ & $R_{\odot}$ & [K] & & & & & Type \\
\hline
\endhead
\hline
\multicolumn{9}{l}{Table continues next page...}\\
\hline
\endfoot
\hline
\endlastfoot 
EBLM J0008+02 & 1.60(11) & 1.54 & 7030(130) & 0.025(21) & 10.06 & 9.9 & 9.35 & F2\\
EBLM J0017-38 & 1.200(80) & 1.24 & 6140(100) & 0.023(21) & 13.08 & 12.85 & 11.91 & F8\\
EBLM J0021-16 & 1.110(70) & 1.08 & 5820(90) & 0.026(20) & 9.81 & 9.37 & 8.65 & G2\\
EBLM J0027-41 & 1.180(70) & 1.19 & 6060(90) & 0.018(18) & 11.77 & 11.2 & 10.84 & F8\\
EBLM J0035-69 & 1.170(70) & 1.15 & 5990(90) & 0.022(19) & 12.38 & 11.88 & 10.9 & G0\\
EBLM J0040+01 & 0.810(60) & 0.75 & 5050(80) & 0.025(22) & 11.4 & 11.01 & 9.55 & K2\\
EBLM J0042-17 & 1.100(70) & 1.06 & 5790(90) & 0.028(20) & 10.35 & 10.01 & 9.08 & G2\\
EBLM J0048-66 & 1.250(90) & 1.29 & 6250(70) & 0.017(15) & 11.63 & 11.26 & 10.59 & F8\\
EBLM J0057-19 & 1.090(80) & 1.04 & 5760(110) & 0.024(23) & 11.62 & 11.15 & 10.43 & G2\\
EBLM J0104-38 & 1.38(11) & 1.38 & 6470(130) & 0.024(21) & 11.26 & 11.07 & 10.37 & F5\\
EBLM J0109-67 & 1.170(80) & 1.16 & 6010(100) & 0.025(20) & 12.73 & 12.21 & 11.32 & G0\\
EBLM J0218-31 & 1.170(80) & 1.15 & 6000(100) & 0.032(22) & 9.96 & 9.57 & 8.78 & G0\\
EBLM J0228+05 & 1.53(11) & 1.46 & 6830(140) & 0.023(24) & 10.24 & 10 & 9.48 & F2\\
EBLM J0239-20 & 1.160(80) & 1.12 & 5950(130) & 0.030(25) & 10.64 & 10.19 & 9.59 & G0\\
EBLM J0247-51 & 1.26(11) & 1.29 & 6260(170) & 0.032(25) & 9.56 & 9.23 & 8.6 & F8\\
EBLM J0310-31 & 1.26(10) & 1.29 & 6270(130) & 0.026(21) & 9.34 & 8.96 & 8.37 & F8\\
EBLM J0315-24 & 1.310(90) & 1.33 & 6350(100) & 0.021(19) & 11.33 & 10.91 & 10.5 & F5\\
EBLM J0326-09 & 1.160(80) & 1.13 & 5960(120) & 0.036(28) & 12.68 & 12.26 & 11.45 & G0\\
EBLM J0339+03 & 1.33(18) & 1.34 & 6380(370) & 0.101(59) & 11.39 & 11.18 & 10.37 & F5\\
EBLM J0351-07 & 1.29(11) & 1.32 & 6320(160) & 0.043(30) & 10.78 & 10.53 & 9.77 & F8\\
EBLM J0353+05 & 1.19(13) & 1.21 & 6100(270) & 0.078(60) & 11.18 & 10.85 & 9.74 & F8\\
EBLM J0353-16 & 1.170(80) & 1.17 & 6020(110) & 0.026(23) & 10.52 & 10.1 & 9.54 & G0\\
EBLM J0400-51 & 1.26(10) & 1.29 & 6270(110) & 0.028(21) & 12.35 & 11.77 & 11.26 & F8\\
EBLM J0425-46 & 1.190(80) & 1.2 & 6080(100) & 0.021(18) & 10.98 & 10.55 & 9.86 & F8\\
EBLM J0432-33 & 1.320(90) & 1.34 & 6370(100) & 0.018(19) & 11 & 10.56 & 10.2 & F5\\
EBLM J0440-48 & 1.29(10) & 1.32 & 6320(110) & 0.027(21) & 11.62 & 11.62 & 10.57 & F8\\
EBLM J0443-06 & 1.39(13) & 1.38 & 6490(210) & 0.047(40) & 11.61 & 11.41 & 10.54 & F5\\
EBLM J0454-09 & 1.18(12) & 1.2 & 6070(250) & 0.052(32) & 12.3 & 12.01 & 11.04 & F8\\
EBLM J0500-46 & 1.190(80) & 1.2 & 6080(100) & 0.021(19) & 12.03 & 11.61 & 10.79 & F8\\
EBLM J0502-38 & 1.26(10) & 1.29 & 6260(110) & 0.028(23) & 12.07 & 12.07 & 11.04 & F8\\
EBLM J0504-09 & 1.48(11) & 1.44 & 6710(170) & 0.060(33) & 11.78 & 11.38 & 10.86 & F5\\
EBLM J0518-39 & 1.47(10) & 1.43 & 6680(120) & 0.032(21) & 10.19 & 9.94 & 9.37 & F5\\
EBLM J0520-06 & 1.18(11) & 1.2 & 6070(230) & 0.083(50) & 11.84 & 11.53 & 10.84 & F8\\
EBLM J0525-55 & 1.31(11) & 1.33 & 6360(130) & 0.030(25) & 11.08 & 10.81 & 10.16 & F5\\
EBLM J0526+04 & 1.17(10) & 1.14 & 5980(200) & 0.058(44) & 12.38 & 11.7 & 10.98 & G0\\
EBLM J0526-34 & 1.35(11) & 1.36 & 6430(140) & 0.034(26) & 11.18 & 10.96 & 10.24 & F5\\
EBLM J0540-17 & 1.200(90) & 1.24 & 6150(130) & 0.029(26) & 11.31 & 10.76 & 10.52 & F8\\
EBLM J0543-57 & 1.23(10) & 1.27 & 6210(160) & 0.051(32) & 11.68 & 11.68 & 10.75 & F8\\
EBLM J0546-18 & 1.210(90) & 1.25 & 6170(160) & 0.062(33) & 12.25 & 12.02 & 11.02 & F8\\
EBLM J0608-59 & 1.200(80) & 1.25 & 6160(110) & 0.034(23) & 11.73 & 11.32 & 10.95 & F8\\
EBLM J0610-52 & 1.47(11) & 1.43 & 6690(160) & 0.040(28) & 11.14 & 10.89 & 10.36 & F5\\
EBLM J0621-46 & 1.19(10) & 1.23 & 6130(170) & 0.052(34) & 11.98 & 11.45 & 11.49 & F8\\
EBLM J0621-50 & 1.23(10) & 1.27 & 6220(170) & 0.055(35) & 11.95 & 11.95 & 10.94 & F8\\
EBLM J0623-27 & 1.180(80) & 1.18 & 6050(110) & 0.031(23) & 11.66 & 11.28 & 10.41 & G0\\
EBLM J0625-43 & 1.160(80) & 1.13 & 5960(140) & 0.039(29) & 12.27 & 11.79 & 10.9 & G0\\
EBLM J0627-67 & 1.34(11) & 1.35 & 6400(150) & 0.047(29) & 11.53 & 11.28 & 10.42 & F5\\
EBLM J0627-59 & 1.160(80) & 1.13 & 5960(130) & 0.044(28) & 12.26 & 11.5 & 10.92 & G0\\
EBLM J0629-67 & 1.45(11) & 1.42 & 6630(130) & 0.025(23) & 12.38 & 12.44 & 11.16 & F5\\
EBLM J0642-60 & 1.66(16) & 1.62 & 7180(300) & 0.058(43) & 10.12 & 9.82 & 9.46 & F0\\
EBLM J0645-61 & 1.45(13) & 1.42 & 6640(230) & 0.070(43) & 10.1 & 9.79 & 9.23 & F5\\
EBLM J0645-26 & 1.25(10) & 1.29 & 6250(140) & 0.030(29) & 12.5 & 12.31 & 11.22 & F8\\
EBLM J0649-27 & 1.45(13) & 1.42 & 6620(220) & 0.047(40) & 10.1 & 9.81 & 9.26 & F5\\
EBLM J0650-34 & 1.74(15) & 1.64 & 7400(320) & 0.053(47) & 10.29 & 10.02 & 9.75 & F0\\
EBLM J0659-61 & 1.160(90) & 1.14 & 5970(180) & 0.028(36) & 11.36 & 10.71 & 10.65 & G0\\
EBLM J0700-30 & 1.33(12) & 1.34 & 6380(190) & 0.032(35) & 11.95 & 11.59 & 10.97 & F5\\
EBLM J0709-52 & 1.11(11) & 1.08 & 5810(240) & 0.090(56) & 12.97 & 13.01 & 11.82 & G2\\
EBLM J0801+02 & 1.47(11) & 1.43 & 6670(140) & 0.032(24) & 11.94 & 11.48 & 11.22 & F5\\
EBLM J0851+05 & 1.220(90) & 1.26 & 6190(120) & 0.025(23) & 12.73 & 12.5 & 11.54 & F8\\
EBLM J0855+04 & 1.52(10) & 1.46 & 6820(150) & 0.031(26) & 9.18 & 8.92 & 8.4 & F2\\
EBLM J0941-31 & 1.19(10) & 1.23 & 6120(170) & 0.045(36) & 11.09 & 10.75 & 10.04 & F8\\
EBLM J0948-08 & 1.41(12) & 1.4 & 6530(160) & 0.034(25) & 9.32 & 9.32 & 8.41 & F5\\
EBLM J0954-23 & 1.44(11) & 1.42 & 6600(160) & 0.037(27) & 10.71 & 10.44 & 9.77 & F5\\
EBLM J0954-45 & 1.69(19) & 1.63 & 7270(430) & 0.100(68) & 9.83 & 9.57 & 9.11 & F0\\
EBLM J0955-39 & 1.23(13) & 1.27 & 6210(250) & 0.065(52) & 13.05 & 12.79 & 11.75 & F8\\
EBLM J1007-40 & 1.52(15) & 1.46 & 6800(290) & 0.070(53) & 10.8 & 10.53 & 9.91 & F2\\
EBLM J1008-29 & 1.71(12) & 1.63 & 7330(160) & 0.026(26) & 10.61 & 10.3 & 10.09 & F0\\
EBLM J1013+01 & 0.960(80) & 0.9 & 5460(120) & 0.036(30) & 11.88 & 11.78 & 10.15 & G8\\
EBLM J1014-07 & 1.30(12) & 1.32 & 6330(180) & 0.033(25) & 9.71 & 9.39 & 8.83 & F8\\
EBLM J1023-43 & 1.58(12) & 1.51 & 6960(200) & 0.041(35) & 12 & 11.73 & 11.09 & F2\\
EBLM J1034-29 & 1.44(11) & 1.42 & 6610(160) & 0.035(26) & 10.74 & 10.4 & 9.92 & F5\\
EBLM J1037-25 & 1.26(10) & 1.29 & 6260(140) & 0.040(27) & 10.17 & 9.83 & 9.13 & F8\\
EBLM J1037-45 & 1.30(13) & 1.32 & 6330(240) & 0.120(51) & 12.73 & 12.62 & 11.47 & F8\\
EBLM J1038-37 & 1.170(80) & 1.14 & 5980(140) & 0.030(28) & 13.26 & 13.49 & 12.52 & G0\\
EBLM J1104-43 & 1.43(13) & 1.41 & 6570(230) & 0.052(43) & 11.63 & 11.38 & 10.68 & F5\\
EBLM J1105-13 & 1.33(11) & 1.34 & 6380(160) & 0.036(28) & 10.29 & 9.88 & 9.43 & F5\\
EBLM J1116-32 & 1.170(80) & 1.15 & 5990(130) & 0.058(30) & 12.1 & 11.58 & 11.09 & G0\\
EBLM J1116-01 & 1.23(10) & 1.27 & 6220(160) & 0.039(29) & 12.77 & 12.34 & 11.76 & F8\\
EBLM J1141-37 & 1.22(10) & 1.27 & 6200(170) & 0.044(37) & 9.58 & 9.23 & 8.47 & F8\\
EBLM J1146-42 & 1.35(14) & 1.36 & 6430(260) & 0.093(50) & 10.29 & 9.96 & 9.21 & F5\\
EBLM J1201-36 & 1.19(10) & 1.21 & 6100(170) & 0.050(34) & 10.82 & 10.44 & 9.81 & F8\\
EBLM J1208-29 & 1.60(11) & 1.54 & 7020(170) & 0.028(28) & 10.1 & 10.1 & 9.38 & F2\\
EBLM J1219-39 & 0.950(70) & 0.89 & 5440(100) & 0.026(26) & 10.32 & 10.32 & 8.9 & G8\\
EBLM J1301-37 & 1.31(11) & 1.33 & 6360(150) & 0.026(29) & 12.09 & 11.9 & 11.16 & F5\\
EBLM J1305-31 & 1.100(80) & 1.06 & 5790(110) & 0.027(26) & 11.94 & 11.79 & 10.89 & G2\\
EBLM J1420-07 & 1.50(11) & 1.45 & 6760(170) & 0.037(26) & 9.75 & 9.46 & 8.96 & F2\\
EBLM J1431-11 & 1.140(90) & 1.11 & 5900(170) & 0.060(37) & 12.75 & 12.39 & 11.45 & G0\\
EBLM J1433-43 & 1.34(14) & 1.35 & 6410(240) & 0.062(46) & 11.45 & 11.26 & 10.4 & F5\\
EBLM J1436-13 & 1.190(90) & 1.21 & 6090(150) & 0.048(34) & 12.52 & 12.2 & 11.35 & F8\\
EBLM J1500-33 & 1.23(12) & 1.27 & 6210(230) & 0.045(44) & 12.61 & 12.62 & 11.05 & F8\\
EBLM J1509-10 & 1.22(11) & 1.26 & 6190(190) & 0.065(39) & 10.94 & 12.09 & 10.82 & F8\\
EBLM J1525-36 & 1.17(10) & 1.15 & 5990(210) & 0.044(44) & 11.7 & 11.27 & 10.43 & G0\\
EBLM J1559-05 & 1.19(15) & 1.22 & 6110(310) & 0.114(68) & 9.7 & 9.3 & 8.36 & F8\\
EBLM J1630+10 & 1.070(80) & 0.99 & 5710(130) & 0.047(30) & 12.01 & 11.55 & 10.41 & G5\\
EBLM J1928-38 & 0.980(80) & 0.91 & 5500(150) & 0.054(40) & 11.21 & 11.21 & 9.9 & G8\\
EBLM J1934-42 & 0.970(70) & 0.91 & 5480(110) & 0.025(29) & 12.42 & 12.23 & 11.23 & G8\\
EBLM J1944-20 & 1.43(12) & 1.41 & 6590(170) & 0.047(31) & 12.68 & 12.59 & 11.8 & F5\\
EBLM J1947-23 & 1.86(14) & 1.66 & 7730(270) & 0.055(39) & 8.81 & 8.65 & 8.27 & A7\\
EBLM J2011-71 & 1.41(13) & 1.4 & 6520(220) & 0.039(32) & 9.3 & 8.97 & 8.31 & F5\\
EBLM J2025-45 & 0.960(60) & 0.9 & 5470(80) & 0.022(21) & 11.16 & 10.7 & 9.69 & G8\\
EBLM J2027+03 & 1.43(12) & 1.41 & 6570(180) & 0.053(35) & 11.46 & 11.17 & 10.38 & F5\\
EBLM J2040-41 & 1.130(80) & 1.1 & 5870(110) & 0.028(24) & 11.5 & 11.03 & 10.52 & G2\\
EBLM J2043-18 & 1.210(90) & 1.26 & 6180(130) & 0.028(26) & 12.68 & 12.63 & 11.47 & F8\\
EBLM J2046-40 & 1.070(80) & 1 & 5720(120) & 0.031(26) & 11.49 & 11 & 10.45 & G5\\
EBLM J2046+06 & 1.28(11) & 1.31 & 6300(160) & 0.039(34) & 9.87 & 9.48 & 8.93 & F8\\
EBLM J2101-45 & 1.29(10) & 1.32 & 6320(110) & 0.022(21) & 10.5 & 10.24 & 9.38 & F8\\
EBLM J2104-46 & 0.990(70) & 0.92 & 5520(100) & 0.034(24) & 13.36 & 13.12 & 11.72 & G8\\
EBLM J2107-39 & 1.200(80) & 1.24 & 6140(100) & 0.025(21) & 12.04 & 11.75 & 10.85 & F8\\
EBLM J2122-32 & 1.19(11) & 1.2 & 6080(210) & 0.052(45) & 10.63 & 10.23 & 9.6 & F8\\
EBLM J2153-55 & 1.040(70) & 0.94 & 5630(90) & 0.024(21) & 12.68 & 12.77 & 11.54 & G5\\
EBLM J2207-41 & 1.210(90) & 1.26 & 6180(110) & 0.025(21) & 10.39 & 10.07 & 9.46 & F8\\
EBLM J2210-48 & 1.37(11) & 1.37 & 6460(160) & 0.026(23) & 8.78 & 8.47 & 7.89 & F5\\
EBLM J2217-04 & 0.950(80) & 0.89 & 5440(130) & 0.052(33) & 12.18 & 12.02 & 10.75 & G8\\
EBLM J2232-31 & 1.25(10) & 1.29 & 6250(120) & 0.022(19) & 10.36 & 10.04 & 9.44 & F8\\
EBLM J2236-36 & 1.240(90) & 1.28 & 6230(100) & 0.028(20) & 11 & 10.71 & 9.92 & F8\\
EBLM J2308-46 & 1.23(10) & 1.27 & 6210(140) & 0.023(20) & 11.36 & 11.14 & 10.48 & F8\\
EBLM J2330-61 & 1.190(80) & 1.22 & 6110(90) & 0.018(18) & 12.54 & 11.79 & 11.29 & F8\\
EBLM J2349-32 & 1.190(80) & 1.22 & 6110(100) & 0.022(17) & 11.53 & 11.14 & 10.53 & F8\\
EBLM J2353-10 & 1.29(10) & 1.32 & 6320(130) & 0.025(22) & 10.99 & 11.4 & 10.78 & F8\\

\end{longtable}
\end{landscape}

\rowcolors{2}{gray!25}{white}

\onecolumn

\begin{landscape}
\tiny
\begin{longtable}{cc|ccc|ccc}
\caption{Parameters of the secondary stars. The magnitudes are estimates calculated using the Baraffe models. They are indicative only, meant to help prepare observations, such as secondary eclipses, and transforming our single line into double line binaries.}\label{tab:secondary}\\
\hline
\multicolumn{2}{c|}{} & \multicolumn{3}{c|}{1 Gyr Age}  & \multicolumn{3}{c}{5 Gyr Age}  \\
name & $m_{\rm B}$ & Vmag & Rmag & Jmag & Vmag & Rmag & Jmag\\
& [$M_{\odot}$] & & & & & & \\ 
\hline
\endfirsthead
\hline 
\multicolumn{2}{c|}{} & \multicolumn{3}{c|}{1 Gyr Age}  & \multicolumn{3}{c}{5 Gyr Age}  \\
name & $m_{\rm B}$ & Vmag & Rmag & Jmag & Vmag & Rmag & Jmag\\
& [$M_{\odot}$] & & & & & & \\ 
\hline
\endhead
\hline
\multicolumn{8}{l}{Table continues next page...}\\
\hline
\endfoot
\hline
\endlastfoot 
EBLM J0008+02 & 0.183(28) & 20.38 & 19.12 & 15.93 & 20.03 & 18.88 & 15.82\\
EBLM J0017-38 & 0.184(24) & 21.89 & 20.64 & 17.47 & 21.57 & 20.42 & 17.37\\
EBLM J0021-16 & 0.194(22) & 17.95 & 16.78 & 13.72 & 17.81 & 16.69 & 13.68\\
EBLM J0027-41 & 0.528(65) & 16.92 & 16.08 & 13.92 & 16.88 & 16.04 & 13.89\\
EBLM J0035-69 & 0.198(20) & 20.19 & 19.05 & 16.04 & 20.12 & 19.01 & 16.02\\
EBLM J0040+01 & 0.102(10) & 21.35 & 19.4 & 15.22 & 21.3 & 19.35 & 15.2\\
EBLM J0042-17 & 0.642(56) & 13.55 & 12.79 & 11.09 & 13.46 & 12.71 & 11.03\\
EBLM J0048-66 & 0.447(55) & 17.54 & 16.63 & 14.24 & 17.51 & 16.6 & 14.21\\
EBLM J0057-19 & 0.136(19) & 21.76 & 20.1 & 16.34 & 20.99 & 19.56 & 16.09\\
EBLM J0104-38 & 0.274(33) & 19.48 & 18.44 & 15.68 & 19.43 & 18.4 & 15.63\\
EBLM J0109-67 & 0.689(68) & 15.3 & 14.56 & 13.06 & 15.19 & 14.47 & 12.99\\
EBLM J0218-31 & 0.359(36) & 16.35 & 15.38 & 12.79 & 16.31 & 15.34 & 12.75\\
EBLM J0228+05 & 0.180(23) & 20.52 & 19.23 & 16 & 20.11 & 18.95 & 15.88\\
EBLM J0239-20 & 0.170(29) & 20.04 & 18.67 & 15.33 & 19.46 & 18.28 & 15.16\\
EBLM J0247-51 & 0.231(38) & 18.03 & 16.94 & 14.05 & 18 & 16.92 & 14.03\\
EBLM J0310-31 & 0.408(41) & 15.94 & 15 & 12.5 & 15.91 & 14.96 & 12.47\\
EBLM J0315-24 & 0.258(45) & 19.71 & 18.66 & 15.85 & 19.67 & 18.62 & 15.81\\
EBLM J0326-09 & 0.193(38) & 20.93 & 19.76 & 16.69 & 20.78 & 19.66 & 16.65\\
EBLM J0339+03 & 0.242(51) & 19.64 & 18.57 & 15.71 & 19.6 & 18.54 & 15.68\\
EBLM J0351-07 & 0.242(40) & 19.05 & 17.97 & 15.12 & 19.01 & 17.94 & 15.09\\
EBLM J0353+05 & 0.179(26) & 19.42 & 18.12 & 14.89 & 19.01 & 17.85 & 14.77\\
EBLM J0353-16 & 0.222(21) & 18.7 & 17.6 & 14.69 & 18.67 & 17.58 & 14.67\\
EBLM J0400-51 & 0.266(52) & 20.14 & 19.1 & 16.31 & 20.1 & 19.05 & 16.26\\
EBLM J0425-46 & 0.627(51) & 14.83 & 14.06 & 12.29 & 14.75 & 13.99 & 12.25\\
EBLM J0432-33 & 0.260(35) & 19.31 & 18.26 & 15.45 & 19.27 & 18.22 & 15.41\\
EBLM J0440-48 & 0.190(39) & 20.74 & 19.53 & 16.43 & 20.53 & 19.39 & 16.37\\
EBLM J0443-06 & 0.58(11) & 16.47 & 15.67 & 13.7 & 16.41 & 15.62 & 13.67\\
EBLM J0454-09 & 0.71(10) & 14.42 & 13.7 & 12.26 & 14.31 & 13.6 & 12.19\\
EBLM J0500-46 & 0.182(21) & 20.6 & 19.32 & 16.12 & 20.22 & 19.07 & 16\\
EBLM J0502-38 & 0.307(54) & 19.35 & 18.35 & 15.67 & 19.29 & 18.29 & 15.61\\
EBLM J0504-09 & 0.268(54) & 20 & 18.95 & 16.17 & 19.95 & 18.91 & 16.13\\
EBLM J0518-39 & 0.220(37) & 19.16 & 18.06 & 15.14 & 19.13 & 18.04 & 15.12\\
EBLM J0520-06 & 0.50(11) & 17.19 & 16.32 & 14.08 & 17.16 & 16.29 & 14.05\\
EBLM J0525-55 & 0.270(41) & 19.24 & 18.19 & 15.42 & 19.19 & 18.15 & 15.37\\
EBLM J0526+04 & 0.193(30) & 21.02 & 19.85 & 16.78 & 20.87 & 19.75 & 16.74\\
EBLM J0526-34 & 0.338(38) & 18.56 & 17.58 & 14.96 & 18.52 & 17.53 & 14.91\\
EBLM J0540-17 & 0.171(22) & 21.34 & 19.98 & 16.64 & 20.76 & 19.58 & 16.47\\
EBLM J0543-57 & 0.160(25) & 21.64 & 20.18 & 16.72 & 20.92 & 19.71 & 16.51\\
EBLM J0546-18 & 0.231(40) & 20 & 18.92 & 16.03 & 19.97 & 18.89 & 16\\
EBLM J0608-59 & 0.325(29) & 19.31 & 18.31 & 15.67 & 19.25 & 18.26 & 15.61\\
EBLM J0610-52 & 0.60(12) & 16.35 & 15.57 & 13.68 & 16.28 & 15.51 & 13.64\\
EBLM J0621-46 & 0.221(58) & 20.95 & 19.85 & 16.93 & 20.92 & 19.83 & 16.91\\
EBLM J0621-50 & 0.420(60) & 18.37 & 17.45 & 14.98 & 18.34 & 17.41 & 14.95\\
EBLM J0623-27 & 0.308(38) & 18.31 & 17.31 & 14.63 & 18.25 & 17.25 & 14.57\\
EBLM J0625-43 & 0.291(42) & 19.06 & 18.04 & 15.33 & 19 & 17.99 & 15.27\\
EBLM J0627-67 & 0.229(27) & 19.76 & 18.67 & 15.77 & 19.72 & 18.64 & 15.75\\
EBLM J0627-59 & 0.389(47) & 18.27 & 17.32 & 14.78 & 18.24 & 17.28 & 14.75\\
EBLM J0629-67 & 0.432(60) & 18.82 & 17.91 & 15.47 & 18.79 & 17.87 & 15.44\\
EBLM J0642-60 & 0.282(48) & 18.79 & 17.76 & 15.02 & 18.74 & 17.71 & 14.97\\
EBLM J0645-61 & 0.138(24) & 21.3 & 19.66 & 15.93 & 20.51 & 19.11 & 15.68\\
EBLM J0645-26 & 0.389(48) & 18.7 & 17.76 & 15.22 & 18.67 & 17.71 & 15.18\\
EBLM J0649-27 & 0.371(61) & 17.4 & 16.44 & 13.87 & 17.36 & 16.39 & 13.83\\
EBLM J0650-34 & 0.308(52) & 18.91 & 17.91 & 15.23 & 18.86 & 17.85 & 15.17\\
EBLM J0659-61 & 0.457(65) & 17.6 & 16.7 & 14.33 & 17.57 & 16.66 & 14.3\\
EBLM J0700-30 & 0.480(65) & 17.81 & 16.92 & 14.61 & 17.78 & 16.89 & 14.58\\
EBLM J0709-52 & 0.361(44) & 19 & 18.03 & 15.45 & 18.96 & 17.99 & 15.41\\
EBLM J0801+02 & 0.202(37) & 21.5 & 20.38 & 17.41 & 21.48 & 20.37 & 17.4\\
EBLM J0851+05 & 0.160(33) & 22.4 & 20.95 & 17.48 & 21.68 & 20.47 & 17.27\\
EBLM J0855+04 & 0.185(41) & 19.12 & 17.87 & 14.71 & 18.81 & 17.66 & 14.61\\
EBLM J0941-31 & 0.218(31) & 19.24 & 18.14 & 15.21 & 19.21 & 18.12 & 15.19\\
EBLM J0948-08 & 0.675(96) & 13.32 & 12.58 & 11.01 & 13.22 & 12.49 & 10.95\\
EBLM J0954-23 & 0.107(14) & 23.02 & 21.12 & 17.01 & 22.8 & 20.91 & 16.91\\
EBLM J0954-45 & 0.412(62) & 17.26 & 16.33 & 13.84 & 17.23 & 16.29 & 13.81\\
EBLM J0955-39 & 0.226(36) & 20.84 & 19.75 & 16.84 & 20.81 & 19.72 & 16.82\\
EBLM J1007-40 & 0.377(68) & 18.07 & 17.11 & 14.55 & 18.03 & 17.07 & 14.52\\
EBLM J1008-29 & 0.368(41) & 18.94 & 17.97 & 15.4 & 18.9 & 17.93 & 15.36\\
EBLM J1013+01 & 0.168(27) & 19.81 & 18.42 & 15.05 & 19.19 & 18.01 & 14.87\\
EBLM J1014-07 & 0.241(39) & 18.11 & 17.03 & 14.17 & 18.07 & 17 & 14.14\\
EBLM J1023-43 & 0.85(14) & 14.63 & 14.1 & 13.02 & 14.43 & 13.92 & 12.88\\
EBLM J1034-29 & 0.151(34) & 21.55 & 20.02 & 16.44 & 20.71 & 19.47 & 16.21\\
EBLM J1037-25 & 0.260(38) & 17.84 & 16.78 & 13.98 & 17.79 & 16.74 & 13.94\\
EBLM J1037-45 & 0.279(82) & 19.92 & 18.89 & 16.14 & 19.87 & 18.84 & 16.09\\
EBLM J1038-37 & 0.173(23) & 22.8 & 21.45 & 18.14 & 22.26 & 21.09 & 17.98\\
EBLM J1104-43 & 0.40(10) & 18.58 & 17.64 & 15.11 & 18.55 & 17.6 & 15.08\\
EBLM J1105-13 & 0.149(25) & 21.01 & 19.47 & 15.87 & 20.17 & 18.91 & 15.62\\
EBLM J1116-32 & 0.590(75) & 16.47 & 15.68 & 13.76 & 16.4 & 15.62 & 13.72\\
EBLM J1116-01 & 0.208(26) & 20.71 & 19.59 & 16.64 & 20.68 & 19.58 & 16.62\\
EBLM J1141-37 & 0.354(50) & 16.17 & 15.2 & 12.6 & 16.13 & 15.15 & 12.56\\
EBLM J1146-42 & 0.539(69) & 15.39 & 14.56 & 12.45 & 15.34 & 14.51 & 12.41\\
EBLM J1201-36 & 0.101(12) & 22.95 & 20.99 & 16.8 & 22.93 & 20.97 & 16.79\\
EBLM J1208-29 & 0.248(56) & 18.94 & 17.87 & 15.03 & 18.9 & 17.84 & 14.99\\
EBLM J1219-39 & 0.100(11) & 21.24 & 19.26 & 15.03 & 21.24 & 19.26 & 15.03\\
EBLM J1301-37 & 0.261(40) & 20.21 & 19.16 & 16.35 & 20.17 & 19.12 & 16.31\\
EBLM J1305-31 & 0.287(28) & 18.81 & 17.78 & 15.06 & 18.75 & 17.73 & 15\\
EBLM J1420-07 & 0.225(46) & 18.69 & 17.6 & 14.7 & 18.66 & 17.58 & 14.67\\
EBLM J1431-11 & 0.117(17) & 23.77 & 21.95 & 17.97 & 23.3 & 21.57 & 17.78\\
EBLM J1433-43 & 0.395(78) & 18.18 & 17.24 & 14.71 & 18.15 & 17.2 & 14.68\\
EBLM J1436-13 & 0.489(73) & 17.71 & 16.83 & 14.54 & 17.68 & 16.8 & 14.51\\
EBLM J1500-33 & 0.344(61) & 18.81 & 17.83 & 15.21 & 18.76 & 17.77 & 15.16\\
EBLM J1509-10 & 0.670(86) & 14.87 & 14.12 & 12.54 & 14.77 & 14.03 & 12.48\\
EBLM J1525-36 & 0.207(24) & 19.42 & 18.31 & 15.35 & 19.39 & 18.29 & 15.33\\
EBLM J1559-05 & 0.161(31) & 18.78 & 17.34 & 13.89 & 18.08 & 16.87 & 13.68\\
EBLM J1630+10 & 0.238(22) & 18.54 & 17.46 & 14.59 & 18.51 & 17.43 & 14.57\\
EBLM J1928-38 & 0.268(21) & 17.67 & 16.62 & 13.84 & 17.62 & 16.58 & 13.8\\
EBLM J1934-42 & 0.178(19) & 20.66 & 19.35 & 16.1 & 20.21 & 19.05 & 15.97\\
EBLM J1944-20 & 0.505(99) & 18.74 & 17.88 & 15.64 & 18.71 & 17.84 & 15.61\\
EBLM J1947-23 & 0.231(60) & 18.54 & 17.46 & 14.57 & 18.51 & 17.43 & 14.54\\
EBLM J2011-71 & 0.285(42) & 17.02 & 16 & 13.26 & 16.97 & 15.94 & 13.21\\
EBLM J2025-45 & 0.218(22) & 18.03 & 16.93 & 14 & 18 & 16.91 & 13.98\\
EBLM J2027+03 & 0.291(50) & 18.88 & 17.86 & 15.14 & 18.82 & 17.8 & 15.09\\
EBLM J2040-41 & 0.165(15) & 21.02 & 19.61 & 16.2 & 20.36 & 19.17 & 16.01\\
EBLM J2043-18 & 0.271(33) & 20.17 & 19.13 & 16.35 & 20.12 & 19.08 & 16.31\\
EBLM J2046-40 & 0.193(14) & 19.99 & 18.82 & 15.75 & 19.84 & 18.72 & 15.71\\
EBLM J2046+06 & 0.192(23) & 19.01 & 17.82 & 14.74 & 18.83 & 17.71 & 14.69\\
EBLM J2101-45 & 0.523(43) & 15.7 & 14.85 & 12.68 & 15.67 & 14.82 & 12.65\\
EBLM J2104-46 & 0.328(41) & 18.98 & 17.99 & 15.35 & 18.93 & 17.94 & 15.3\\
EBLM J2107-39 & 0.253(38) & 19.6 & 18.54 & 15.71 & 19.56 & 18.5 & 15.68\\
EBLM J2122-32 & 0.593(57) & 14.92 & 14.14 & 12.23 & 14.86 & 14.08 & 12.19\\
EBLM J2153-55 & 0.316(30) & 19.12 & 18.12 & 15.46 & 19.06 & 18.06 & 15.4\\
EBLM J2207-41 & 0.121(12) & 22.13 & 20.35 & 16.41 & 21.58 & 19.92 & 16.2\\
EBLM J2210-48 & 0.362(69) & 16.08 & 15.11 & 12.53 & 16.04 & 15.07 & 12.49\\
EBLM J2217-04 & 0.208(22) & 19.22 & 18.1 & 15.15 & 19.19 & 18.09 & 15.13\\
EBLM J2232-31 & 0.215(39) & 18.9 & 17.8 & 14.86 & 18.88 & 17.78 & 14.85\\
EBLM J2236-36 & 0.267(47) & 18.66 & 17.61 & 14.83 & 18.61 & 17.57 & 14.78\\
EBLM J2308-46 & 0.181(40) & 21.11 & 19.83 & 16.62 & 20.73 & 19.58 & 16.51\\
EBLM J2330-61 & 0.615(67) & 16.4 & 15.63 & 13.81 & 16.33 & 15.56 & 13.77\\
EBLM J2349-32 & 0.195(31) & 20.09 & 18.93 & 15.89 & 19.98 & 18.86 & 15.86\\
EBLM J2353-10 & 0.228(35) & 19.77 & 18.68 & 15.78 & 19.74 & 18.66 & 15.76\\

\end{longtable}
\end{landscape}

\begin{figure}
\begin{center}
\includegraphics[width=0.49\textwidth,trim={0 0 0 0},clip]{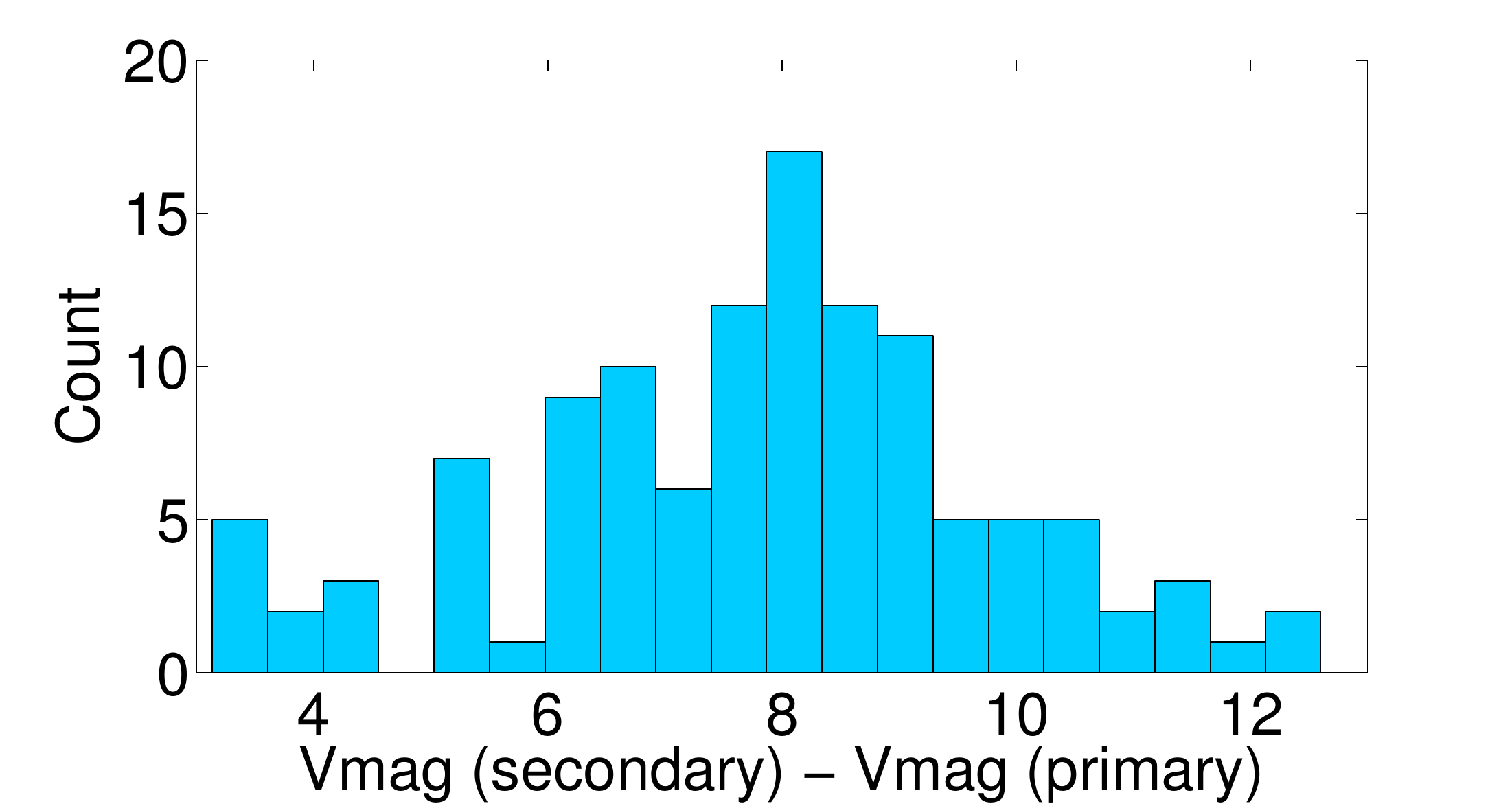}
\caption{Histogram of the difference between the secondary and primary visual magnitudes, using the data provided in Table~\ref{tab:obs}. Primary Vmag values come from the NOMAD survey, except for three binaries where they are calculated using models from \citet{Baraffe:1998ly} at an age of 1 Gyr. All secondary Vmag values are calculated using Baraffe models with a 1 Gyr age.}\label{fig:Vmag_difference_histogram}
\end{center}
\end{figure}

\section{Observational parameters}

\rowcolors{2}{gray!25}{white}

\onecolumn

\begin{landscape}
\tiny
\begin{longtable}{cccccc|ccccccc}
\caption{Observational parameters of the systems we observed.}\label{tab:obs}\\
\hline
\multicolumn{6}{c|}{System}  & \multicolumn{6}{c}{Observations}  \\
name & RA & dec & $P$ & $T_{\rm 0, pri}$ & $T_{\rm 0, sec}$  & $\sigma_{\rm 1800s}$ & $\sigma_{\rm median}$ & $\sigma_{\rm add}$  & timespan & num. & flag\\
& [hr] & [deg] & [day] & [BJD-2,455,000] & [BJD-2,455,000] &  [m/s] & [m/s] & [m/s] & [yr] & obs. & \\ 
\hline
\endfirsthead
\hline 
\multicolumn{6}{c|}{System} &  \multicolumn{6}{c}{Observations}  \\
name & RA & dec & $P$ & $T_{\rm 0, pri}$ & $T_{\rm 0, sec}$ & $\sigma_{\rm 1800s}$ & $\sigma_{\rm median}$ & $\sigma_{\rm add}$  & timespan & num. & flag\\
& [hr] & [deg] & [day] & [BJD-2,455,000] & [BJD-2,455,000] &  [m/s] & [m/s] & [m/s] & [yr] & obs. & \\ 
\hline
\endhead
\hline
\multicolumn{13}{l}{Table continues next page...}\\
\hline
\endfoot
\hline
\endlastfoot 
EBLM J0008+02 & 00 08 57.97 & +02 56 42.0 & 4.7223 & 1709.053 & 1707.1606 & 15 & 18 & 13 & 3.96 & 25 & drift\\
EBLM J0017-38 & 00 17 48.30 & -38 06 39.2 & 6.3401 & 1791.7449 & 1788.5748 & -1 & 136 & 173 & 2.33 & 13 & \\
EBLM J0021-16 & 00 21 00.77 & -16 07 28.5 & 5.9673 & 1069.3159 & 1066.3322 & 9 & 12 & 27 & 5.34 & 34 & active\\
EBLM J0027-41 & 00 27 16.51 & -41 34 17.7 & 4.928 & 1311.5215 & 1309.0894 & 82 & 115 & 69 & 7.41 & 14 & \\
EBLM J0035-69 & 00 35 40.39 & -69 48 52.2 & 8.4146 & 1232.2357 & 1229.3053 & 67 & 77 & 0 & 4.26 & 21 & \\
EBLM J0040+01 & 00 40 01.50 & +01 05 40.3 & 7.2348 & 1400.9647 & 1397.6659 & 11 & 18 & 8 & 4.85 & 20 & \\
EBLM J0042-17 & 00 42 34.21 & -17 17 53.1 & 10.3475 & 777.227 & 772.1594 & 7 & 10 & 0 & 5.36 & 17 & \\
EBLM J0048-66 & 00 48 21.46 & -66 09 36.8 & 6.6493 & 1316.3741 & 1313.1821 & 38 & 83 & 74 & 2.92 & 18 & \\
EBLM J0057-19 & 00 57 58.94 & -19 49 47.5 & 4.3005 & 632.2772 & 630.1269 & 32 & 60 & 61 & 6.48 & 18 & \\
EBLM J0104-38 & 01 04 19.13 & -38 18 30.7 & 8.2561 & 824.3094 & 828.4318 & 25 & 35 & 0 & 7.5 & 16 & drift\\
EBLM J0109-67 & 01 09 12.87 & -67 55 08.3 & 9.03 & 1017.5335 & 1013.1116 & 48 & 105 & 0 & 2.33 & 21 & \\
EBLM J0218-31 & 02 18 13.24 & -31 05 17.3 & 8.8841 & 1128.6773 & 1124.2352 & 10 & 18 & 17 & 7.5 & 45 & drift\\
EBLM J0228+05 & 02 28 08.87 & +05 35 47.7 & 6.6347 & 1787.7005 & 1784.3832 & 28 & 34 & 0 & 2.49 & 15 & \\
EBLM J0239-20 & 02 39 29.28 & -20 02 24.0 & 2.7787 & 459.7136 & 458.3243 & 46 & 75 & 90 & 3.29 & 21 & drift\\
EBLM J0247-51 & 02 47 20.50 & -51 27 10.4 & 4.0079 & 1322.1115 & 1320.1076 & 34 & 46 & 106 & 1.45 & 19 & \\
EBLM J0310-31 & 03 10 22.62 & -31 07 35.7 & 12.6428 & 1668.2438 & 1672.1333 & 5 & 5 & 4 & 1.23 & 15 & \\
EBLM J0315-24 & 03 15 26.53 & -24 15 45.5 & 3.1905 & 688.6659 & 687.0706 & 74 & 119 & 135 & 5.09 & 21 & \\
EBLM J0326-09 & 03 26 45.00 & -09 20 31.6 & 2.4004 & 1608.8353 & 1607.6351 & 123 & 134 & 928 & 4.04 & 14 & \\
EBLM J0339+03 & 03 39 09.63 & +03 05 37.5 & 3.5807 & 1043.6536 & 1041.8633 & 53 & 98 & 46 & 3.16 & 15 & \\
EBLM J0351-07 & 03 51 00.54 & -07 05 54.9 & 4.0809 & 642.7425 & 644.7646 & 27 & 66 & 59 & 3.67 & 21 & \\
EBLM J0353+05 & 03 53 08.94 & +05 36 33.3 & 6.862 & 1108.5117 & 1111.9414 & 10 & 16 & 0 & 5.97 & 51 & drift\\
EBLM J0353-16 & 03 53 54.52 & -16 57 15.3 & 11.7612 & 870.8052 & 864.9543 & 8 & 12 & 0 & 6.29 & 29 & \\
EBLM J0400-51 & 04 00 56.86 & -51 07 27.5 & 2.6921 & 1593.3853 & 1592.0393 & 106 & 112 & 456 & 2.02 & 13 & \\
EBLM J0425-46 & 04 25 31.70 & -46 13 07.7 & 16.5879 & 1716.2169 & 1708.4069 & 13 & 15 & 20 & 1.55 & 14 & \\
EBLM J0432-33 & 04 32 58.79 & -33 29 47.9 & 5.3055 & 987.8293 & 985.1766 & 40 & 84 & 14 & 1.99 & 21 & \\
EBLM J0440-48 & 04 40 14.58 & -48 17 52.6 & 2.543 & 1006.8321 & 1005.5606 & 80 & 113 & 451 & 2 & 21 & \\
EBLM J0443-06 & 04 43 01.76 & -06 25 50.7 & 3.1119 & 992.7277 & 991.2755 & 93 & 134 & 475 & 4.69 & 20 & \\
EBLM J0454-09 & 04 54 11.23 & -09 29 53.4 & 5.0135 & 417.0402 & 414.5335 & 45 & 107 & 0 & 4.87 & 19 & \\
EBLM J0500-46 & 05 00 32.88 & -46 11 21.3 & 8.2844 & 2211.6977 & 2208.7698 & 41 & 48 & 0 & 1.3 & 13 & \\
EBLM J0502-38 & 05 02 38.60 & -38 43 31.0 & 3.2563 & 1172.8914 & 1171.2632 & 93 & 133 & 407 & 2.85 & 16 & \\
EBLM J0504-09 & 05 04 34.94 & -09 13 29.2 & 2.699 & 1422.9315 & 1421.582 & 93 & 140 & 192 & 6.17 & 22 & \\
EBLM J0518-39 & 05 18 46.47 & -39 03 16.3 & 3.6498 & 580.8828 & 579.0579 & 28 & 40 & 57 & 3.61 & 21 & \\
EBLM J0520-06 & 05 20 59.46 & -06 42 16.6 & 2.1315 & 796.9447 & 795.8789 & 91 & 105 & 3182 & 3.78 & 14 & \\
EBLM J0525-55 & 05 25 24.71 & -55 01 11.6 & 4.8001 & 2062.9363 & 2060.5363 & 43 & 80 & 0 & 1.24 & 14 & \\
EBLM J0526+04 & 05 26 04.85 & +04 51 35.4 & 4.031 & 1328.6471 & 1326.6316 & 46 & 87 & 0 & 4.92 & 14 & \\
EBLM J0526-34 & 05 26 39.07 & -34 36 59.4 & 10.1909 & 697.4621 & 701.7735 & 17 & 20 & 0 & 6.4 & 21 & \\
EBLM J0540-17 & 05 40 43.58 & -17 32 44.8 & 6.0049 & 1849.1862 & 1852.1875 & 14 & 16 & 0 & 1.95 & 18 & drift\\
EBLM J0543-57 & 05 43 51.45 & -57 09 48.5 & 4.4638 & 904.4484 & 902.2165 & 20 & 27 & 6 & 5.65 & 35 & triple\\
EBLM J0546-18 & 05 46 04.85 & -18 17 55.2 & 3.1919 & 612.8086 & 611.2126 & 95 & 133 & 339 & 3.06 & 21 & \\
EBLM J0608-59 & 06 08 31.95 & -59 32 28.1 & 14.6085 & 561.1967 & 567.8254 & 26 & 26 & 0 & 5.67 & 21 & \\
EBLM J0610-52 & 06 10 53.61 & -52 53 46.5 & 2.417 & 653.5863 & 652.3778 & 85 & 116 & 1186 & 3.18 & 19 & \\
EBLM J0621-46 & 06 21 53.54 & -46 13 10.1 & 1.5508 & 701.6539 & 700.8784 & 102 & 117 & 3214 & 4.13 & 19 & \\
EBLM J0621-50 & 06 21 56.64 & -50 55 32.4 & 4.9638 & 838.9965 & 836.5145 & 60 & 69 & 40 & 6.11 & 25 & \\
EBLM J0623-27 & 06 23 11.42 & -27 43 45.0 & 5.7779 & 2042.8657 & 2040.1509 & 41 & 78 & 18 & 1.98 & 14 & \\
EBLM J0625-43 & 06 25 16.09 & -43 47 12.0 & 3.969 & 706.5484 & 704.5639 & 64 & 121 & 159 & 2.94 & 21 & \\
EBLM J0627-67 & 06 27 31.40 & -67 46 18.9 & 9.4689 & 771.9299 & 767.2872 & 25 & 37 & 43 & 4.05 & 24 & \\
EBLM J0627-59 & 06 27 47.56 & -59 12 57.4 & 5.7296 & 2040.9632 & 2038.0984 & 70 & 114 & 185 & 1.5 & 13 & \\
EBLM J0629-67 & 06 29 14.15 & -67 25 11.3 & 18.2903 & 1134.3974 & 1142.5091 & 54 & 115 & 388 & 0.81 & 15 & drift\\
EBLM J0642-60 & 06 42 03.34 & -60 40 13.0 & 5.0115 & 1104.2588 & 1101.753 & 46 & 77 & 65 & 3.24 & 16 & \\
EBLM J0645-61 & 06 45 15.79 & -61 05 29.4 & 4.4537 & 896.69 & 894.9977 & 38 & 68 & 207 & 2.79 & 36 & \\
EBLM J0645-26 & 06 45 37.92 & -26 42 42.3 & 7.5648 & 668.8738 & 672.471 & 53 & 97 & 77 & 4.88 & 22 & \\
EBLM J0649-27 & 06 49 06.01 & -27 20 58.3 & 4.308 & 803.8123 & 801.6583 & 34 & 44 & 29 & 5.2 & 20 & \\
EBLM J0650-34 & 06 50 29.08 & -34 36 17.7 & 8.9577 & 1101.6304 & 1097.1516 & 78 & 93 & 4356 & 6.51 & 13 & \\
EBLM J0659-61 & 06 59 07.78 & -61 50 24.1 & 4.2356 & 1710.3217 & 1712.4299 & 66 & 92 & 0 & 3.31 & 19 & drift\\
EBLM J0700-30 & 07 00 42.36 & -30 43 09.0 & 6.5456 & 1975.9857 & 1972.7129 & 43 & 117 & 76 & 1.88 & 13 & \\
EBLM J0709-52 & 07 09 15.32 & -52 55 18.0 & 9.108 & 1691.6174 & 1688.5013 & 92 & 151 & 337 & 1.5 & 16 & \\
EBLM J0801+02 & 08 01 18.10 & +02 34 09.0 & 3.3488 & 1712.5164 & 1710.842 & 84 & 143 & 68 & 3.18 & 13 & \\
EBLM J0851+05 & 08 51 54.02 & +05 42 30.5 & 2.5537 & 930.632 & 929.3551 & 97 & 136 & 1162 & 4.05 & 16 & \\
EBLM J0855+04 & 08 55 27.48 & +04 50 04.7 & 2.227 & 842.875 & 843.9002 & 49 & 74 & 178 & 2.87 & 22 & \\
EBLM J0941-31 & 09 41 16.76 & -31 49 10.2 & 5.5456 & 341.5389 & 339.4668 & 45 & 59 & 43 & 5.03 & 21 & \\
EBLM J0948-08 & 09 48 49.45 & -08 29 36.4 & 5.3798 & 873.4589 & 870.9369 & 12 & 21 & 17 & 5.96 & 26 & drift\\
EBLM J0954-23 & 09 54 52.89 & -23 19 55.7 & 7.5746 & 1357.5882 & 1361.3132 & 15 & 19 & 29 & 3.94 & 21 & \\
EBLM J0954-45 & 09 54 58.68 & -45 17 26.2 & 8.0726 & 741.6541 & 738.3248 & 24 & 27 & 47 & 7.54 & 23 & \\
EBLM J0955-39 & 09 55 18.26 & -39 52 58.9 & 5.3136 & 459.6361 & 456.9793 & 53 & 130 & 57 & 6.65 & 23 & \\
EBLM J1007-40 & 10 07 10.08 & -40 28 16.6 & 3.936 & 600.7162 & 598.7482 & 50 & 79 & 149 & 2.93 & 21 & \\
EBLM J1008-29 & 10 08 33.54 & -29 35 57.5 & 10.4009 & 1864.5846 & 1859.3842 & 43 & 58 & 52 & 2.7 & 13 & \\
EBLM J1013+01 & 10 13 50.84 & +01 59 28.1 & 2.8923 & 730.2204 & 728.7743 & 32 & 81 & 207 & 5.2 & 21 & \\
EBLM J1014-07 & 10 14 45.10 & -07 13 33.5 & 4.5575 & 823.7153 & 821.8634 & 31 & 46 & 41 & 5.45 & 24 & drift\\
EBLM J1023-43 & 10 23 58.03 & -43 25 26.6 & 3.6841 & 717.0993 & 715.2572 & 106 & 112 & 1515 & 3.88 & 16 & \\
EBLM J1034-29 & 10 34 18.90 & -29 48 55.3 & 2.1743 & 842.0065 & 840.9194 & 68 & 91 & 168 & 4.96 & 24 & \\
EBLM J1037-25 & 10 37 06.93 & -25 34 17.6 & 4.9366 & 769.231 & 766.8676 & 27 & 35 & 34 & 5.97 & 20 & \\
EBLM J1037-45 & 10 37 27.52 & -45 21 48.3 & 1.5939 & 701.3871 & 700.5902 & 93 & 107 & 9144 & 4.8 & 13 & \\
EBLM J1038-37 & 10 38 24.51 & -37 50 18.1 & 5.0217 & 1380.0845 & 1382.5916 & 35 & 131 & 0 & 3.89 & 13 & drift\\
EBLM J1104-43 & 11 04 34.88 & -43 14 25.1 & 1.7616 & 744.5903 & 743.7095 & 84 & 114 & 1095 & 3.92 & 18 & \\
EBLM J1105-13 & 11 05 27.67 & -13 53 02.1 & 3.9343 & 579.1865 & 577.2194 & 26 & 53 & 128 & 3.22 & 17 & \\
EBLM J1116-32 & 11 16 08.54 & -32 39 07.7 & 4.7456 & 546.9853 & 544.6125 & 53 & 101 & 71 & 6.79 & 22 & \\
EBLM J1116-01 & 11 16 44.43 & -01 52 07.5 & 7.3758 & 883.6445 & 879.9566 & 66 & 129 & 43 & 3.1 & 14 & \\
EBLM J1141-37 & 11 41 12.18 & -37 47 29.6 & 5.1477 & 908.7803 & 906.2064 & 18 & 28 & 65 & 6.57 & 21 & \\
EBLM J1146-42 & 11 46 50.49 & -42 36 59.4 & 10.4664 & 1453.7398 & 1458.9303 & 7 & 18 & 6 & 3.34 & 13 & triple\\
EBLM J1201-36 & 12 01 46.86 & -36 26 49.0 & 9.1131 & 1593.4912 & 1597.2211 & 16 & 19 & 0 & 3.73 & 15 & \\
EBLM J1208-29 & 12 08 41.33 & -29 39 46.8 & 2.676 & 742.9323 & 741.7599 & 61 & 69 & 1536 & 5.67 & 20 & \\
EBLM J1219-39 & 12 19 21.03 & -39 51 25.6 & 6.76 & 742.0458 & 738.8906 & 8 & 11 & 0 & 6.57 & 22 & \\
EBLM J1301-37 & 13 01 01.17 & -37 58 40.9 & 6.5498 & 1082.8007 & 1085.0376 & 73 & 147 & 394 & 5.76 & 13 & \\
EBLM J1305-31 & 13 05 05.91 & -31 26 13.3 & 10.6191 & 1348.5758 & 1353.6605 & 31 & 53 & 0 & 5.08 & 17 & \\
EBLM J1420-07 & 14 20 47.49 & -07 36 33.5 & 2.7039 & 749.1641 & 750.2994 & 60 & 82 & 296 & 4.1 & 20 & \\
EBLM J1431-11 & 14 31 52.15 & -11 18 40.4 & 4.4501 & 1038.1617 & 1035.9366 & 46 & 109 & 94 & 2.27 & 19 & \\
EBLM J1433-43 & 14 33 45.03 & -43 00 40.0 & 3.0825 & 1471.4781 & 1469.9368 & 83 & 117 & 192 & 2.61 & 16 & \\
EBLM J1436-13 & 14 36 46.42 & -13 32 35.5 & 3.9975 & 652.8486 & 650.8499 & 108 & 160 & 283 & 1.99 & 22 & \\
EBLM J1500-33 & 15 00 57.38 & -33 26 20.7 & 3.7382 & 584.4377 & 582.6763 & 104 & 147 & 191 & 3.98 & 25 & \\
EBLM J1509-10 & 15 09 40.97 & -10 52 10.3 & 6.8678 & 482.7147 & 479.2808 & 39 & 100 & 58 & 2.85 & 20 & \\
EBLM J1525-36 & 15 25 29.47 & -36 24 16.6 & 9.0089 & 508.6974 & 504.193 & 66 & 83 & 0 & 3.34 & 22 & \\
EBLM J1559-05 & 15 59 19.87 & -05 33 38.2 & 3.7601 & 1173.7 & 1171.82 & 26 & 50 & 106 & 4.86 & 18 & \\
EBLM J1630+10 & 16 30 25.64 & +10 09 29.8 & 10.9638 & 692.2887 & 697.2967 & 27 & 41 & 0 & 5.36 & 20 & drift\\
EBLM J1928-38 & 19 28 58.85 & -38 08 27.2 & 23.3227 & 1786.5841 & 1797.4422 & 13 & 14 & 0 & 1.93 & 17 & \\
EBLM J1934-42 & 19 34 25.69 & -42 23 11.6 & 6.3524 & 1887.4302 & 1884.254 & 24 & 79 & 0 & 1.26 & 14 & \\
EBLM J1944-20 & 19 44 00.13 & -20 51 05.1 & 2.7408 & 1043.408 & 1042.0376 & 116 & 145 & 1674 & 5.16 & 13 & \\
EBLM J1947-23 & 19 47 10.76 & -23 22 52.1 & 1.9196 & 1190.1808 & 1189.221 & 69 & 71 & 237 & 6.17 & 16 & drift\\
EBLM J2011-71 & 20 11 19.73 & -71 40 02.4 & 5.8727 & 1779.1339 & 1782.0374 & 4 & 6 & 0 & 2.09 & 23 & triple\\
EBLM J2025-45 & 20 25 25.48 & -45 49 45.0 & 6.192 & 1257.8296 & 1254.8425 & 13 & 25 & 0 & 5.45 & 36 & active\\
EBLM J2027+03 & 20 27 38.77 & +03 14 26.3 & 3.8397 & 1612.9585 & 1611.0386 & 60 & 95 & 86 & 3.9 & 15 & \\
EBLM J2040-41 & 20 40 41.58 & -41 31 59.8 & 14.4562 & 1716.1185 & 1710.5662 & 37 & 29 & 0 & 2.39 & 16 & \\
EBLM J2043-18 & 20 43 41.26 & -18 15 56.4 & 6.9114 & 1448.8266 & 1445.3709 & 79 & 136 & 80 & 3.92 & 15 & \\
EBLM J2046-40 & 20 46 38.09 & -40 32 19.2 & 37.0143 & 1273.6148 & 1282.1019 & 9 & 23 & 0 & 3.84 & 29 & triple\\
EBLM J2046+06 & 20 46 43.88 & +06 18 09.7 & 10.1078 & 846.0198 & 850.3199 & 16 & 20 & 0 & 6.08 & 14 & \\
EBLM J2101-45 & 21 01 02.24 & -45 06 57.4 & 25.5769 & 1519.1914 & 1507.7925 & 20 & 27 & 0 & 4.21 & 20 & \\
EBLM J2104-46 & 21 04 51.47 & -46 19 33.8 & 4.3573 & 400.2385 & 398.0653 & 43 & 92 & 0 & 5.05 & 20 & drift\\
EBLM J2107-39 & 21 07 11.07 & -39 45 58.7 & 3.9618 & 558.5945 & 556.6136 & 77 & 142 & 171 & 5.26 & 20 & \\
EBLM J2122-32 & 21 22 57.86 & -32 29 17.1 & 18.4214 & 1990.8124 & 1996.5523 & 13 & 10 & 0 & 1.14 & 13 & \\
EBLM J2153-55 & 21 53 16.54 & -55 59 07.2 & 8.5448 & 1256.3929 & 1252.1205 & 49 & 126 & 0 & 4.11 & 16 & \\
EBLM J2207-41 & 22 07 28.13 & -41 48 55.7 & 14.7748 & 1873.4177 & 1880.5037 & 9 & 9 & 0 & 2.17 & 13 & \\
EBLM J2210-48 & 22 10 48.74 & -48 53 26.3 & 2.8201 & 1736.2791 & 1734.8691 & 22 & 45 & 53 & 3.9 & 15 & drift\\
EBLM J2217-04 & 22 17 58.13 & -04 51 52.6 & 8.1553 & 868.0918 & 864.1819 & 42 & 57 & 0 & 5.29 & 15 & \\
EBLM J2232-31 & 22 32 30.29 & -31 14 39.1 & 3.1415 & 1750.7231 & 1749.1523 & 74 & 100 & 271 & 3.89 & 13 & drift\\
EBLM J2236-36 & 22 36 40.32 & -36 56 38.7 & 3.0672 & 588.0486 & 586.515 & 66 & 97 & 119 & 4.19 & 18 & \\
EBLM J2308-46 & 23 08 45.66 & -46 06 36.6 & 2.1992 & 712.6746 & 711.575 & 85 & 113 & 472 & 5.03 & 19 & \\
EBLM J2330-61 & 23 30 35.02 & -61 58 27.5 & 7.4572 & 1308.3033 & 1304.5747 & 50 & 136 & 101 & 3.69 & 16 & \\
EBLM J2349-32 & 23 49 15.23 & -32 46 17.5 & 3.5497 & 532.834 & 531.0591 & 41 & 74 & 113 & 5.1 & 20 & \\
EBLM J2353-10 & 23 53 46.73 & -10 53 05.9 & 4.5345 & 1632.1894 & 1629.9222 & 55 & 97 & 69 & 2.79 & 15 & \\

\end{longtable}
\end{landscape}

\section{Binary orbits and residuals}\label{app:orbits_and_residuals}
[Figures were removed for submission to arXiv, but will be visible on the version at the journal]

\section{Triple star systems}\label{app:triple}

\begin{landscape}
\begin{table}
\caption{Orbital parameters from the selected models for k2 fits}\label{tab:triple}
\centering 
\tiny
\begin{tabular}{cccccccccccccc}
\rowcolor{gray!50}
\hline
name & $P$ & $a\sin i_{\rm C}$ & $K$ & $e$ & $\omega$ & $T{\rm peri}$ & $f(m)$ & $m_{\rm A}$ & $m_{\rm B}$ & $m_{\rm C} \sin i_{\rm C}$ \\
 & [day] & [AU] & [km/s] &  & [deg] & [BJD-2,455,000] & [$10^{-3} M_{\odot}$] & [$M_{\odot}$] & [$M_{\odot}$] & [$M_{\odot}$] \\
\hline
EBLM J0543-57 inner binary &4.4638343(29)&0.0592(18)&16.6460(60)&$<$0.0018&$--$&903.33248(40)&2.1332(23)&1.23(10)&0.160(25)&--\\
EBLM J0543-57 tertiary orbit &3062(91)&4.98(25)&4.080(53)&0.426(12)&23.4(2.2)&2396(40)&15.9(1.1)&-- &-- &0.366(30)\\
\hline
EBLM J1146-42 inner binary &10.46644(16)&0.1158(43)&34.418(68)&0.0598(28)&96.1(3.4)&1453.898(98)&43.98(65)&1.35(14)&0.539(69)&--\\
EBLM J1146-42 tertiary orbit &259.83(24)&1.049(38)&7.76(14)&0.2194(88)&39.5(3.6)&1569.9(98)&11.68(64)&-- &-- &0.393(36)\\
\hline
EBLM J2011-71 inner binary &5.8727000(59)&0.0760(26)&23.6638(22)&0.03099(15)&-106.45(24)&1781.7851(40)&8.0513(61)&1.41(13)&0.285(42)&--\\
EBLM J2011-71 tertiary orbit &662.5(2.2)&1.815(64)&1.9869(31)&0.1008(36)&34.5(1.6)&1874.2(40)&0.5303(43)&-- &-- &0.1207(85)\\
\hline
EBLM J2046-40 inner binary &37.01426(33)&0.2350(58)&11.986(12)&0.47316(56)&155.771(61)&1276.0866(48)&4.515(28)&1.070(80)&0.193(14)&--\\
EBLM J2046-40 tertiary orbit &5583.663562(16)&7.3(1.6)&3.78(34)&0.500(44)&134(11)&2262(48)&20(11)&-- &-- &0.379(89)\\

\hline 

\hline 
\end{tabular}
\end{table}
\end{landscape}

\begin{figure}
\begin{center}
\begin{subfigure}[t]{0.49\textwidth}
\subcaption*{EBLM J0543-57 inner: $m_{\rm B} = 0.16M_{\odot}$, $P = 4.464$ d, $e = 0$}
\includegraphics[width=\textwidth,trim={0 0.6cm 0 0},clip]{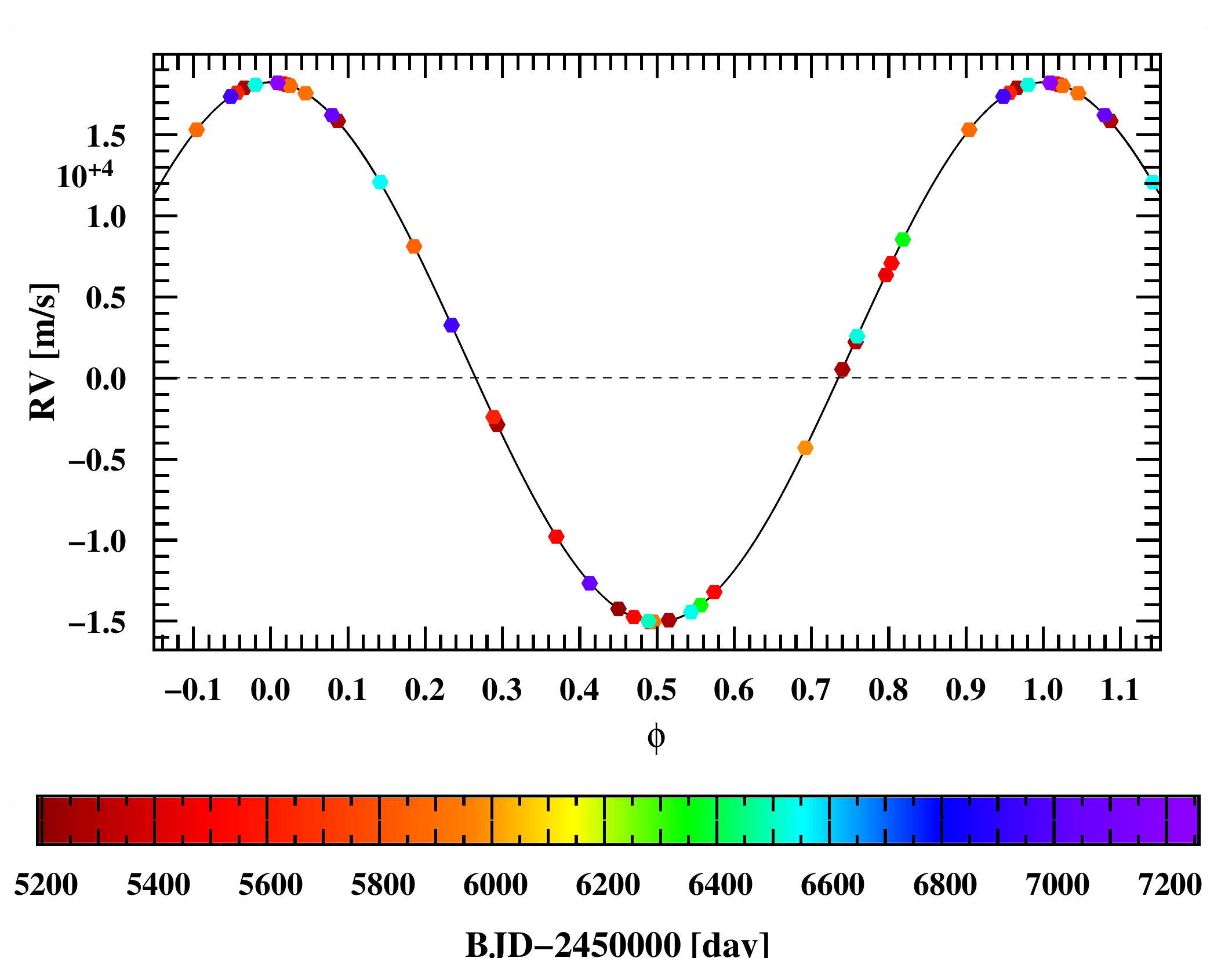}
\end{subfigure}
\begin{subfigure}[t]{0.49\textwidth}
\subcaption*{EBLM J0543-57 outer: $m_{\rm C} = 0.366M_{\odot}$, $P = 3062.039$ d, $e = 0.426$}
\includegraphics[width=\textwidth,trim={0 0.6cm 0 0},clip]{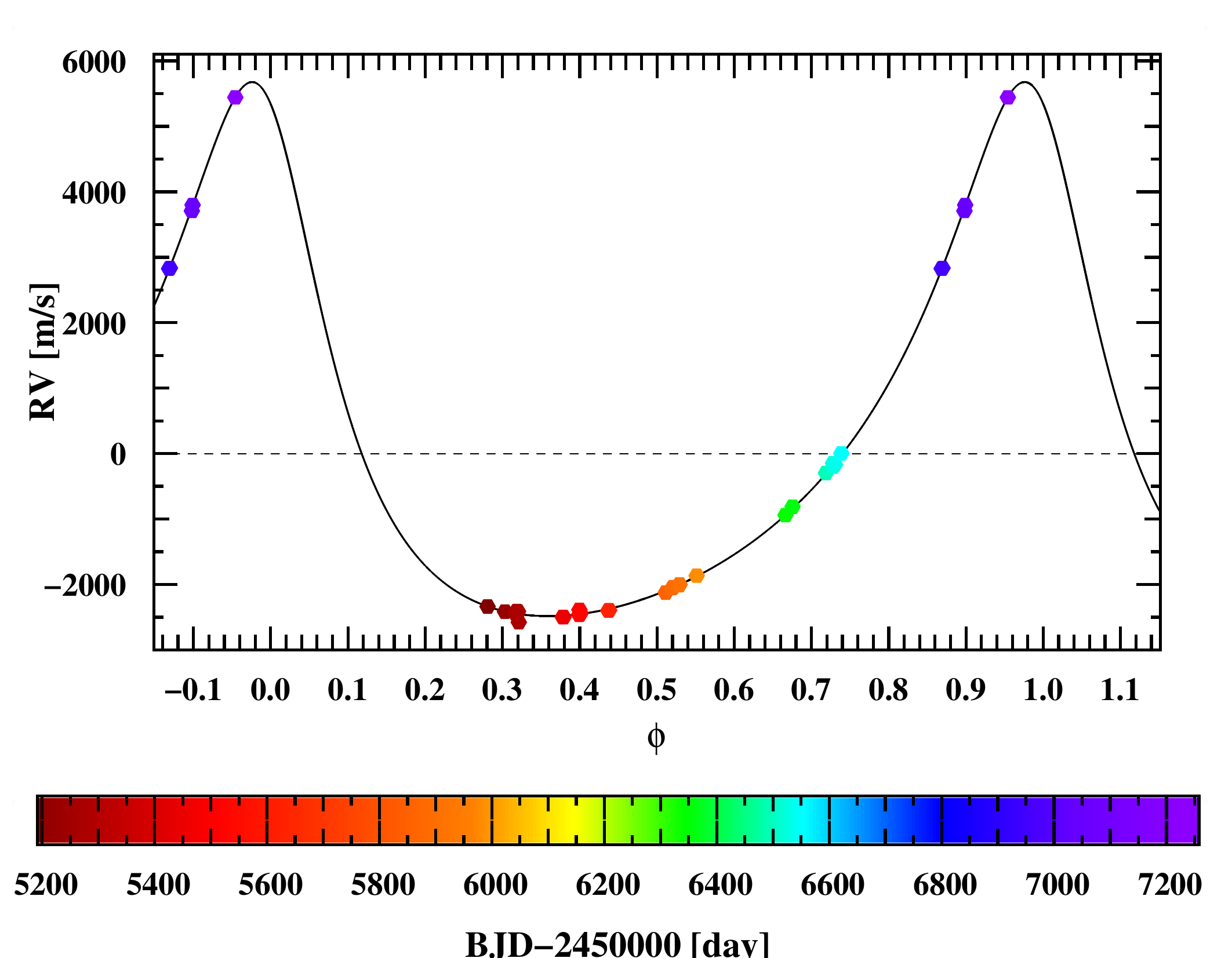}
\end{subfigure}
\begin{subfigure}[t]{0.49\textwidth}
\includegraphics[width=\textwidth,trim={0 0 2cm 0},clip]{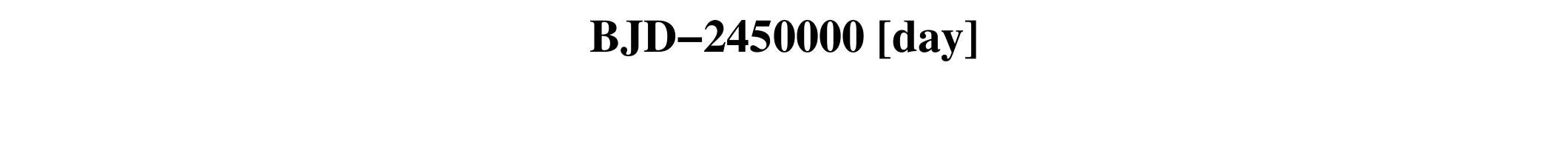}
\end{subfigure}
\begin{subfigure}[t]{0.49\textwidth}
\includegraphics[width=\textwidth,trim={0 0 2cm 0},clip]{bjd_bar.pdf}
\end{subfigure}
\begin{subfigure}[t]{0.49\textwidth}
\subcaption*{EBLM J0543-57 RVs over time and residuals}
\includegraphics[width=\textwidth,trim={0 10cm 0 1.2cm},clip]{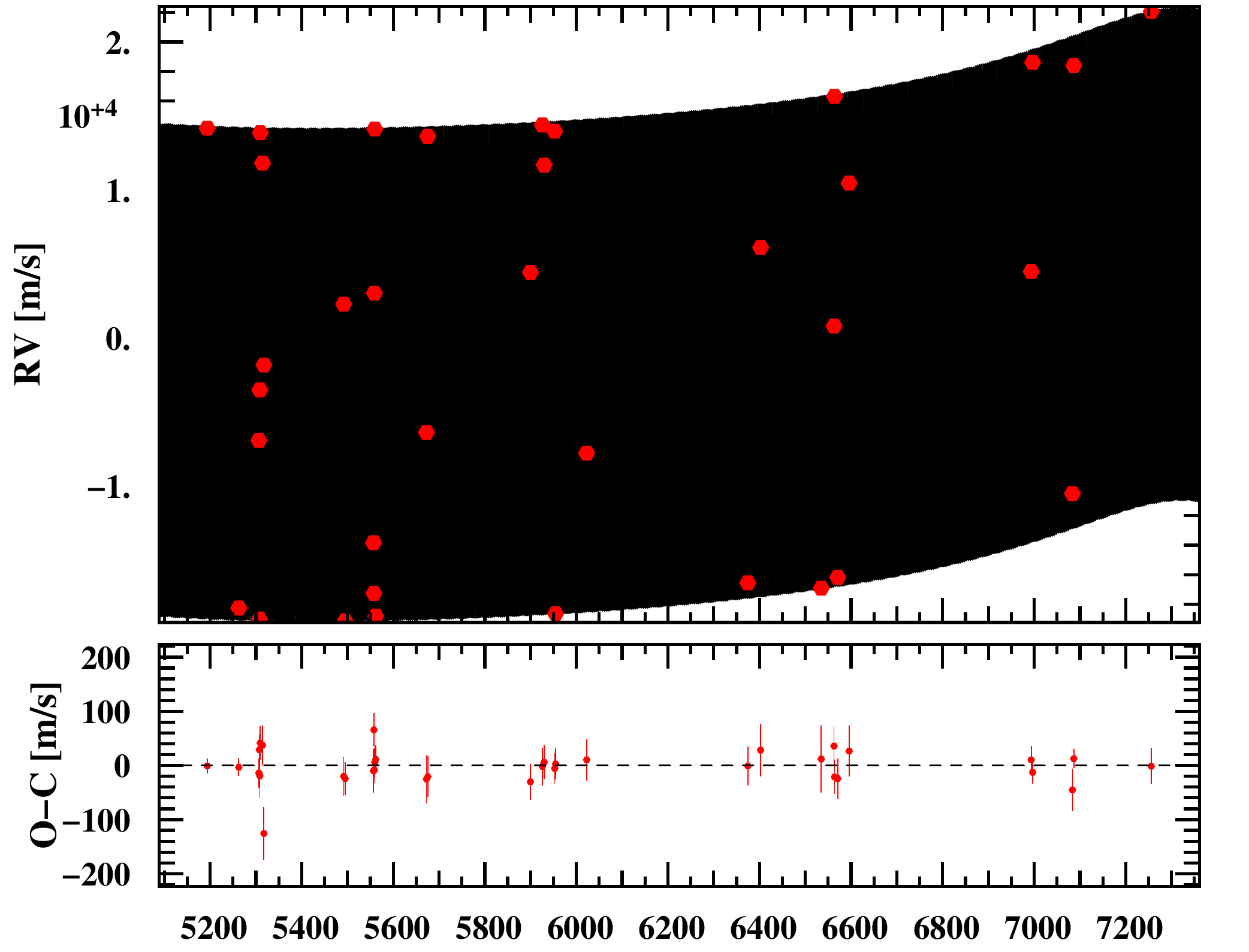}
\end{subfigure}
\begin{subfigure}[t]{0.49\textwidth}
\subcaption*{EBLM J0543-57 inner and outer orbit diagrams}
\includegraphics[width=\textwidth,trim={0 1cm 0 4cm},clip]{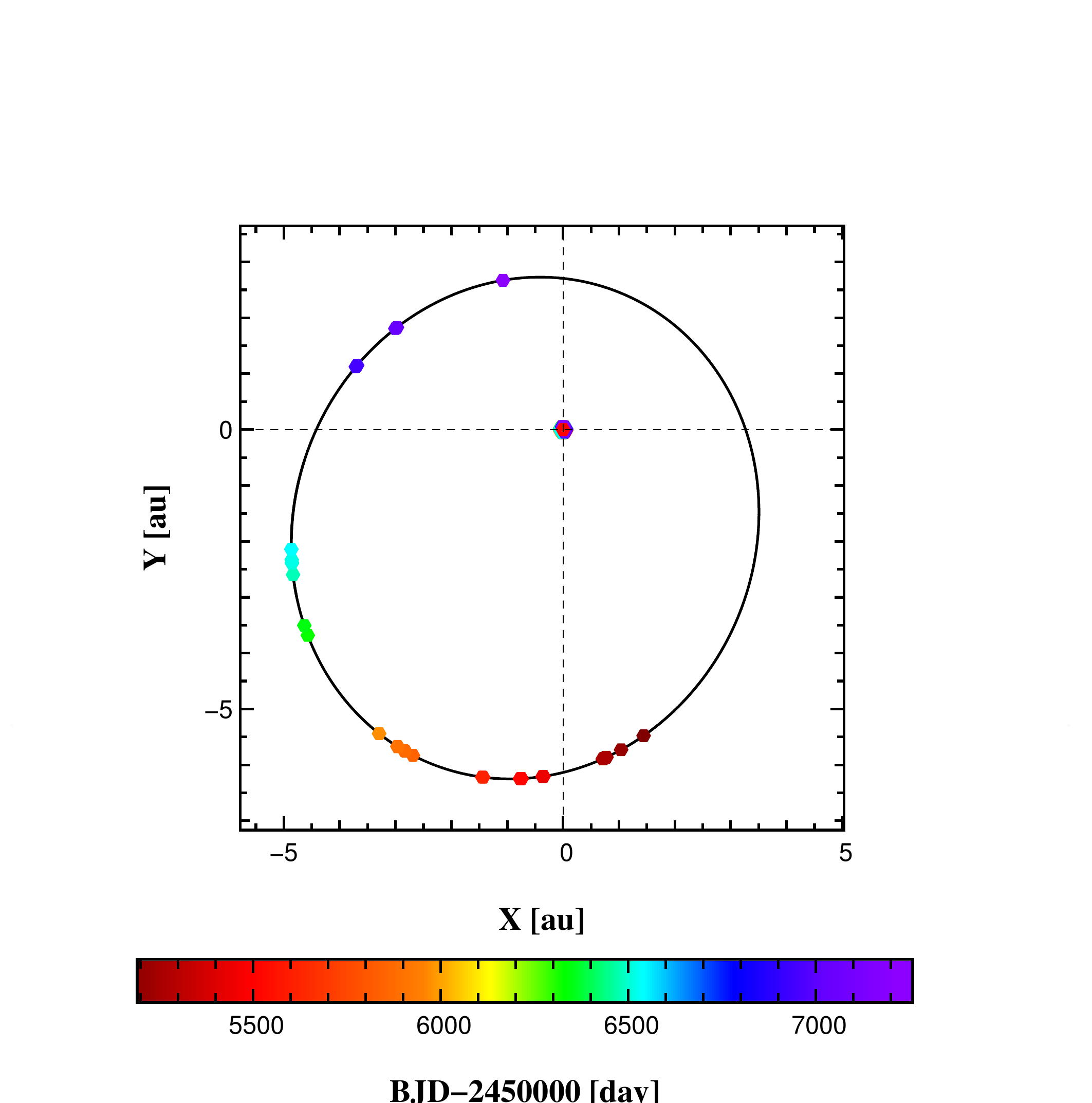}
\end{subfigure}
\begin{subfigure}[t]{0.49\textwidth}
\includegraphics[width=\textwidth,trim={0 0 2cm 0},clip]{bjd_bar.pdf}
\end{subfigure}
\begin{subfigure}[t]{0.49\textwidth}
\includegraphics[width=\textwidth,trim={0 0 2cm 0},clip]{bjd_bar.pdf}
\end{subfigure}
\end{center}
\end{figure}
\begin{figure}
\begin{center}
\begin{subfigure}[t]{0.49\textwidth}
\subcaption*{EBLM J1146-42 inner: $m_{\rm B} = 0.539M_{\odot}$, $P = 10.466$ d, $e = 0.06$}
\includegraphics[width=\textwidth,trim={0 0.6cm 0 0},clip]{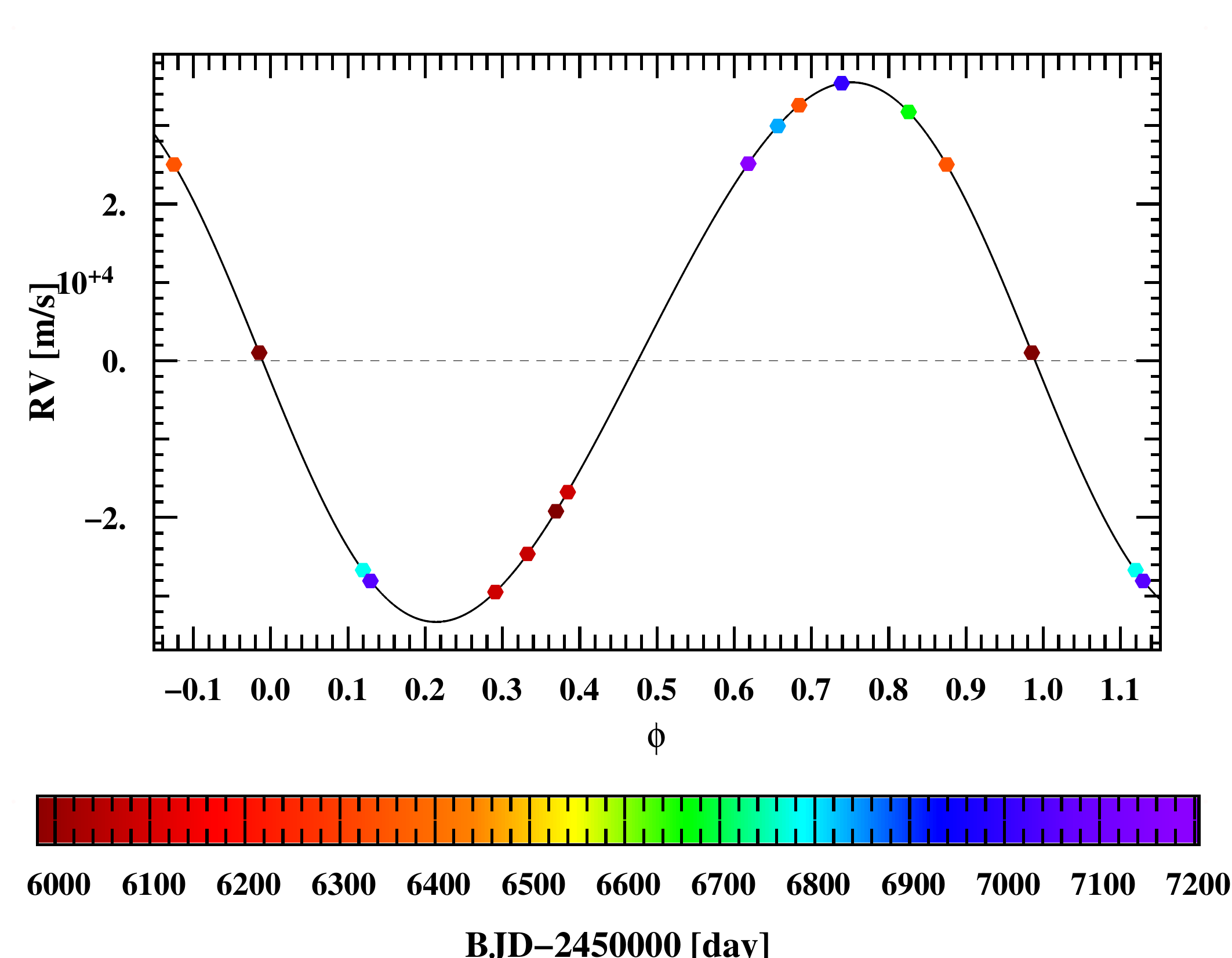}
\end{subfigure}
\begin{subfigure}[t]{0.49\textwidth}
\subcaption*{EBLM J1146-42 outer: $m_{\rm C} = 0.393M_{\odot}$, $P = 259.835$ d, $e = 0.219$}
\includegraphics[width=\textwidth,trim={0 0.6cm 0 0},clip]{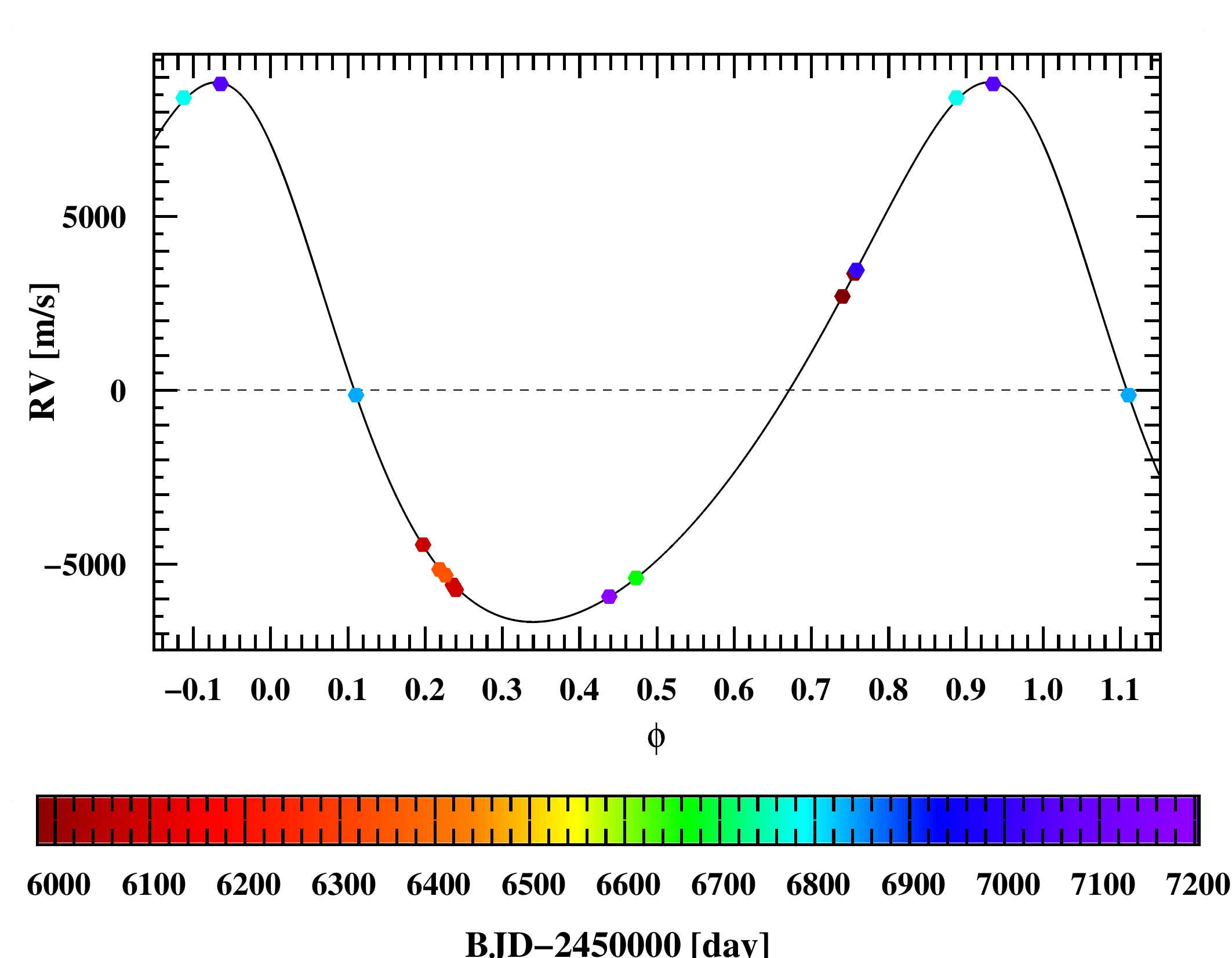}
\end{subfigure}
\begin{subfigure}[t]{0.49\textwidth}
\includegraphics[width=\textwidth,trim={0 0 2cm 0},clip]{bjd_bar.pdf}
\end{subfigure}
\begin{subfigure}[t]{0.49\textwidth}
\includegraphics[width=\textwidth,trim={0 0 2cm 0},clip]{bjd_bar.pdf}
\end{subfigure}
\begin{subfigure}[t]{0.49\textwidth}
\subcaption*{EBLM J1146-42 RVs over time and residuals}
\includegraphics[width=\textwidth,trim={0 10cm 0 1.2cm},clip]{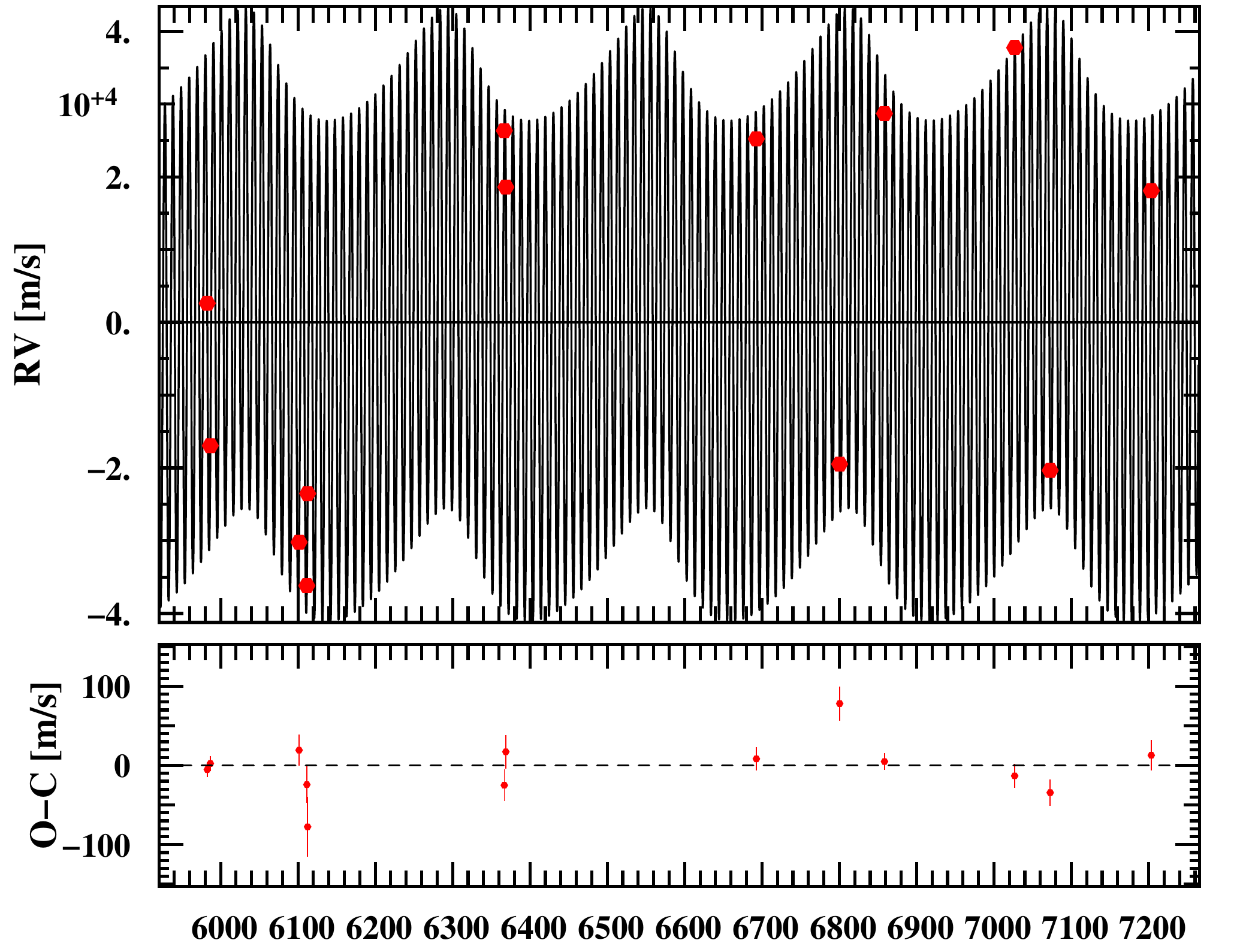}
\end{subfigure}
\begin{subfigure}[t]{0.49\textwidth}
\subcaption*{EBLM J1146-42 inner and outer orbit diagrams}
\includegraphics[width=\textwidth,trim={0 1cm 0 4cm},clip]{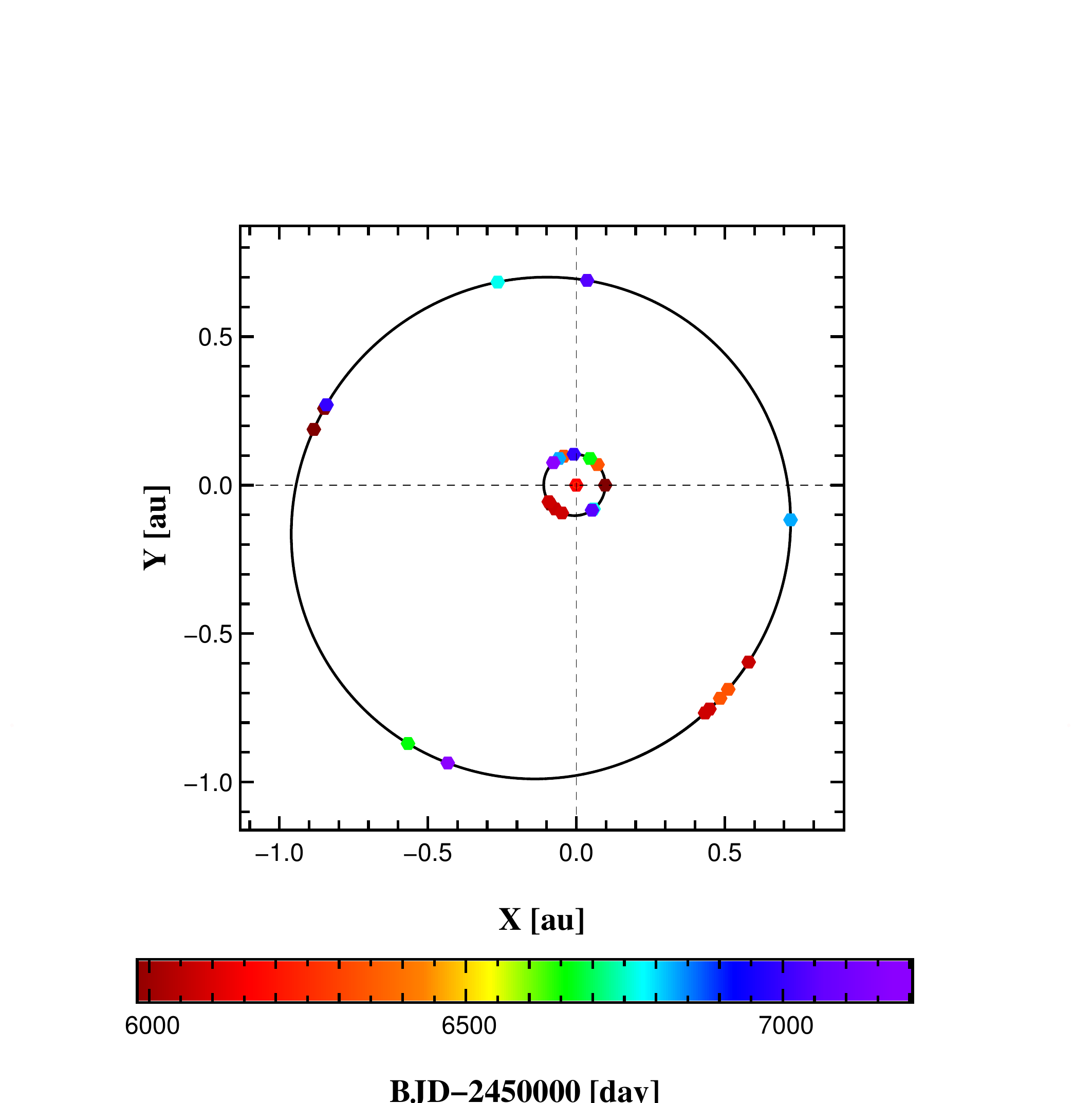}
\end{subfigure}
\begin{subfigure}[t]{0.49\textwidth}
\includegraphics[width=\textwidth,trim={0 0 2cm 0},clip]{bjd_bar.pdf}
\end{subfigure}
\begin{subfigure}[t]{0.49\textwidth}
\includegraphics[width=\textwidth,trim={0 0 2cm 0},clip]{bjd_bar.pdf}
\end{subfigure}
\end{center}
\end{figure}
\begin{figure}
\begin{center}
\begin{subfigure}[t]{0.49\textwidth}
\subcaption*{EBLM J2011-71 inner: $m_{\rm B} = 0.285M_{\odot}$, $P = 5.873$ d, $e = 0.031$}
\includegraphics[width=\textwidth,trim={0 0.6cm 0 0},clip]{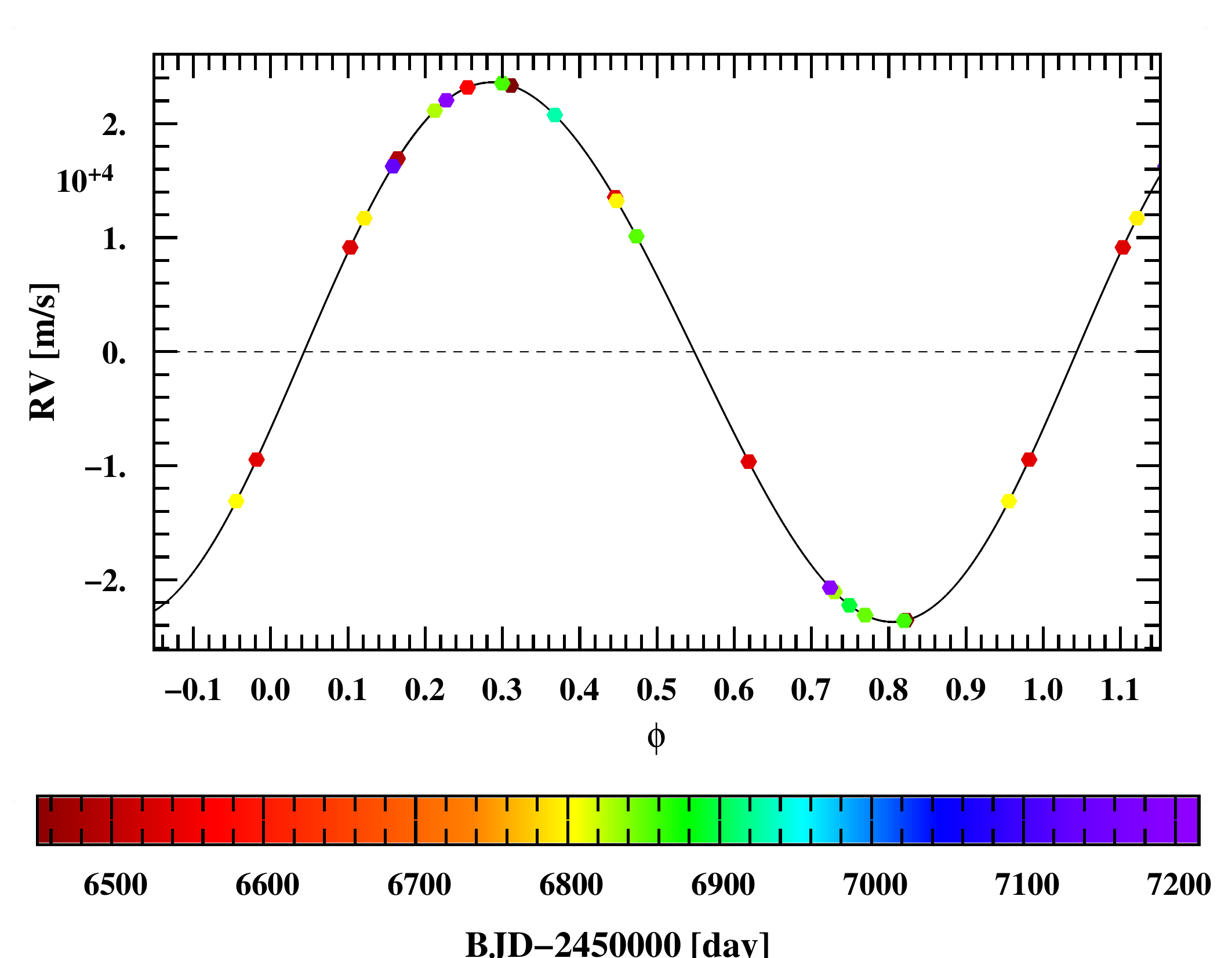}
\end{subfigure}
\begin{subfigure}[t]{0.49\textwidth}
\subcaption*{EBLM J2011-71 outer: $m_{\rm C} = 0.121M_{\odot}$, $P = 662.54$ d, $e = 0.101$}
\includegraphics[width=\textwidth,trim={0 0.6cm 0 0},clip]{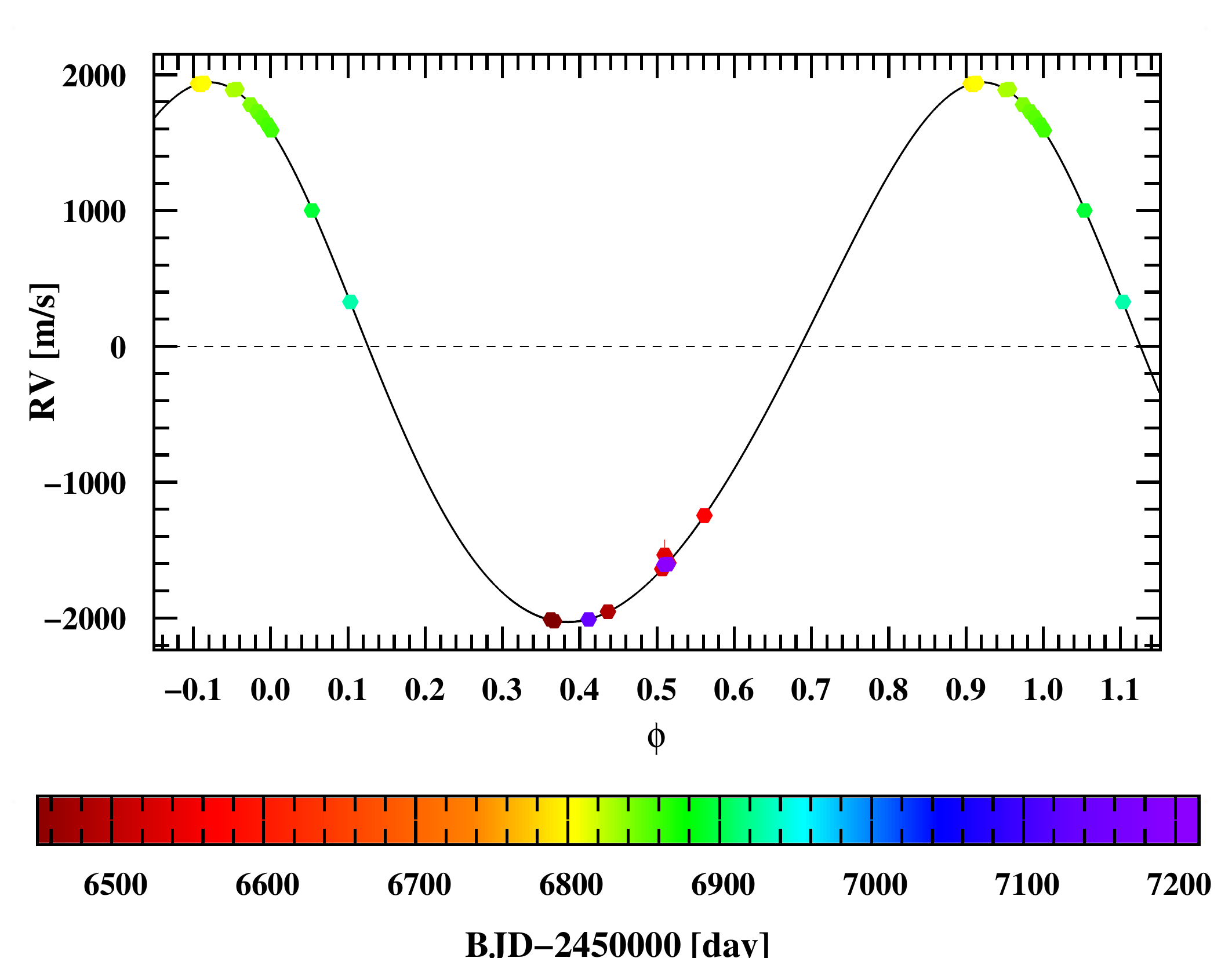}
\end{subfigure}
\begin{subfigure}[t]{0.49\textwidth}
\includegraphics[width=\textwidth,trim={0 0 2cm 0},clip]{bjd_bar.pdf}
\end{subfigure}
\begin{subfigure}[t]{0.49\textwidth}
\includegraphics[width=\textwidth,trim={0 0 2cm 0},clip]{bjd_bar.pdf}
\end{subfigure}
\begin{subfigure}[t]{0.49\textwidth}
\subcaption*{EBLM J2011-71 RVs over time and residuals}
\includegraphics[width=\textwidth,trim={0 10cm 0 1.2cm},clip]{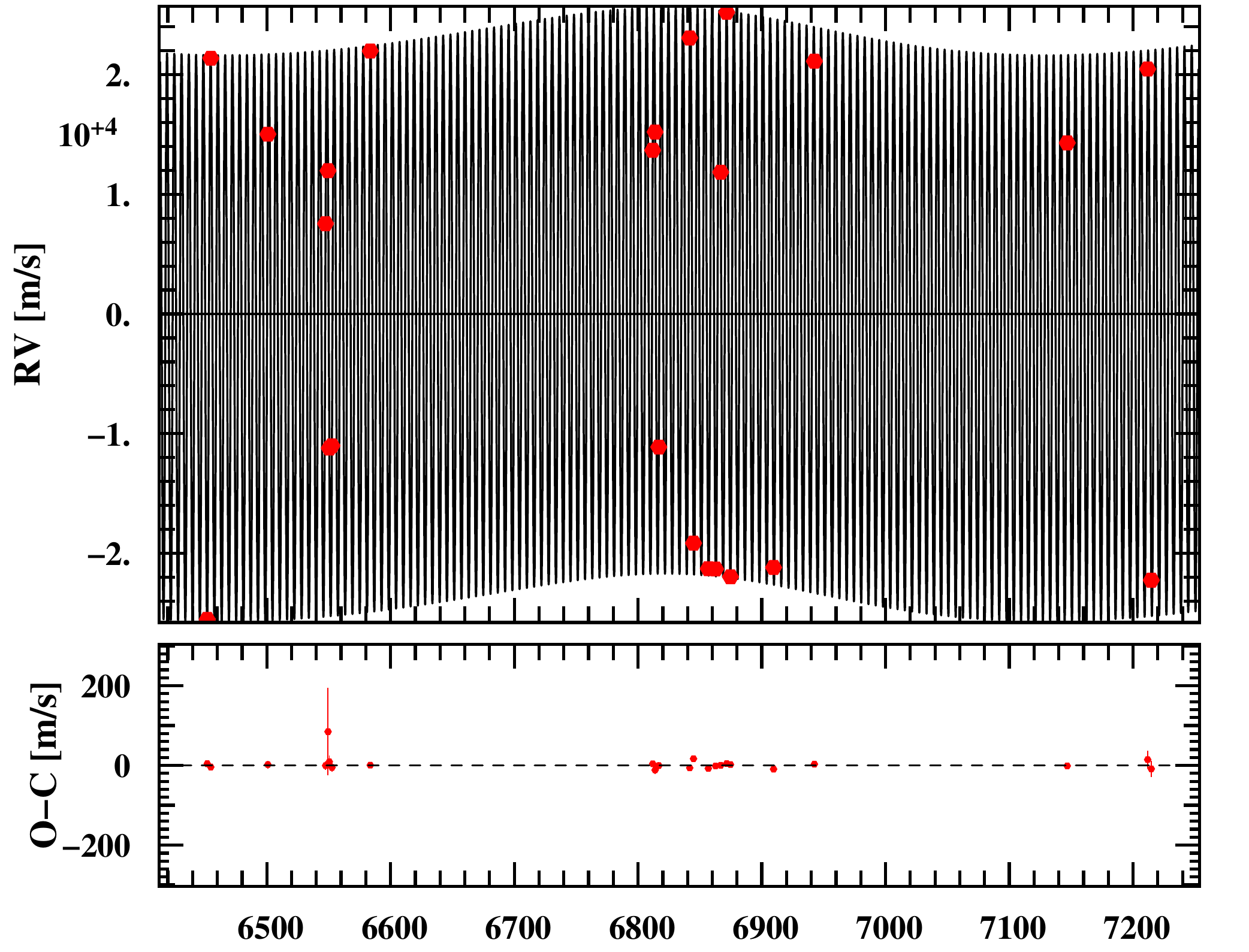}
\end{subfigure}
\begin{subfigure}[t]{0.49\textwidth}
\subcaption*{EBLM J2011-71 inner and outer orbit diagrams}
\includegraphics[width=\textwidth,trim={0 1cm 0 4cm},clip]{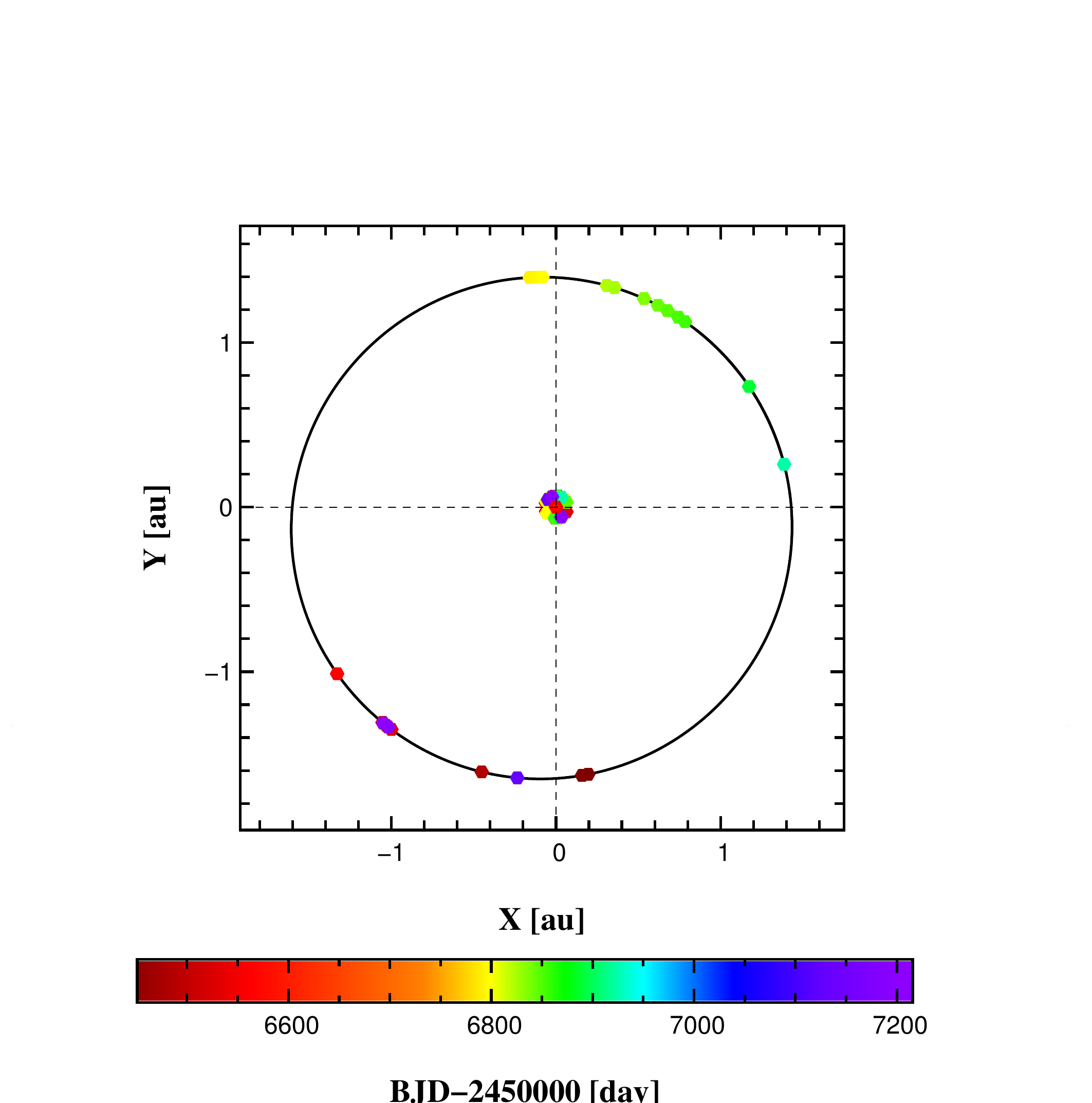}
\end{subfigure}
\begin{subfigure}[t]{0.49\textwidth}
\includegraphics[width=\textwidth,trim={0 0 2cm 0},clip]{bjd_bar.pdf}
\end{subfigure}
\begin{subfigure}[t]{0.49\textwidth}
\includegraphics[width=\textwidth,trim={0 0 2cm 0},clip]{bjd_bar.pdf}
\end{subfigure}
\end{center}
\end{figure}
\begin{figure}
\begin{center}
\begin{subfigure}[t]{0.49\textwidth}
\subcaption*{EBLM J2046-40 inner: $m_{\rm B} = 0.193M_{\odot}$, $P = 37.014$ d, $e = 0.473$}
\includegraphics[width=\textwidth,trim={0 0.6cm 0 0},clip]{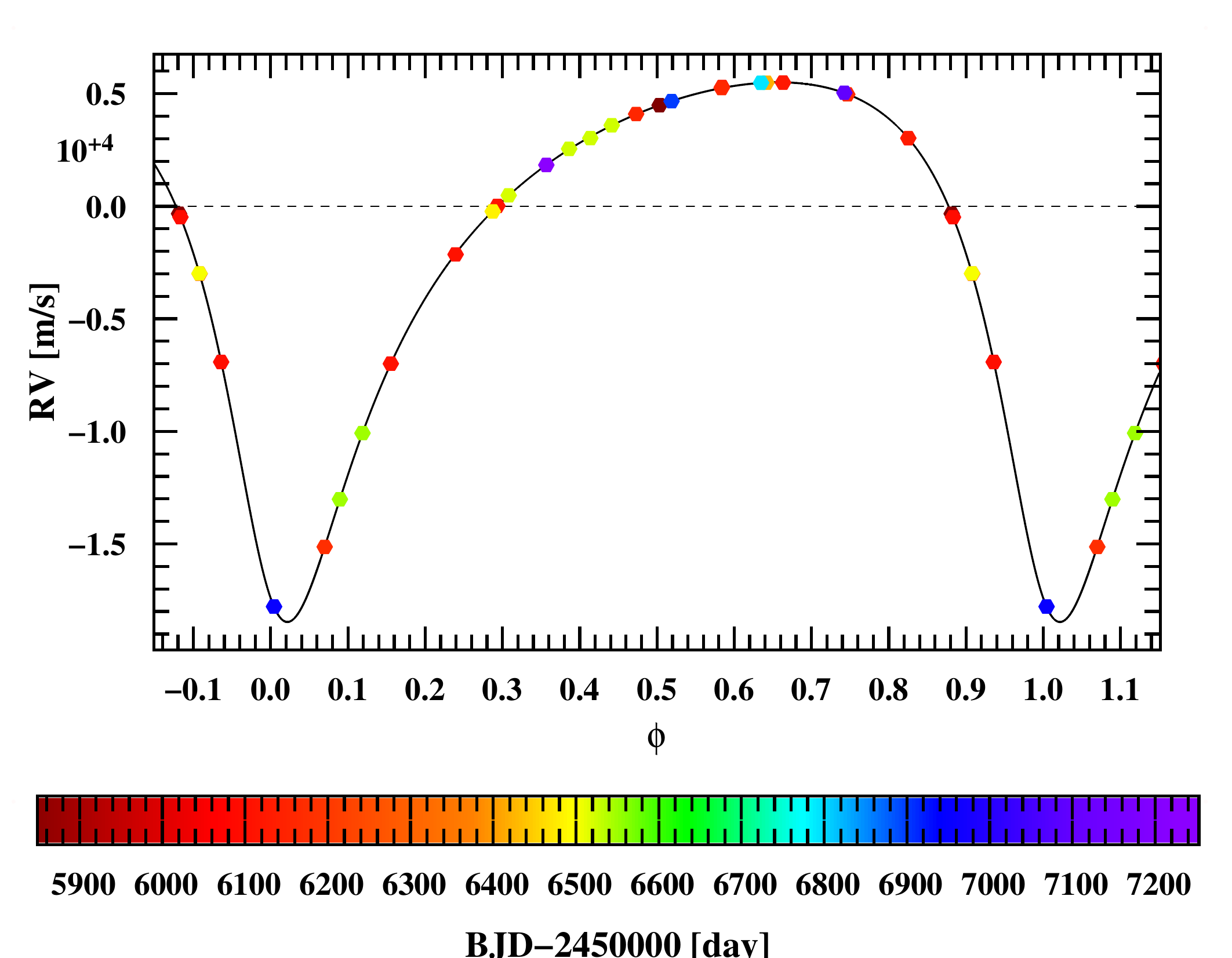}
\end{subfigure}
\begin{subfigure}[t]{0.49\textwidth}
\subcaption*{EBLM J2046-40 outer: $m_{\rm C} = 0.379M_{\odot}$, $P = 5583.664$ d, $e = 0.5$}
\includegraphics[width=\textwidth,trim={0 0.6cm 0 0},clip]{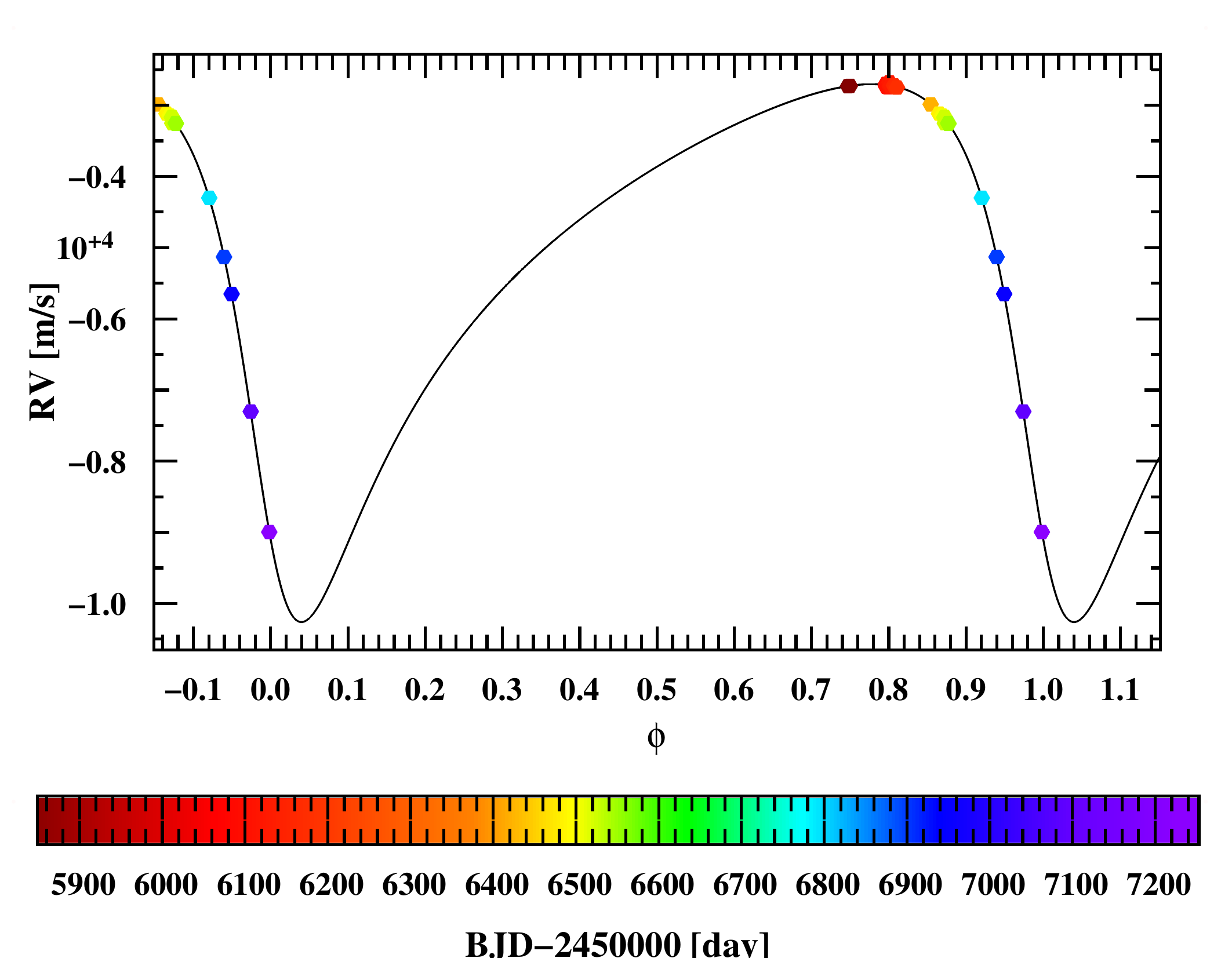}
\end{subfigure}
\begin{subfigure}[t]{0.49\textwidth}
\includegraphics[width=\textwidth,trim={0 0 2cm 0},clip]{bjd_bar.pdf}
\end{subfigure}
\begin{subfigure}[t]{0.49\textwidth}
\includegraphics[width=\textwidth,trim={0 0 2cm 0},clip]{bjd_bar.pdf}
\end{subfigure}
\begin{subfigure}[t]{0.49\textwidth}
\subcaption*{EBLM J2046-40 RVs over time and residuals}
\includegraphics[width=\textwidth,trim={0 10cm 0 1.2cm},clip]{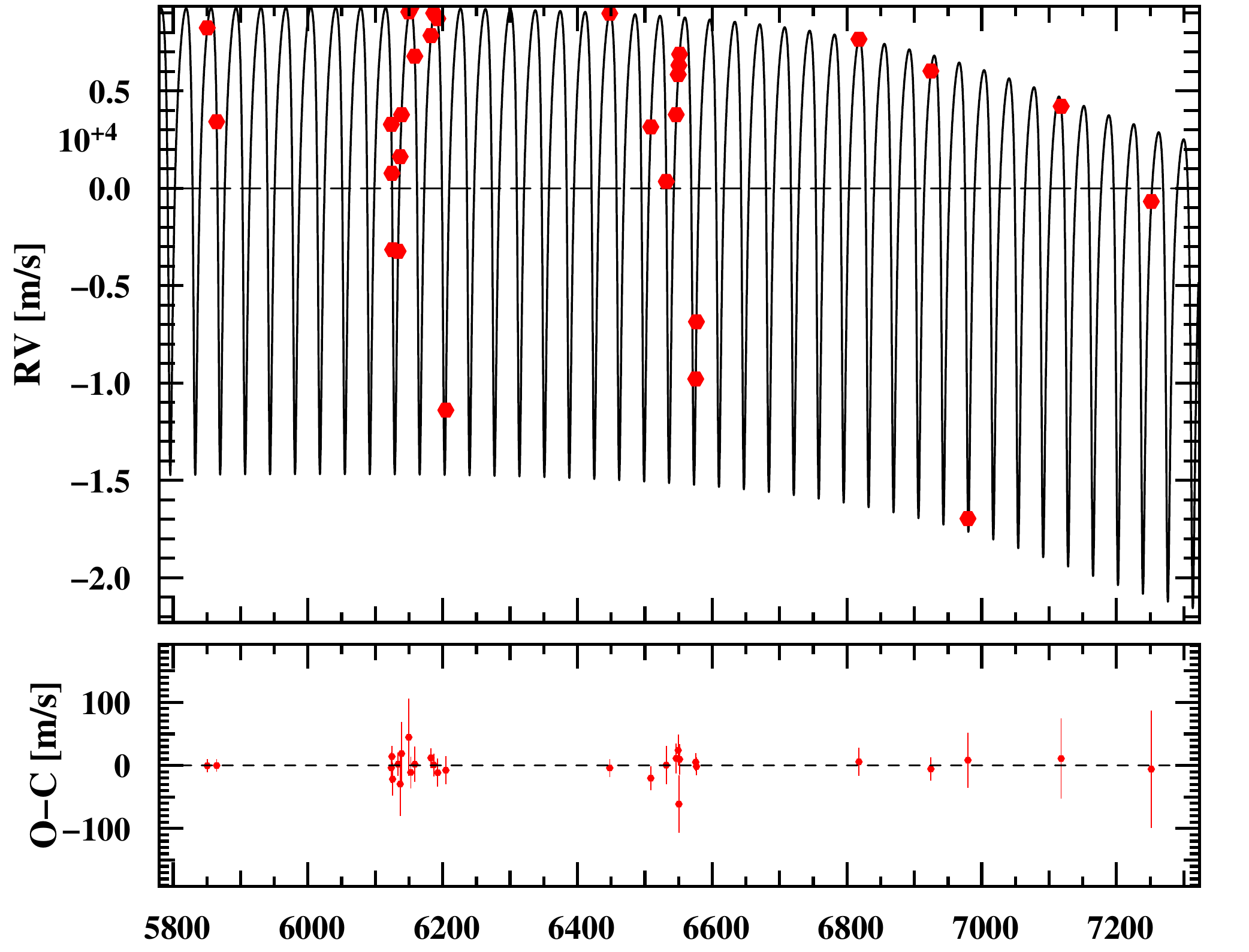}
\end{subfigure}
\begin{subfigure}[t]{0.49\textwidth}
\subcaption*{EBLM J2046-40 inner and outer orbit diagrams}
\includegraphics[width=\textwidth,trim={0 1cm 0 4cm},clip]{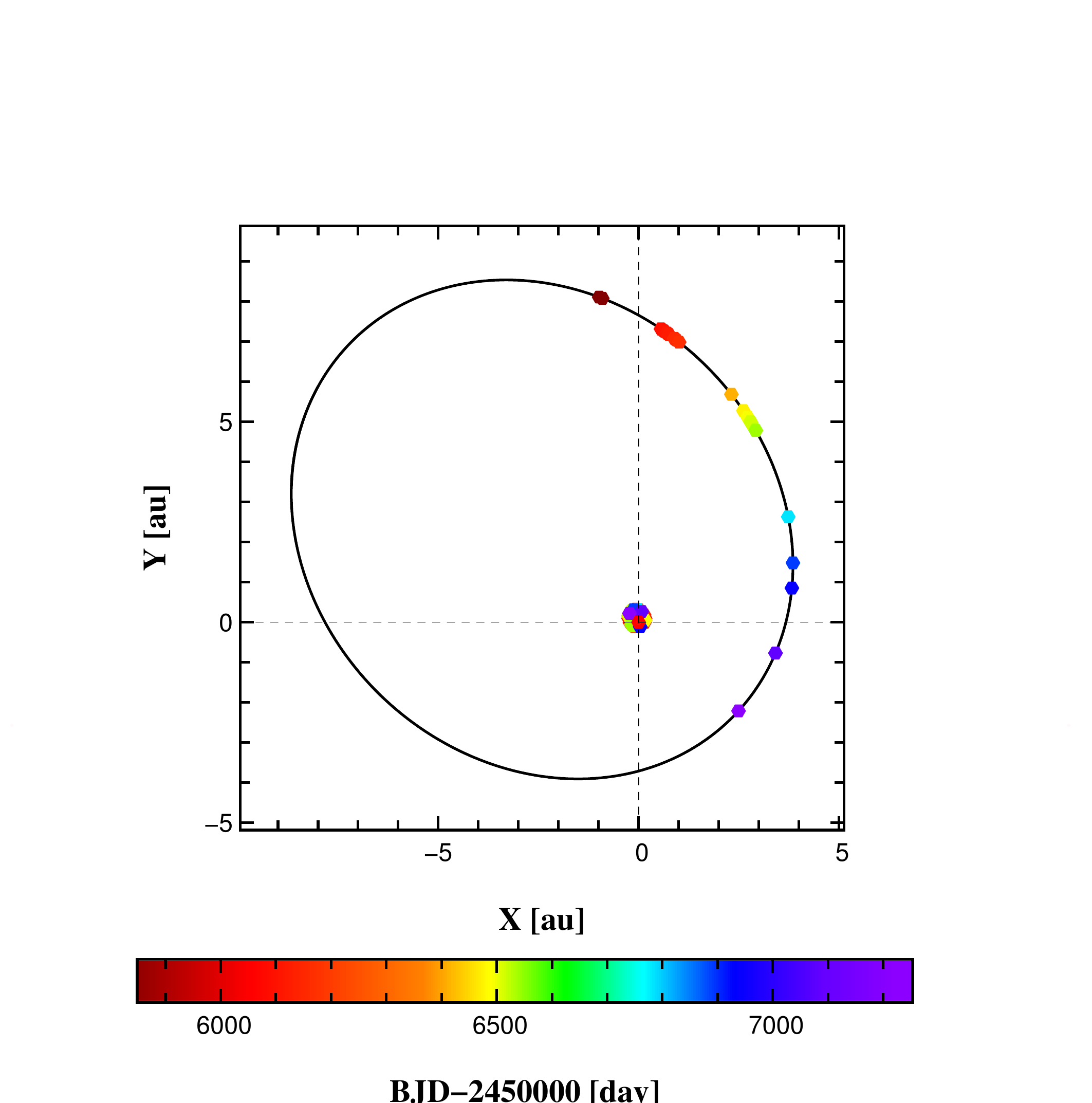}
\end{subfigure}
\begin{subfigure}[t]{0.49\textwidth}
\includegraphics[width=\textwidth,trim={0 0 2cm 0},clip]{bjd_bar.pdf}
\end{subfigure}
\begin{subfigure}[t]{0.49\textwidth}
\includegraphics[width=\textwidth,trim={0 0 2cm 0},clip]{bjd_bar.pdf}
\end{subfigure}
\end{center}
\end{figure}

\end{document}